\documentclass[10pt]{article}
\usepackage[margin=1in]{geometry}

\usepackage{color}
\usepackage{graphicx} % for pdf, bitmapped graphics files
\usepackage{amssymb}  % assumes amsmath package installed
\usepackage{amsthm}
\usepackage{amsmath,mathtools}

\allowdisplaybreaks[1]

\usepackage{float}
\usepackage{mwe}
\usepackage[latin1]{inputenc}
\usepackage{enumerate}
\usepackage[shortlabels]{enumitem}
\usepackage{algorithm}
\usepackage{algpseudocode}
\usepackage{caption,subcaption}%
\usepackage{needspace}
\usepackage[normalem]{ulem}
\numberwithin{equation}{section}
\usepackage{biblatex}
\usepackage[colorlinks=true, allcolors=blue]{hyperref}

\theoremstyle{plain}  % Bold name, italics font
\newtheorem{theorem}{Theorem}[section]

\newtheorem{lemma}[theorem]{Lemma}
\newtheorem{corollary}[theorem]{Corollary}
\newtheorem{proposition}[theorem]{Proposition}

\newcommand{\vast}{\bBigg@{4}}
\newcommand{\Vast}{\bBigg@{5}}
\newcommand\numberthis{\addtocounter{equation}{1}\tag{\theequation}}
\newcommand{\ER}{Erd\H{o}s-R\'{e}nyi }
\newcommand{\p}{\mathbb{P}}
\newcommand{\E}{\mathbb{E}}
\newcommand{\G}{\mathcal{G}}
\newcommand{\Gb}{\mathbb{G}}

\newcommand\distrib{\mathrel{\overset{\makebox[0pt]{\mbox{\normalfont\small\sffamily d}}}{=}}}

\def \R{ \mathbb{R}}
\def \N{ \mathbb{Z}_{\geq 0}}
\def \A{ \mathcal{A}}
\def \P{ \mathcal{P}}
\def \T{ \mathcal{T}}
\def \C{ \mathcal{C}}
\def \U{ \mathcal{U}}

\def \l{\left}
\def \r{\right}

\def \Parisi {\mathcal{P}}
\def \P {\mathbf{P}}
\def \Pt {\tilde{\mathbf{P}}}

\def \Ft {\tilde{\mathbf{F}}}
\def \q {\mathbf{q}}
\def \qt {\tilde{\mathbf{q}}}
\def \u {\mathbf{u}}
\def \ut {\tilde{\mathbf{u}}}

\def \rrt {\tilde{\mathbf{r}}}
\def \alphaa{\pmb{\alpha}}
\def \X{\mathbf{X}}
\def \x{\mathbf{x}}
\def \Y{\mathbf{Y}}

\def \Z{\mathbf{Z}}
\def \V{\mathcal{V}}
\def \plim{\normalfont \underset{{ n\to \infty}}{\text{ p-lim }}}
\newcommand{\plimd}{\mathop{\mathrm{p\text{-}lim}}\limits_{\substack{n\to\infty\\ \text{then } d\to\infty}}}
\newcommand{\plimnd}{\mathop{\mathrm{p\text{-}lim}}\limits_{\substack{n\to\infty\\ \text{then } d\to\infty}}}

\def \plimsup{\normalfont \underset{{ n\to \infty}}{\text{ p-limsup }}}
\def \b{\mathbf}

\title{\LARGE \textbf{Algorithmic Universality,  Low-Degree Polynomials, and Max-Cut in Sparse Random Graphs \\} }
\author{Houssam El Cheairi\thanks{MIT. Email: \texttt{houssamc@mit.edu}}, David Gamarnik\thanks{MIT. Email : \texttt{gamarnik@mit.edu}}}

\begin{document}
\date{\today}
\maketitle

\def \b{\mathbf}

\begin{abstract}
Universality, namely distributional invariance, is a   well-known property for many random structures. For example, it is known to hold for a broad range of variational problems with random input. Much less is known about the algorithmic  universality of specific methods for solving such variational problems.  Namely, whether algorithms tuned to specific variational tasks produce the same asymptotic behavior across different input distributions with matching moments.

In this paper, we establish algorithmic universality for a class of models, which includes spin glass models and constraint satisfaction problems on sparse graphs, provided that an algorithm can be coded as a low-degree polynomial (LDP). We illustrate this specifically for the case of the Max-Cut problem in sparse \ER graph $\Gb(n,d/n)$. We use the fact that the Approximate Message Passing (AMP) algorithm, which is an effective algorithm  for finding near-ground states of the Sherrington-Kirkpatrick (SK) model, is well approximated by an LDP.  We then establish our main universality result:  the performance of the LDP based algorithms exhibiting a certain connectivity  property, is the same  in the  mean-field (SK) and in the random graph $\Gb(n,d/n)$ setting, up to an appropriate  rescaling.  The main technical challenge we address in this paper is showing that the output of an LDP algorithm on $\Gb(n,d/n)$ is truly discrete, namely, that it is close to the set of points in the binary cube. This is achieved by establishing universality of coordinate-wise statistics of the LDP output across disorder ensembles, which implies that proximity to the cube transfers from the Gaussian to the sparse graph setting. Our result in particular recovers an analogue of the main result in~\cite{el2023local}. There, the authors construct a local near-optimal algorithm for the Max-Cut problem in regular graphs with large girth. Here, we establish a similar result for the $\Gb(n,d/n)$ random graph model.

\end{abstract}

\tableofcontents

%things to check
% 1. The constant case convention
% 2. Check impact of Lemma 2.6 modification

\section{Introduction and Problem Statement}\label{chapter2:section:1}

Universality in random structures refers to the phenomenon where macroscopic observables of a system do not depend on the precise distribution of the underlying randomness. Specifically, when basic parameters such as the first two moments are matched, large-scale quantities have the same limiting behavior. Classical examples of universality include the Central Limit Theorem~\cite{feller1971introduction}, as well as the Wigner semicircle law for random matrices~\cite{guionnet2010}. In this paper, we discuss universality for a broad class of discrete optimization problems on random structures. Namely, we consider optimizing Hamiltonians $H(\pmb{\sigma}, \X)$ over a search space $\mathbf{\Sigma}$, where $\b{X}$ represents a disorder matrix. When $\b{X}$ is sampled from a distribution $\mathbb{Q}$, the optimum value is a random variable denoted by ${\rm OPT}(\mathbb{Q}):= \max_{\pmb{\sigma} \in \mathbf{\Sigma}}H(\pmb{\sigma}, \X)$. Universality of values can be equivalently defined in this setting as ${\rm OPT}(\mathbb{Q}_1) \approx {\rm OPT}(\mathbb{Q}_2)$ when $\mathbb{Q}_1$ and $\mathbb{Q}_2$ have matching first and second moments (mild assumptions on higher moments might also be required). Here, $\approx$ means that the first-order asymptotic terms of both optima match.  To specify our discussion, we consider the task of optimizing the  quadratic Hamiltonian $H(\pmb{\sigma}, \X) = \pmb{\sigma}^\top \b{X} \pmb{\sigma}$ on the discrete hypercube $\b{\Sigma} = \{-1, 1\}^n$. 

Let $\mathbf{X}$ be an $n \times n$ random symmetric matrix belonging to the Gaussian Orthogonal Ensemble, denoted as ${\rm GOE}(n)$. Specifically, the entries are independent (up to symmetry) with mean zero, variance $1/n$ for off-diagonal entries, and variance $2/n$ for diagonal entries. When the diagonal terms $\b{X}_{ii}$ are set to zero instead of $\mathcal{N}(0, 2/n)$, we denote this disorder by ${\rm GOE}^0(n)$. Maximizing $H(\pmb{\sigma}, \X)$ with $\b{X}$ sampled from the ${\rm GOE}(n)$ ensemble corresponds to finding the ground state of the Sherrington-Kirkpatrick (SK) model. It is well-established that the normalized optimum $\max_{\pmb{\sigma}\in \b{\Sigma}} H(\pmb{\sigma}, \mathbf{X}) / 2n$ converges in probability as $n \to \infty$ to the \emph{Parisi} constant $\mathcal{P}^* \approx 0.76$~\cite{parisi1979infinite,talagrand2006parisi, panchenko2013sherrington}. The same result holds trivially when $\b{X}$ is sampled from ${\rm GOE}^0(n)$, as the contribution from diagonal terms  ($\sum_{i\in[n]} \b{X}_{ii}/2n$) in the Hamiltonian   vanishes for both ${\rm GOE}(n)$ and ${\rm GOE}^0(n)$. Moreover, the previous limit extends to disorders beyond Gaussian ensembles. If $\mathbf{Y}$ is sampled from a distribution matching the first two moments of $\mathbf{X}$ and satisfies mild tail conditions (e.g., a finite third moment), the normalized optimum $\max_{\pmb{\sigma}\in \b{\Sigma}} H(\pmb{\sigma}, \mathbf{Y})/2n$ converges in probability to the same limit $\mathcal{P}^*$. This result can be established via a standard Lindeberg interpolation argument~\cite{chatterjee2006generalization}, namely, by introducing a softmax smooth approximation of   the maximum $f(\b{X}) := \log\l(\sum_{\pmb{\sigma}\in \b{\Sigma}} \exp(\beta \pmb{\sigma}^\top \b{X} \pmb{\sigma})\r)/\beta$ for large inverse temperature parameter $\beta>0$. This interpolation argument allows us to extend the limit $\mathcal{P}^*$ to broader distributions $\Y$, including dense graphs $\Gb(n, c)$ where $c$ is a constant independent of $n$. Furthermore, the same principle also applies in more intricate settings when we consider sparse distributions $\Gb(n, d/n)$ with large $d$, and $\Gb(n, \tau_n/n)$ with $\lim_n \tau_n =\infty$. By bounding the interpolation error and applying standard concentration results, it can be shown that the limiting value remains $\mathcal{P}^*$ once the first and second moments are matched. For completeness, we provide a self-contained formal proof of this result in Section~\ref{chapter2:section:gound_state}.

To distinguish between the \ER graph model $\Gb(n, p)$ and its centered, normalized disorder variant, we use the notation $\G(n, t/n)$ to denote the distribution over symmetric random matrices $\b{Y}\in \mathbb{R}^{n\times n}$ with null diagonal terms $\b{Y}_{ii}=0$, and  independent, identically distributed upper triangular terms given by
\begin{align}
    \b{Y}_{ij} = -{\frac{1}{\sqrt{t (1- \frac{t}{n})}}}\l({\rm Bernoulli}\l(\frac{t}{n}\r)-\frac{t}{n}\r).\label{chapter2:eq:X-centered-rescaled}
\end{align}
Here, the minus sign is added to relate maximizing $H(\pmb{\sigma}, \b{Y})$ to finding extremal cuts on $\Gb(n, t/n)$ (more on this below). Two asymptotic regimes are of particular interest to us: the sparse regime where $t=d$ for some large $n$-independent constant $d$, and the diverging degree regime where $t=\tau_n$ with $\lim_{n} \tau_n = \infty$. Universality of the ground state energy in the former setting yields 
\begin{align}
 \plimd \frac{1}{2n}\max_{\pmb{\sigma} \in \{\pm 1\}^n}\pmb{\sigma}^\top\Y \pmb{\sigma}=\Parisi^*, \label{chapter2:eq:spin-sparse}
\end{align}
where $\plimd$ denotes the limit in probability when $n\to\infty$ first, followed by $d\to\infty$ (informally $1\ll d\ll n$). In the setting $t=\tau_n$,  the same result holds under the standard limit in probability $\plim$. 

We now show that optimizing the Hamiltonian $H$ is directly equivalent to finding extremal cuts in graphs. Consider the Max-Cut problem on a given graph $G=(V,E)$ with node set $V=\{1,2,\ldots,n\}$,  and edge set $E$. This can be formulated as the following constrained optimization task
\begin{equation}
\begin{aligned}
& \text{maximize}
& & \sum_{(i,j)\in E} \frac{1-\pmb{\sigma}_i \pmb{\sigma}_j}{2}   \\
& \text{subject to}
& &\pmb{\sigma} \in \{-1, 1\}^n.
\end{aligned}
\end{equation}
Observe that the objective function above can be expressed as $|E|/2-(1/4)H(\pmb{\sigma},\b{A}_G)$,
where $\b{A}_{G}$ is the adjacency matrix of $G$ with entries given by $[\b{A}_G]_{ij} = 1_{(i, j) \in E}$. Consequently, the Max-Cut problem can be reformulated as
\begin{equation}
\begin{aligned}
& \text{maximize}
& & \frac{|E|}{2} -\frac{1}{4} H(\pmb{\sigma}, \b{A}_G)    \\
& \text{subject to}
& &\pmb{\sigma} \in \{-1, 1\}^n.
\end{aligned}
\end{equation}
Thus, estimating the value of Max-Cut is equivalent to minimizing the Hamiltonian $H(\pmb{\sigma},\mathbf{A}_G)$ over the discrete hypercube $\{-1, 1\}^n$. In the particular case where $G$ is sampled from $\Gb(n, t/n)$, note that $\b{A}_G = \frac{t}{n}\b{J} - \sqrt{t(1-\frac{t}{n})} \b{Y}$, where $\b{J}_{ij} = 1_{i\neq j}$, and $\b{Y}$ is sampled from $\G(n, t/n)$. Therefore
\begin{align*}
    \frac{H(\pmb{\sigma}, \b{A}_G)}{n} &=t\l(\frac{2\sum_{1\leq i < j \leq n} \pmb{\sigma}_i \pmb{\sigma}_j}{n^2}\r) - \sqrt{t\l(1 - \frac{t}{n}\r)} \frac{\pmb{\sigma}^\top \b{Y} \pmb{\sigma}}{n}\\
    &= t\l(m^2(\pmb{\sigma}) - \frac{1}{n}\r) - \sqrt{t\l(1 - \frac{t}{n}\r)} \frac{H(\pmb{\sigma}, \b{Y})}{n},
\end{align*}
where $m(\pmb{\sigma}) \triangleq \sum_{i\in [n]} \pmb{\sigma}_i /n$ is the magnetization. When the search space is constrained to balanced cuts, i.e., cuts with $m(\pmb{\sigma})=0$, then, minimizing $H(\pmb{\sigma}, \b{A}_G)$ is equivalent to maximizing $H(\pmb{\sigma}, \b{Y})$. In fact, it is enough to restrict ourselves to balanced cuts as the discrepancy between unconstrained optimal cuts and balanced optimal cuts can be bounded with high probability (w.h.p.) as $t\to \infty$  by $n\sqrt{t}\varepsilon_t$, where $\lim_{t\to \infty}\varepsilon_t = 0$ (see Theorem 1.6 in~\cite{dembo2017extremal}). In particular,   identity (\ref{chapter2:eq:spin-sparse})  implies the following asymptotic value with high probability  for the Max-Cut on $\Gb(n,d/n)$ in the double limit regime $n\to \infty$ followed by $d\to \infty$ 
\begin{align*}
    \frac{{\text{MaxCut}}(\Gb(n, d/n))}{n} = \frac{d}{4} + \Parisi^* \sqrt{\frac{d}{4}}+ o_d(\sqrt{d}),
\end{align*}
and a similar result can be obtained on $\Gb(n, \tau_n/n)$ with $\lim_n \tau_n=\infty$. Henceforth, we restrict our attention  in the remainder of this paper to optimizing $H(\pmb{\sigma}, \cdot)$ over the Bernoulli ensembles $\G(n, t/n)$ with $d\leq t \leq \tau_n$, and the Gaussian ensemble ${\rm GOE}^0(n)$.

\subsection{Algorithmic Universality. The Main Result}
The previous section illustrates the universality of extremal cut values on random graphs. A totally different question is the universality of algorithms, which we now discuss. Suppose one runs an algorithm ${\rm ALG}$ for some input $\X=\X(\mathbb{Q})$ distributed  as $\mathbb{Q}$ and obtains a solution $\pmb{\sigma}_{\rm ALG}(\X)$. Suppose now $\Y$ is generated according to distribution $\mathbb{Q}'$ which is different from $\mathbb{Q}$, but shares the first two moments with $\mathbb{Q}$. Is it the case that 
 \begin{align}\label{chapter2:eq:universality}
 H(\pmb{\sigma}_{\rm ALG}(\Y), \b{Y})\approx H(\pmb{\sigma}_{\rm ALG}(\X), \b{X})~?
 \end{align}
 In particular, is $\pmb{\sigma}_{\rm ALG}(\Y)$  near optimal when $\pmb{\sigma}_{\rm ALG}(\X)$ is? Evidence supporting this conjecture is found in~\cite{el2023local}. Specifically, the authors of the latter work adapted the algorithmic framework developed in~\cite{montanari2019optimization} for the SK model, to the same model but defined now on sparse regular locally tree-like graphs. Both the original and the adapted algorithm are based on the so-called Approximate Message Passing (AMP) scheme (see the next section), which was found to be very effective in a broad range of models~\cite{feng2022unifying}. In~\cite{el2023local}, the authors reproduce the AMP scheme and  prove that the output of their algorithm is near optimal in the double limit $n\to\infty$ followed by $d\to\infty$.  In particular, (\ref{chapter2:eq:universality}) is validated in this setting for $\mathbb{Q}={\rm GOE}(n)$ and $\mathbb{Q}'$ the distribution of sparse regular tree-like graphs with large  degree $d$. This, however, required reproducing the optimality proof for their algorithm in the new setting $\mathbb{Q}'$. A similar universality result was established recently in the quantum setting in~\cite{basso2022performance}. There, the universality was also verified directly by computing the performance of an algorithm for the SK model on the one hand, and  sparse random graphs on the other hand,  and verifying that the results match. However, if the algorithmic universality (\ref{chapter2:eq:universality}) was known to hold for these models, then the need for reproducing the proof would have been  eliminated.
%%%%%%%%%%%%%%%% 
In this paper, we establish the algorithmic universality (\ref{chapter2:eq:universality}) for the specific class of algorithms known as Low-Degree Polynomials (LDP), exhibiting certain  connectivity properties to be defined.  Our universality result is established specifically for the SK and both the distribution of sparse graphs with large average degree $\G(n, d/n)$ and the distribution of graphs with diverging average degree $\G(n,\tau_n/n)$ (i.e., $\omega(1)\leq \tau_n\leq o(n)$), but we believe that our technique can be easily extended to a broader range of Hamiltonians $H$. We will elaborate on this after we state our main result below informally.  Next, we use the  fact that AMP can be approximated by an LDP-based algorithm. While variants of this are known in the literature \cite{montanari2022equivalence, ivkov2024semidefinite}, we prove this fact in this paper for completeness. Combining these two results, we obtain an LDP-based algorithm for solving the Max-Cut problem to near optimality on sparse graphs.

A key technical component of our analysis is a universality result at the level of coordinate-wise statistics of algorithm outputs. Namely, for tree-structured LDP algorithms, we establish universality of empirical averages of pseudo-Lipschitz test functions applied to the output coordinates across Gaussian and sparse random graph ensembles.  This result does not concern the Hamiltonian, which cannot be expressed as an empirical average, as it involves output interaction terms. Instead, its relevance to the Max-Cut problem is that it implies universality of the proximity of LDP outputs to the cube $[-1,1]^n$. Since the squared distance from $[-1,1]^n$ is a separable, coordinate-wise functional, closeness to the cube established in the Gaussian setting automatically transfers to sparse random graph models, which is crucial for our rounding argument.

Our analysis unfortunately does not carry over to regular graphs.  Specifically, while for a regular graph, we may also naturally  define rescaling variables $\b{Y}_{ij} = \frac{1_{(i,j)\in E} - \frac{d}{n-1}}{\sqrt{\frac{nd}{n-1}(1- \frac{d}{n-1})}}$, these terms are no longer independent, which creates difficulties in obtaining tight bounds on  moments $\E[\prod_{(i,j) \in E } {\b{Y}}_{ij}]$ for  subsets of edges $E$.

Before we state our main result informally, we comment on some existing literature on algorithmic universality, which to the best of our knowledge is rather limited. It is known that AMP is a universal algorithm when the denoisers are polynomials~\cite{bayati2015universality}. More recently, this universality was extended to Lipschitz denoisers~\cite{chen2021universality}. However, these results are established for dense AMP settings and $n$-independent disorder distributions. Therefore, they do not apply to sparse random graph models such as $\G(n,d/n)$ or $\G(n,\tau_n/n)$, nor to the AMP setting considered in~\cite{el2023local}. Similarly, universality of the Incremental AMP algorithm we consider in this paper is shown in~\cite{montanari2019optimization}, specifically for \ER graphs with constant edge probability, and the ${\rm GOE}(n)$ ensemble.

While the above references deal with standard random matrix ensembles, a recent work~\cite{dudeja2023universality} establishes universality for a class of semi-random matrices, which involve  less randomness. Their results apply to a restricted sub-class of memory-free AMP algorithms and rely on structural assumptions on the disorder matrix. In particular, their definition of semi-random matrices excludes sparse random graph ensembles such as $\G(n, d/n)$ and $\G(n,\tau_n/n)$ whenever $\tau_n=o(n^{1-\delta})$ for some $\delta\in (0, 1)$, since the  $\ell_\infty$ norm of these matrices scales as $d^{-1/2}$ and $\tau_n^{-1/2}$ , whereas their framework requires these norms to be bounded for any fixed $\varepsilon>0$ by $O(n^{-1/2 + \varepsilon})$.

Universality of AMP for some $n$-dependent distributions was established in~\cite{wang2024universality}. There, the authors show that state evolutions are universal for a broad class of generalized Wigner ensembles, including sparse random graphs $\Gb(n, \tau_n/n)$ with $\tau_n\gtrsim \log n$. Their results were derived by expanding the AMP iterations as products of random matrices and diagonal tensors, and using a moments-based method. In contrast, our universality results hold in the sparser settings $\Gb(n, d/n)$, and $\Gb(n, \tau_n/n)$ with $1\ll \tau_n \leq \log n$, and are derived for a broader class of algorithms that can be represented as tree-structured LDPs. Moreover, our proof technique is conceptually different, as we analyze the discrepancies from single entry perturbations instead of comparing the moments arising over two distinct disorder matrices. We leave it as a separate interesting question as to whether one can find a simple and direct argument proving universality of AMP in our $n$-dependent regimes.

As mentioned earlier, algorithmic universality was also established in the quantum setting for the SK and $\G(n,d/n)$ models  in the same context of ground-state and Max-Cut problems, for the class of quantum algorithms known as Quantum Approximate Optimization Algorithms (QAOA)~\cite{basso2022performance}. This was achieved by a direct computation of the performance of this algorithm in the two settings and verifying that the outputs are nearly identical. It is an interesting challenge to establish the same universality but from general principles. Unfortunately, the quantum nature of the QAOA algorithm precludes setting it up as an LDP, though a step in this direction was taken in~\cite{anshu2023concentration}. 

Regarding non-algorithmic/non-variational universality results, the literature is naturally much broader. Close to our setting,  universality for low-degree polynomials was  established in~\cite{mossel2010noise} in the context of the so-called Majority is Stablest conjecture. There the universality was also proven in a simpler $n$-independent setting, and for the case when the LDP is the value function itself  rather than an algorithmic tool (that is, not in a variational setting). One can view our result as a generalization of results in~\cite{mossel2010noise} to the more challenging $n$-dependent distributional setting.  Finally, a reference loosely related to algorithmic universality is~\cite{deift2014universality}, where the universality of the stopping time of a numerical procedure for solving a differential equation system is established. 

We now describe our main result informally. We begin by describing the class of LDP-based algorithms.  This is a class of algorithms viewed as mappings $\R^{n\times n}\to \{-1, 1\}^n$ constructed as follows. One fixes a vector $p=(p_i, i\in [n])$ of $n$ multivariate polynomials $p_i, i\in [n]$ in variables $\b{x}_{ij}, 1\le i< j\le n$.  The degree $\Delta={\rm Deg}(p)$ of this polynomial is the maximum  degree among all monomials appearing in any $p_i$. Let $\x=(\x_{ij}, 1\leq i<j\le n)$.  The vector $p$ induces a mapping $\R^{n\times n}\to \R^n$ via $\x\to p(\x)=(p_i(\x), 1\le i\le n)$.

The projection $\R^n\to \{-1, 1\}^n$ is obtained first by truncating values outside $[-1,1]$ to $\pm 1$ and then ``pushing'' values inside $(-1,1)$ to $\pm 1$ sequentially while increasing the objective value (we specify these operations in the next section). The resulting map is denoted by $\pmb{\sigma}({\rm ALG}(p),\x)$. We say that $p$ is a low-degree (LDP)   connected tree-based polynomial if (a) $\Delta$ is bounded by a $d$- and $n$-independent  constant, (b) the degree of every variable $\b{x}_{ij}$ in every monomial is at most $2$,  and (c) the graphs on the node set $[n]$ induced by the monomials in $p_i, i\in [n]$ are connected trees. Part  (c) will be made more precise in the next section.  We now state informally the main result of this paper. %We denote by $\plim$ convergence in probability as $n\to \infty$.

\begin{theorem}[Informal]\label{chapter2:thm:main}
Let $\X$ be sampled from the ensemble ${\rm GOE}^0(n)$, and $\Y$ sampled from $\G(n, t/n)$, where $d\leq t\leq \tau_n$, and $\tau_n$ satisfies $1 \ll \tau_n \ll n$. Let $p=(p_i, i\in [n])$ be a  connected tree-based LDP with $p_i \in \V_i$ for $i\in[n]$, and let $\mathcal{P}$ be a constant. Then, the following equivalence holds
\begin{align*}
\plim {\frac{1}{2n}}\pmb{\sigma}({\rm ALG}(p),\X)^\top \X \pmb{\sigma}({\rm ALG}(p),\X)\geq\mathcal{P}, 
\end{align*}
if and only if 
\begin{alignat}{2}
\plimd {\frac{1}{2n}}\pmb{\sigma}({\rm ALG}(p),\Y)^\top \Y\pmb{\sigma}({\rm ALG}(p),\Y)&\geq\mathcal{P}.\nonumber% \quad &&\text{for} \quad \Y\distrib \G(n, t/n),
%\plim {\frac{1}{2n}}\pmb{\sigma}({\rm ALG}(p),\Y)^\top \Y\pmb{\sigma}({\rm ALG}(p),\Y)&\geq\mathcal{P}, \quad &&\text{for} \quad \Y\distrib \G(n, \tau_n/n).
\end{alignat}
\end{theorem}
Under the following ``no overlap gap'' condition from \cite[Assumption~1]{montanari2019optimization}: letting $\mu_\beta$ denote the Parisi measure of the SK model at inverse temperature $\beta$, there exists $\beta_0<\infty$ such that for all $\beta>\beta_0$ the function $t \mapsto \mu_\beta([0,t])$ is strictly increasing on $[0,q_*(\beta)]$ and $\mu_\beta([0,q_*(\beta)])=1$, approximate message passing algorithms are known to be optimal for SK optimization. This condition is widely believed to hold in the statistical physics literature \cite{mezard1987spin}. Throughout this paper, we suppose it holds. We note that \cite{montanari2019optimization} also refers to this condition as ``full (or continuous) replica symmetry breaking''. We avoid this terminology since ``full-RSB'' is often used elsewhere to mean  that the Parisi measure is not finitely atomic. The above condition is stronger and in particular it implies that $\mu_\beta$ is not finitely atomic. The next result states that roughly speaking, connected tree-based LDPs achieve near optimal value $\mathcal{P}^*$.

\begin{theorem}[Informal]\label{chapter2:thm:2}
For every $\varepsilon>0$ there exists a  connected tree-based LDP $p=(p_i, i\in[n])$ such that 
\begin{align*}
\plim {\frac{1}{2n}}\pmb{\sigma}({\rm ALG}(p),\X)^\top \X \pmb{\sigma}({\rm ALG}(p),\X)\geq \mathcal{P}^*-\varepsilon,
\end{align*}
where $\X$ is as in Theorem~\ref{chapter2:thm:main}.
\end{theorem}
Combining the previous two theorems, we obtain the following corollaries.
\begin{corollary}\label{chapter2:cor:LDP-near-opt}
In the setting of Theorem  \ref{chapter2:thm:main}, for every $\varepsilon>0$, there exists  a connected tree-based LDP such that
\begin{alignat}{2}
\plimd {\frac{1}{2n}}\pmb{\sigma}({\rm ALG}(p), \Y)^\top \Y \pmb{\sigma}({\rm ALG}(p),\Y)&\geq \mathcal{P}^* - \varepsilon,\nonumber % \quad &&\text{if} \quad \Y \distrib \G(n, d/n),\\
%\plim {\frac{1}{2n}}\pmb{\sigma}({\rm ALG}(p), \Y)^\top \Y \pmb{\sigma}({\rm ALG}(p),\Y)&\geq \mathcal{P}^* - \varepsilon, \quad &&\text{if} \quad \Y \distrib \G(n, \tau_n/n),
\end{alignat}
where  $\Y$ is as in Theorem~\ref{chapter2:thm:main}. In particular, if  ${\rm CUT}(\pmb{\sigma}({\rm ALG}(p), {\Y}))$ is the cut value obtained from $\pmb{\sigma}({\rm ALG}(p), {\Y})$, then
\begin{alignat}{2}
    \plimd \frac{{\rm CUT}(\pmb{\sigma}({\rm ALG}(p), {\Y}))}{n\sqrt{t}} - \frac{\sqrt{t}}{4} - \frac{\Parisi^*}{2}  &\geq -\varepsilon.\nonumber %\quad &&\text{if} \quad \Y \distrib \G(n, d/n),\\
    %\plim \frac{{\rm CUT}(\pmb{\sigma}({\rm ALG}(p), {\Y}))}{n\sqrt{\tau_n}} - \frac{\sqrt{\tau_n}}{4} - \frac{\Parisi^*}{2}  &\geq -\varepsilon, \quad &&\text{if} \quad \Y \distrib \G(n, \tau_n/n).
\end{alignat}
\end{corollary}
Next, we describe the proof techniques underlying our main results.  The first element of the proof of  our main result, Theorem~\ref{chapter2:thm:main} is the Lindeberg interpolation method.  This technique  is based on expanding the underlying function using Taylor expansions, taking advantage of the matching of the first two moments, and then showing that the terms depending on higher moments are vanishing.  This needs to be done both for the objective function $H$ evaluated at the solution produced by the LDP, and importantly, for the constraints which ensure that the solution remains close to $[-1,1]^n$.  While it is relatively easy to show that the objective value remains roughly the same in expectation when switching from the Gaussian to the Bernoulli distribution underlying both $\G(n, d/n)$ and $\G(n,\tau_n/n)$ models for the LDP based solutions, upgrading this comparison to a high-probability statement is more challenging. Moreover, it is worth noting that an $o(n)$ bound on the expected optimality gap $\E\l[\max_{\pmb{\sigma} \in \{-1,1\}^n} H(\pmb{\sigma}, \Y)- H(\pmb{\sigma}({\rm ALG}(p), \Y), \Y)\r]$ would already imply an $o(n)$ gap with high probability by Markov's inequality, since $H(\pmb{\sigma}({\rm ALG}(p), \Y), \Y)\leq \max_{\pmb{\sigma} \in \{-1,1\}^n} H(\pmb{\sigma}, \Y)$ pointwise. Separately, establishing that the vector of solutions remains close to $[-1,1]^n$ with high probability is also challenging. We resolve this by introducing sigmoid-type penalty functions, one penalty for deviating under the target lower bound $\mathcal{P}$ in the evaluation of the Hamiltonian, and one penalty for deviating away from the cube $[-1,1]^n$, which will also require a smooth approximation of the distance function from $[-1,1]^n$. We then conduct interpolation analyses on these penalties to obtain the results.   These interpolation arguments also lead to a general universality result for coordinate-wise statistics of LDP outputs (Theorem~\ref{chapter2:lemma:state_ev}), which is then used to control proximity to $[-1,1]^n$.

While the derivation of our bounds is technically involved, it is fairly elementary and simple conceptually. Also, importantly, this analysis is conducted generically,  \emph{regardless} of the details of the algorithm giving rise to the underlying LDP. Namely, the analysis of the algorithmic performance is also  universal in some sense.

We also stress that while our proof is laid out for the special case of the objective function $H$ of the form $H(\pmb{\sigma},\X)=\pmb{\sigma}^\top \X \pmb{\sigma}$, our proof technique applies to a broader set of objectives, so long as they can be  written as polynomials in $\pmb{\sigma}$ themselves (the case above corresponding to the quadratic case). In particular, we anticipate that the same proof goes through for the case of  $p$-spin models and their counterparts without change.

The proof of Theorem~\ref{chapter2:thm:2} is not particularly novel and versions of it can be found in~\cite{montanari2022equivalence, ivkov2024semidefinite}. Although it is quite lengthy we include it for convenience. The novel technical part of this result involves  showing that the approximating polynomials can be taken to satisfy the required graphical properties such as connectivity.

\subsection{Notation}\label{chapter2:section:notation}
We end the section
with some notational conventions. 
We denote matrices and vectors by bold letters. We denote $[n]=\{1,\ldots,  n\}$ for $n \in \mathbb{Z}_{\geq 1}$, and let $[0,n]=\{0\}\cup [n]$. For $\Delta\geq 0, i\in [n]$ we  denote by $\T$ the set of trees on $[n]$. Given $(i,j)\in [n]^2$, we denote by $[ij]$ the elementary matrix $\b{E}_{ij} \in \R^{n\times n}$, i.e., the $n\times n$ matrix with $1$ at entry $(i,j)$ and $0$ elsewhere. We  write $k | \pmb{\alpha}$ for $k \in \mathbb{Z}_{\geq 1}, \pmb{\alpha} \in \mathbb{Z}^{n\times n}$ if and only if $k$ is a divisor of $\pmb{\alpha}_{ij}+\pmb{\alpha}_{ji}$ for all $ 1\leq i<j \leq n$. We denote by $\max(\pmb{\alpha})$ the maximum of $\{ \pmb{\alpha}_{ij}+\pmb{\alpha}_{ji} \mid 1\leq i<j\leq n\}$. We let ${\rm Supp}(\pmb{\alpha})$ denote the support of $\pmb{\alpha}$, namely the set of pairs $(i,j)$ with $\pmb{\alpha}_{ij}>0$. Given  $\pmb{\alpha} \in \mathbb{Z}_{\ge 0}^{n\times n}$ we  denote by $\b{X}^{\pmb{\alpha}}$ the monomial $\prod_{1\leq i<j\leq n} \X_{ij}^{\pmb{\alpha}_{ij}+\pmb{\alpha}_{ji}}$. For a polynomial $p$ in the monomials $\b{X}^{\pmb{\alpha}}$, we denote by $\|p\|_{\rm coef}$ the maximum absolute value of its coefficients.

We use the notation $\text{p-lim}$ to denote  convergence in probability. Specifically, the notation $\plim$ denotes convergence in probability as $n\to \infty$, and $\plimd$ denotes convergence in probability when we first take the limit $n\to \infty$ followed by $d\to \infty$. We abbreviate ``with high probability'' as w.h.p., and use it to indicate that an event $\mathcal{E}_{n, d}$ satisfies $\lim_{d\to \infty} \liminf_{n\to\infty}\p(\mathcal{E}_{n, d})=1$. The asymptotic notation $o(.), O(.), \omega(.),$ and $ \Omega(.)$ hide constants in $n$ and $d$ in the regime $n\to\infty$ followed by $d\to\infty$, where $n$ is the dimension of the disorder matrices we consider and $d$ is a parameter for the sparse regime. In particular, $o(1)$ encodes a sequence $\varepsilon_{n, d}$ satisfying $\lim_{d\to \infty} \lim\sup_{n\to \infty} |\varepsilon_{n,d}|=0$. We use $\tilde o(.), \tilde O(.), \tilde \omega(.), \tilde \Omega(.)$ to hide polylog asymptotic terms in $d$. We use $\lesssim$ to hide constants in $n,d$, and $\lesssim_d $ to hide polylog asymptotic terms in $d$ in inequalities. We write $a_{n,d} \sim b_{n,d}$ if $a_{n,d} = (1+o(1)) b_{n,d}$.  We denote by $\|\cdot\|_p, p\geq 0$ the usual $\ell_p$ norms, by $\|\cdot\|_{\rm fro}$ the Frobenius norm, by $\|\cdot\|$ the operator norm, by $\|\cdot\|_{\infty}$ the infinity norm (for matrices and scalar functions), and by $d(\b{z}, K)$ the $\ell_2$ distance between $z$ and the set $K$, whenever $K\neq \emptyset$. 

Given a random variable $X$ with distribution $\mathcal{D}$, we write $X \distrib \mathcal{D}$.  We denote by ${\rm GOE}^0(n)$ the distribution on $n\times n$ symmetric matrices $\X$ where $\X_{ij}=\X_{ji}$ are independent and identically distributed as $\mathcal{N}(0,\frac{1}{n})$ for $1\leq i<j\leq n$ and $\X$ has a null diagonal. When the diagonal terms are instead Gaussian $\mathcal{N}(0, \frac2{n})$, we denote this distribution by ${\rm GOE}(n)$. Using  either ${\rm GOE}(n)$ or ${\rm GOE}^0(n)$ has no effect on the SK optimization problem as the contribution from diagonal terms in $\pmb{\sigma}^\top \b{X}\pmb{\sigma}/2n$ given by  $\sum_{i\in[n]} \b{X}_{ii}/2n$ scales as $o(1)$ w.h.p. when $\X\distrib {\rm GOE}(n)$.

We denote by  $\Y\distrib\Gb(n,t/n)$ a symmetric matrix with zeros on the diagonal and $\p(\Y_{ij}=1)=t/n$, $\p(\Y_{ij}=0)=1-t/n$, i.i.d. across $1\le i<j\le n$. When $\Y$ is instead defined via (\ref{chapter2:eq:X-centered-rescaled}) we write $ \G(n,t/n)$ for the distribution of $\Y$.

Given $m\in \mathbb{Z}_{\geq 0}$, we denote by $\mathcal{C}^m(\mathcal{D}, \mathbb{R})$ the set of $m$-times continuously differentiable functions from $\mathcal{D}$ to $\R$. Given a differentiable function $f\in \mathcal{C}^m(\mathcal{D}, \mathbb{R})$ and $k\in [m]$ we denote by $f^{(k)}$ the $k$-th derivative of $f$. Given $\ell\in \mathbb{Z}_{\geq 0}$, we say that a function $f:\mathbb{R}^k\to \mathbb{R}$ is pseudo-Lipschitz of order $\ell$ (and write $f\in {\rm PL}(\ell)$) if $|f(\b{x})-f(\b{y})|\leq C(1+\|\b{x}\|_2^{\ell-1} + \|\b{y}\|_2^{\ell-1})\|\b{x}-\b{y}\|_2$ for all $\b{x},\b{y}\in \mathbb{R}^k$, where $C$ is a uniform constant independent of $\b{x}, \b{y}$.

For any matrix $\X\in\R^{n\times n}$ with null diagonal and vector $\x\in\R^n$, we define $\pmb{\sigma}(\x,\X)\in \{-1, 1\}^n$ as follows. First set $\x'=(\x_i', i\in [n])$ via $\x_i'=\min(1,\max(-1,\x_i))$. Then obtain $\x''=\pmb{\sigma}(\x,\X)$ by sequentially replacing $\x_i'\in (-1,1)$ by $\x_i''=\pm 1$ so that the value $(\x')^\top \X \x'$ increases or stays the same (this is achievable since the objective ${\b{x}'}^\top \b{X} \b{x}'$ is linear in each individual coordinate $\b{x}'_i$ once  $(\b{x}'_j)_{j\neq i}$ are fixed).

%\subsection{Approximate message passing algorithms}

\section{Universality of Low-Degree Polynomials. Main Results}
We next formally define the class of Low-Degree Polynomials (LDP) we use and then restate our main results formally. For a symmetric matrix $\b{X}\in \R^{n\times n}$ we  denote by $\R[\b{X}]_{\leq \Delta}$ the set of multivariate polynomials in $\{ \X_{ij}, 1\leq i<j \leq n\}$ with degree at most $\Delta$ and denote the degree of a polynomial $p$ by ${\rm Deg}(p)$. When $\Delta$ is bounded by a constant independent of $n$, we  refer to these polynomials as Low-Degree Polynomials (LDPs). We denote by $\U^{n\times n}$ the set of strictly upper triangular matrices in $\mathbb{Z}_{\geq 0}^{n\times n}$ and introduce
\begin{equation}
    \label{chapter2:U:notation}
    \begin{aligned}
         \U^{n\times n}_{1}  &\triangleq \{\pmb{\alpha} \in \U^{n\times n} \mid \pmb{\alpha}_{ij}=1 , \text{ for some } 1\leq i <j \leq n\},\\
         \U^{n\times n}_{*}  &\triangleq \{\pmb{\alpha} \in \U^{n\times n} \mid \pmb{\alpha}_{ij}\neq1 , \text{ for all } 1\leq i <j \leq n\},
    \end{aligned}
\end{equation}
so that $\U^{n\times n}=\U^{n\times n}_{1} \cup \U^{n\times n}_{*} $. To each polynomial $p$ in $\X$ we  associate its coefficients $c_{p, \pmb{\alpha}}, \ \pmb{\alpha}\in \U^{n\times n} $ so that
$$p(\b{X}) = \sum_{\pmb{\alpha} \in \U^{n \times n}} c_{p, \pmb{\alpha}} \b{X}^{\pmb{\alpha}}.$$
Note that the degree of $p$ is given by $\max \|\pmb{\alpha}\|_1$, where the maximum is over $\pmb{\alpha}$ such that $c_{p,\pmb{\alpha}}\ne 0$. Thus, $p\in \R[\b{X}]_{\leq \Delta}$ if $\|\pmb{\alpha}\|_1\le \Delta$ for all $\pmb{\alpha}$ with $c_{p,\pmb{\alpha}}\ne 0$. Recall the following notation from section~\ref{chapter2:section:notation}
\begin{align*}
    \|p\|_{\rm coef} \triangleq \max_{\pmb{\alpha}\in \U^{n\times n}} |c_{p, \pmb{\alpha}}|.
\end{align*}
We use the coefficient norm $\|p\|_{\mathrm{coef}}$ because many of our estimates bound the contribution of each monomial separately, and these contributions scale linearly with their coefficients. In particular, our combinatorial estimates are driven by the average values of monomials and their scaling in terms of $n$ and $d$, rather than any finer structure of the coefficient vector that would motivate a different norm.

Each $\pmb{\alpha} \in \U^{n\times n}\setminus \{\b{0}\}$ is associated with a graph $G_{\pmb{\alpha}} = (V_{{\pmb{\alpha}}}, E_{{\pmb{\alpha}}})$ which we call a  monomial/factor graph as follows  
\begin{equation}
\begin{aligned}
&V_{{\pmb{\alpha}}} &&\triangleq \{ k\in [n] \mid \exists j\in [n]\setminus\{k\}, \ \pmb{\alpha}_{kj} + \pmb{\alpha}_{jk}>0 \},\\
&E_{{\pmb{\alpha}}} && \triangleq \{ (i,j) \in [n]^2 \mid i\neq j,\ \pmb{\alpha}_{ij}+ \pmb{\alpha}_{ji}>0 \}.
\end{aligned}
\end{equation}
Similarly, we introduce
\begin{align*}
    V(p) \triangleq \bigcap_{\pmb{\alpha}  \in \{ \pmb{\alpha} \in \U^{n\times n}\setminus \{\b{0}\}\mid c_{p, \pmb{\alpha}}\neq 0\}} V_{\pmb{\alpha}}. \numberthis \label{chapter2:notation:Vp}
\end{align*}
We  say that $\b{X}^{\pmb{\alpha}}$ is connected if $G_{\pmb{\alpha}}$ is a connected graph and write $\pmb{\alpha} \in  \C$.  For an LDP $p\in \R[\b{X}]_{\leq \Delta}$, we  say that $p$ is  connected and write $p\in \mathcal{C}$ if  for all $\pmb{\alpha}\in \U^{n\times n}\setminus \{\b{0}\}$ with $c_{p, \pmb{\alpha}}\neq 0$, the monomial $\b{X}^{\pmb{\alpha}}$ is  connected.

Given $\pmb{\alpha} \in \U^{n\times n}\setminus \{\b{0}\}$, we  say that $\b{X}^{\pmb{\alpha}}$ is a tree  and write $\pmb{\alpha}\in\mathcal{T}$  if $G_{\pmb{\alpha}}$ is a tree.  Given an LDP $p\in \R[\b{X}]_{\leq \Delta}$,  we say that $p$ is a tree if for all $\pmb{\alpha} \in \U^{n\times n}\setminus \{\b{0}\}$ with  $c_{p, \pmb{\alpha}}\neq 0$, the monomial $\b{X}^{\pmb{\alpha}}$ is a tree. Note that every tree polynomial $p$ is connected. By convention, the constant monomial/LDP associated to $\pmb{\alpha} = \b{0}$ is assumed to be a connected tree. 

Given a subset $S\subseteq[n]$,  we  use the slight abuse of notation $S \subseteq \pmb{\alpha}$  when  $S \subseteq V_{\pmb{\alpha}}$. We use the notation $\mathcal{V}_S$ to denote the class of LDPs $p$ that are either a constant polynomial, or satisfy $S\subseteq V(p)$. In particular, we say that $p$ \emph{contains} the nodes in $S$ when $p\in \mathcal{V}_S$. When $S=\{i\}$ has a single element, we also use the notation $\mathcal{V}_S = \mathcal{V}_i$. Introduce the sets

\begin{align*}
    \mathcal{T}_{n,2,\Delta} &\triangleq \{\pmb{\alpha} \in \U^{n\times n} \mid  \pmb{\alpha}\in \T,\ \max(\pmb{\alpha}) \leq 2,\ \|\pmb{\alpha}\|_1\le \Delta \},\\
        \mathcal{C}_{n,2,\Delta} &\triangleq \{\pmb{\alpha} \in \U^{n\times n} \mid  \pmb{\alpha}\in \C,\ \max(\pmb{\alpha}) \leq 2,\ \|\pmb{\alpha}\|_1\le \Delta \}.
\end{align*}
We define  $ p^{\rm Tr}$ as the restriction of $p$ to monomials $\b{X}^{\pmb{\alpha}} $ with $ \pmb{\alpha} \in \mathcal{T}_{n, 2, {\rm Deg}(p)}$, i.e.
\begin{align*}
     p^{\rm Tr}(\b{X}) \triangleq \sum_{\pmb{\alpha} \in \mathcal{T}_{n,2,{\rm Deg}(p)} }c_{p, \pmb{\alpha}} \b{X}^{\pmb{\alpha}}.
\end{align*}
We  write $p\in \mathcal{T}_{n,2,\Delta}$ if  $\pmb{\alpha}\in \mathcal{T}_{n,2,\Delta}$ whenever $c_{p,\pmb{\alpha}}\ne 0$. By definition, $p^{\rm Tr}\in\mathcal{T}_{n,2,{\rm Deg}(p)}$ for every polynomial $p$. We will use the notation $\mathcal{T}_{n, 2},\ \mathcal{C}_{n, 2}$ when  the constant $\Delta$  is clear from context. We note that for any LDP $p\in \R[\X]_{\leq \Delta}$, the projection $p^{\rm Tr}$ can be computed in $n^{O(\Delta)}$ time complexity.

Before stating our main results, we briefly discuss the choice of the class of connected LDPs denoted by $\mathcal{C}$ and $\mathcal{C}_{n, 2, \Delta}$, and tree-structured LDPs denoted by $\mathcal{T}_{n, 2, \Delta}$. Clearly, one has $\mathcal{T}_{n, 2, \Delta} \subset \mathcal{C}_{n, 2, \Delta} \subset \mathcal{C}$, and both $\mathcal{T}_{n, 2, \Delta}$ and $\mathcal{C}_{n, 2, \Delta}$ can be used interchangeably throughout this paper without issue. Indeed, tree-structured LDPs are \emph{sufficient} to represent connected LDPs. Namely, Lemma \ref{chapter2:lemma:pre:6} shows that the expected contribution of non-tree monomials in both the Gaussian setting ${\rm GOE}^0(n)$, and sparse settings $\G(n, d/n)$  and $\G(n, \tau_n/n)$ is negligible compared to the contribution from tree terms. Specifically, in the case of Gaussian disorder $\b{X}\distrib {\rm GOE}^0(n)$, and for a connected LDP $p \in \mathcal{C}$  containing a subset of nodes $S$ in all its monomial graphs, we have 
\begin{align*}
    \E[p(\b{X})]= \E[p^{\rm Tr}(\b{X})] + O(\|p\|_{\rm coef} n^{-|S|}). \numberthis \label{chapter2:tree_structured.justification}
\end{align*} 
Moreover, Lemma \ref{chapter2:lemma:+1} shows that $|\E[p^{\rm Tr}(\b{X})]|- |p(\b{0})| = O(\|p\|_{\rm coef} n^{-|S|+1})$. The latter  bound is also optimal: consider the case where $p$ has equal coefficients, satisfies $p(\b{0})=0$, and its monomials have tree factor graphs isomorphic to a single fixed tree, then  $|\E[p^{\rm Tr}(\b{X})]| = \Theta(\|p\|_{\rm coef} n^{-|S|+1})$. Therefore, we can rewrite (\ref{chapter2:tree_structured.justification}) informally as
$\E[p(\b{X})]= (1 + O(n^{-1}))\cdot \E[p^{\rm Tr}(\b{X})]$ for ``nice'' LDPs $p$. In fact, we expect the LDPs obtained from the AMP approximation conducted in this paper to have equal coefficients per factor graph isomorphism class and per exponent structure, as a result of the symmetry of roles between the edge disorders. However, we do not need these  assumptions for our proofs.

Furthermore, the arguments in the proof of Lemma \ref{chapter2:lemma:pre:6} show that it is sufficient to only consider monomials with powers at most $2$ in any of the disorder entries, as the contribution of all other monomials is negligible. Crucially, this property will be helpful in our interpolation analyses in Section \ref{chapter2:section:state_ev} and \ref{chapter2:section:6}. Indeed, suppose $p$ is an LDP in $(\b{x}_{ij}, 1\leq i<j \leq n)$ and $p\in \mathcal{C}_{n, 2, \Delta}$. If we fix all entries except $\b{x}_{i j}$,  then $p(\b{x}) = p(\b{x}_{ij}) = u(\b{x}) + \b{x}_{ij} a(\b{x}) + \b{x}_{ij}^2 b(\b{x})$, where the terms $u(\b{x})$, $a(\b{x})$, and $b(\b{x})$ are LDPs only depending on the (fixed) entries $(\b{x}_{k\ell}, 1\leq k<\ell \leq n, (k, \ell) \neq (i, j))$. In particular, if we switch $\b{x}_{ij}$ to $\b{y}_{ij}$ as part of an interpolation step, the new value of $p$ is  $u(\b{x}) + \b{y}_{ij} a(\b{x}) + \b{y}_{ij}^2 b(\b{x})$. In contrast, working with a general connected LDP $p\in \mathcal{C}$, the same representation is more cumbersome: $p(\b{x}) = u(\b{x}) + \sum_{k \in [{\rm Deg}(p)]}(\b{x}_{ij})^k a_{k}(\b{x}) $, where $(a_k(\b{x}),\ k \in[{\rm Deg}(p)])$ are LDPs.

Finally, we remark that another benefit of restricting our attention to connected LDPs is that connectivity is preserved under several algebraic  manipulations arising throughout this paper. For instance, consider the Hamiltonian term $H(p(\X), \X)$, where $p = (p_i,\ i \in [n])$ is a collection of connected LDPs. The quantity $H(p(\X), \X)$ decomposes as a sum of terms $p_i(\X) \X_{ij} p_j(\X)$. If $p_i$ and $p_j$ contain nodes $i$ and $j$, respectively (i.e., $p_i\in \V_i$ and $p_j\in \V_j$), then the product $w(\b{X}) = p_i(\X) \X_{ij} p_j(\X)$ is itself a connected LDP whose monomial graphs  contain both $i$ and $j$ (i.e., $w \in \V_{\{i, j\}}$). More generally, if $p$ and $q$ are connected LDPs containing a common node $k\in [n]$ (i.e., $p, q \in \V_k$), then $pq$ is a connected LDP containing the same node $k$ (i.e., $pq\in \V_k$). Thus, the class of connected LDPs is closed under these and other operations used in this paper, which will be particularly convenient when constructing and analyzing several expressions that appear in subsequent sections.

We next formally state the main findings of this paper. Our main result regarding the Max-Cut problem, Theorem~\ref{chapter2:thm:main-formal}, relies on establishing that the LDP outputs are close to the hypercube $[-1,1]^n$. To achieve this, we first establish a more general result: universality for the empirical averages of pseudo-Lipschitz functions applied to the coordinates of  tree-structured LDPs. While Theorem~\ref{chapter2:thm:main-formal} is our primary variational result, Theorem~\ref{chapter2:lemma:state_ev} below provides the necessary distributional control, ensuring that state evolution predictions transfer from Gaussian to sparse ensembles.

\begin{theorem}[Universality of LDP Coordinate Statistics]\label{chapter2:lemma:state_ev}
    Let $\b{X}\distrib {\rm GOE}^0(n)$ and $\b{Y}\distrib \G(n, t/n)$ where $d\leq t \leq \tau_n$ and $\tau_n$ satisfies $\omega(1)\leq\tau_n\leq o(n)$.   Suppose $\chi: \mathbb{R}\to \mathbb{R}$ is a pseudo-Lipschitz function of order $k\in \mathbb{Z}_{\geq 1}$, i.e., $|\chi(x) - \chi(y)| \leq L|x-y|(1+ |x|^{k-1} + |y|^{k-1})$ for all $x,y \in \R$, where $L>0$ and $k$ are constants independent  of $n, d, x$ and $y$. Let $(p_i,\ i\in [n])$ be a collection of LDPs satisfying
    \begin{itemize}
        \item There exists a constant $C>0$ independent of $n$ and $d$ such that $\forall i\in [n],\ \|p_i\|_{\rm coef} \leq C$.
        \item There exists a constant $\Delta>0$ independent of $n$ and $d$ such that $\forall i \in [n],\ p_i \in \mathcal{T}_{n, 2, \Delta}$.
        \item For all $i\in [n]$, the LDP $p_i$ contains node $i$ (i.e., $p_i\in \V_i$).
    \end{itemize}
    Finally, let $\kappa\in \mathbb{R}$ be a scalar. Then, the following two statements are equivalent
    \begin{align*}
        \plim \frac{1}{n} \sum_{i=1}^{n} \chi(p_i(\b{X})) &= \kappa, \numberthis \label{chapter2:lemma:state_ev.1}\\
        \plimnd \frac{1}{n} \sum_{i=1}^{n} \chi(p_i(\b{Y})) &= \kappa. \numberthis \label{chapter2:lemma:state_ev.2}
    \end{align*}
\end{theorem}
We remark that while Theorem~\ref{chapter2:lemma:state_ev} is formulated as an equivalence result, its practical utility relies on the fact that the behavior in the Gaussian setting (\ref{chapter2:lemma:state_ev.1}) is already well-understood. Specifically, the limits of empirical averages  are fully characterized by the standard state evolution framework in the analysis of AMP on Gaussian matrices. In particular, we show in Proposition  \ref{chapter2:prop:appendix} that state evolution holds for the specific LDPs we obtain for finding near-ground states of the SK Hamiltonian. Consequently, Theorem~\ref{chapter2:lemma:state_ev}  effectively establishes that these state evolution predictions transfer automatically to the sparse ensembles $\G(n,d/n)$ and $\G(n,\tau_n/n)$ with $\omega(1)\leq \tau_n \leq o(n)$, thereby verifying the validity of state evolution in these sparse regimes without requiring a separate derivation. 

Furthermore, by selecting the specific test function $\chi(x) \triangleq d(x, [-1, 1])^2$, Theorem~\ref{chapter2:lemma:state_ev} readily implies the universality of the proximity of LDP outputs to $[-1,1]^n$. This guarantees that if the LDP algorithm finds a solution close to the hypercube in the Gaussian setting, it does so in the sparse  settings as well. The latter will be a key requirement later for our rounding arguments of LDP outputs in Section \ref{chapter2:section:rounding}.  With this distributional foundation established, we now restate Theorem~\ref{chapter2:thm:main} formally, recalling the rounding operator $\pmb{\sigma}(\x,\X)$ defined in the notation section.

\begin{theorem}[Universality of LDP Performance]\label{chapter2:thm:main-formal}
Let $\Delta>0$, and for each $n$ consider a vector of polynomials  $p=(p_i,\ i\in [n])$ such that $p_i\in\mathcal{T}_{n,2,\Delta}$, $\|p_i\|_{\rm coef}\le c$ for some constant $c=c(\Delta)$, and $p_i \in \V_i $ for all  $i\in[n]$.  Let $\mathcal{P}$ be a constant, and consider the following statements: 
\begin{enumerate}[label=(\roman*), itemsep=0pt, topsep=2pt, parsep=0pt]
    \item \label{chapter2:thm:main-formal.1}   $\X\distrib{\rm GOE}^0(n)$, and
        \begin{align}
        &\plim {\frac{1}{2n}}p(\X)^\top \X p(\X)\geq \mathcal{P}, \label{chapter2:eq:P-limit}\\
        &\plim {\frac{d(p(\X), [-1,1]^n)}{\sqrt{n}}}=0. \label{chapter2:eq:box-limit}
        \end{align}
    \item \label{chapter2:thm:main-formal.2}  $\Y \distrib \G(n, t/n)$, where $d\leq t\leq \tau_n$ and $\omega(1)\leq\tau_n\leq o(n)$, and  
    \begin{align}
    &\plimnd {\frac{1}{2n}}p(\Y)^\top \Y p(\Y)\geq\mathcal{P},   \label{chapter2:eq:P-limit-Y.1}\\
    &\plimnd {\frac{d(p(\Y), [-1,1]^n)}{\sqrt{n}}}=0.  \label{chapter2:eq:box-limit-Y.1}
    \end{align}
    % \item \label{chapter2:thm:main-formal.3}  $\Y \distrib \G(n, \tau_n/n)$, where $\tau_n$ satisfies $\tau_n=\omega(1)$ and $\tau_n=o(n)$, and
    % \begin{align}
    % &\plim {\frac{1}{2n}}p(\Y)^\top \Y p(\Y)\geq\mathcal{P},   \label{chapter2:eq:P-limit-Y.2}\\
    % &\plim {\frac{d(p(\Y), [-1,1]^n)}{\sqrt{n}}}=0.  \label{chapter2:eq:box-limit-Y.2}
    % \end{align}
    \item \label{chapter2:thm:main-formal.d}  $\Y \distrib \G(n, t/n)$, where $d\leq t\leq \tau_n$ and $\omega(1)\leq\tau_n\leq o(n)$, and  
    \begin{align}
        \plimd \frac{1}{2n} \pmb{\sigma}(p(\Y), \Y)^\top \Y \pmb{\sigma}(p(\Y), \Y) \geq \mathcal{P}.  \label{chapter2:eq:P-limit-Y-final.d}
    \end{align}
    % \item \label{chapter2:thm:main-formal.tau}  $\Y\distrib \G(n, \tau_n/n)$, where $\tau_n$ satisfies $\tau_n=\omega(1)$ and $\tau_n=o(n)$, and
    % \begin{align}
    %     \plim \frac{1}{2n} \pmb{\sigma}(p(\Y), \Y)^\top \Y \pmb{\sigma}(p(\Y), \Y) \geq \mathcal{P}.  \label{chapter2:eq:P-limit-Y-final.tau}
    % \end{align}
\end{enumerate}
    Then we have the following
    \begin{enumerate}[label=(\alph*)]
        \item \label{chapter2:thm:main-formal.a} Statements \ref{chapter2:thm:main-formal.1} and \ref{chapter2:thm:main-formal.2}  are equivalent.
        \item \label{chapter2:thm:main-formal.b} If  \ref{chapter2:thm:main-formal.2} holds, then \ref{chapter2:thm:main-formal.d} also holds.
        %\item \label{chapter2:thm:main-formal.c} If  \ref{chapter2:thm:main-formal.3} holds, then \ref{chapter2:thm:main-formal.tau} also holds.
    \end{enumerate}
\end{theorem}
Next, we move to stating formally our third result regarding approximating AMP via LDP. For this, we begin by introducing AMP formally.  There is a multitude of ways to describe AMP iterations, and each AMP algorithm is typically problem specific. For our purposes, we will use an appropriate variant of the general AMP iteration given in  (6) in \cite{montanari2019optimization} (namely, we consider the setting where the denoisers do not depend on $\b{y}$), which we restate here. Consider a sequence of \textit{denoisers} $f^t: \R^{t+1} \to \R$ and an initialization $\b{u}^0 \in \R^n$. The AMP iterations are given by

\begin{equation}
\begin{aligned}
& \b{u}^{t+1} &&= \mathbf{X} f^t (\b{u}^0, \ldots,\b{u}^t) - \sum_{j=1}^{t} b_{t,j} f^{j-1} (\b{u}^0,\ldots,\b{u}^{j-1}),\\
& b_{t, j} &&= \frac{1}{n} \sum_{i=1}^{n} \frac{\partial f^t}{\partial \b{u}_i^j}(\b{u}_i^0,\ldots,\b{u}_i^t),
\end{aligned}
\label{chapter2:amp}
\end{equation}
where $f^j=0$ for $j<0$, and $f^t(\b{u}^0,\ldots,\b{u}^t)$ is applied coordinate-wise.  The AMP iterations satisfy a certain \textit{state evolution} property formulated in terms of the  centered Gaussian process $(U_j), j\ge 0$ with covariance specified recursively as
$\b{Q} = (\b{Q}_{k+1,j+1})_{k,j\geq 0}$ with 
\begin{align}
    \b{Q}_{k+1,j+1} = \E [f^k(U_0,\ldots,U_k) f^j(U_0,\ldots,U_j)], \quad \forall k,j \geq 0. \label{chapter2:eq:covariance-Q}
\end{align}
The output of the AMP is a vector $\b{v}$ given by
\begin{align*}
    \b{v} = \sqrt{\delta} \sum_{k=1}^{\lfloor \bar q  / \delta \rfloor} f^k(\b{u}^0, \ldots,\b{u}^k), \numberthis \label{chapter2:candidate}
\end{align*}
where $\delta$ and $\bar q$ are constant parameters driving the optimality gap, and the number of iterations of the AMP satisfies $T\geq \lfloor \bar q  / \delta \rfloor$. The final output of the AMP algorithm with rounding is then given by $\pmb{\sigma}(\b{v},\X)$. The main result of \cite{montanari2019optimization} asserts  the near optimality of this algorithm for appropriate choice of denoisers $f^t$. Specifically, it asserts  the existence for any $\varepsilon>0$ of  denoisers $f^t$ and parameters $\bar q, \delta$, so that w.h.p. 
\begin{align}
 {\frac{1}{2n}} \b{v}^\top \X \b{v}&\ge \mathcal{P}^*-\varepsilon, \label{chapter2:eq:v:obj}\\
 \frac{d(\b{v}, [-1,1]^n)}{\sqrt{n}} &\leq \varepsilon \label{chapter2:eq:v:box}.
\end{align}
This version of AMP is called Incremental AMP, which for  the remainder of this paper, we call ${\rm IAMP}$. Naturally, the ${\rm IAMP}$ algorithm depends on the parameter $\varepsilon>0$, and this dependency will be clear from context whenever we refer to ${\rm IAMP}$. We will show that the output $\b{v}$ of  ${\rm IAMP}$ can be well approximated by LDPs, and, as a result, the implied objective value can be reached via LDPs as well. This is the essence of our intermediate result, Theorem~\ref{chapter2:thm:2}, stated  formally below.

\begin{theorem}\label{chapter2:thm:2-formal}
Given $\varepsilon>0$ and an integer $T$, 
consider a sequence of denoisers $f^t$ with $t\in [T]$ satisfying,
\begin{enumerate}
    \item There exists a constant $L$ such that $f^t$ are $L$-Lipschitz.
    \item $f^t$ have weak derivatives that are either pseudo-Lipschitz or indicators.
    \item The covariance matrix $\b{Q}_{\leq t}$ defined in (\ref{chapter2:eq:covariance-Q})  satisfies $\b{Q}_{\leq t}\succeq \b{I}_{t}$ and $ \|\b{Q}_{\leq t}\|_{\infty}\leq 2$
    for all  $t\in [T]$.
\end{enumerate}
Suppose there exists a constant $M>0$ independent of $n$ such that $\|\b{u}^0\|_{\infty} \leq M$ w.h.p. Then, there exist constants $\Delta=\Delta(\varepsilon, T),\  c=c(\varepsilon, T)$, and a sequence $p=(p_i,\ i\in [n])$, with $p_i\in \mathcal{T}_{n,2,\Delta}$,  such that w.h.p. 
\begin{align*}
{\frac{\| \b{v}-p(\X)\|_2}{\sqrt{n}}} &\leq \varepsilon,\\
 \frac{|\b{v}^\top \X \b{v} - p(\X)^\top \X p(\X)|}{n} &\leq \varepsilon, 
\end{align*}
and $p$ satisfies $\forall i\in[n],\ p_i \in \V_i$ and $\|p_i\|_{\rm coef}\leq c$. As a result, for every $\varepsilon>0$ there exists $\Delta$ and a sequence $p=(p_i,\ i\in [n])$ with $p_i\in\mathcal{T}_{n,2,\Delta}$ and $p_i\in \V_i$  such that w.h.p.
\begin{align*}
{\frac{1}{2n}} p(\X)^\top \X p(\X) &\geq \mathcal{P}^*-\varepsilon,\\
\frac{d(p(\X), [-1,1]^n)}{\sqrt{n}} &\leq \varepsilon.
\end{align*}
\end{theorem}
We will show that the IAMP in~\cite{montanari2019optimization} can be adapted to satisfy the assumptions of the theorem above. Near optimality of the LDP after rounding, namely Corollary~\ref{chapter2:cor:LDP-near-opt} is now  an immediate corollary of Theorems~\ref{chapter2:thm:main-formal} and \ref{chapter2:thm:2-formal}, and is  restated as follows. 

\begin{corollary}\label{chapter2:cor:LDP-near-opt-formal}
Let $\b{Y}\distrib \G(n, t/n)$ with $d\leq t\leq \tau_n $ and $\omega(1)\leq\tau_n \leq o(n)$. For every $\varepsilon>0$ there exists a sequence  $p=(p_i,\  i\in [n])$, with $p_i\in\mathcal{T}_{n,2,\Delta}$ and $p_i \in \V_i$  such that
\begin{align*}
\plimnd {\frac{1}{2n}}\pmb{\sigma}(p(\Y),\Y)^\top \Y \pmb{\sigma}(p(\Y),\Y) &\geq \mathcal{P}^*-\varepsilon.
\end{align*}

\end{corollary}

\section{Preliminary Results}\label{chapter2:section:4}
In this section, we establish some preliminary technical results.

% \begin{lemma}[Fa\`a di Bruno \cite{arbogast1800calcul}]\label{chapter2:lemma:pre:-1}
% Let $ g \in \mathcal{C}^{m}(\mathcal{D}_g, \R),\ f\in \mathcal{C}^{m}(\mathcal{D}_f, \R)$  be $m$-times differentiable scalar functions with ${\rm Im}(g) \subset \mathcal{D}_f$, and let $w(z)=f(g(z))$. The following holds for all $z\in \mathcal{D}_g$ and $k\leq m$
% \begin{align*}
%     w^{(k)}(z) = \sum_{\substack{\ell_1+ 2\ell_2+\cdots+k\ell_k = k\\ \ell_1,\ldots,\ell_k\geq 0}} \frac{k!}{\ell_1! \ldots \ell_k!} \cdot f^{(\ell_1 + \cdots + \ell_k)}(g(z)) \cdot \prod_{j=1}^{k} \l(\frac{g^{(j)}(z)}{j!}\r)^{\ell_j}.
% \end{align*}
% \end{lemma}

\subsection{Analytic Preliminaries}
We begin by establishing some non-probabilistic bounds on functions and their derivatives appearing in our analysis. The next lemma introduces sigmoid-type penalty functions and derives basic bounds on their derivatives. These functions will play a key role in subsequent interpolation analyses, allowing us to transfer high-probability bounds across different disorder distributions.
\begin{lemma}\label{chapter2:lemma:pre:4}
    Given $a,b\in \R$ let
    \begin{align*}
    \phi :\quad  &\R \to \R , \quad z\mapsto   \frac{e^z}{1+e^z},\\
    \psi_{a,b} :\quad &\R \to \R ,\quad z\mapsto  \phi\l(a z - b\r).
\end{align*}
Then
\begin{enumerate}
    \item $\psi_{a,b}$ is $|a|$-Lipschitz,
    \item $\|\psi_{a, b}^{(j)}\|_{\infty} \leq |a|^{j}, \forall j\in \{1,2,3\}$,
\end{enumerate}
\end{lemma}
\begin{proof}
The first three derivatives of $\phi$ are
\begin{align*}
    \phi^{(1)}(z) &= \frac{e^z}{(1+e^z)^2},\\
    \phi^{(2)}(z) &= \frac{e^z(1-e^z)}{(1+e^z)^3},\\
    \phi^{(3)}(z) &= \frac{e^z(-4e^z + e^{2z} +1)}{(1+e^z)^4}.
\end{align*}
In particular, elementary calculus yields $\|\phi^{(j)}\|_{\infty}\leq 1$ for $j\leq 3$.
    \begin{enumerate}
        \item The first claim follows from noting that $\psi_{a,b}^{(1)}(z) = a \phi^{(1)}(az -b) $, and thus $\|\psi_{a,b}^{(1)}\|_{\infty} \leq |a|$.
        \item The second claim follows from $\psi^{(j)}(z)= a^{j} \phi^{(j)}(az-b)$ combined with the previous bounds on $\|\phi^{(j)}\|_{\infty}$ for $j\leq 3$.
    \end{enumerate}
\end{proof}

The next lemma derives bounds on derivatives of smooth functions composed with quadratic polynomials. These estimates will later be used to control Taylor remainders arising in the interpolation analysis.
\begin{lemma}\label{chapter2:claim:5.12}
Suppose $u,a,b \in \R$ are scalars, $v (z)=u  + a z  + b z^2$ is a quadratic, and let $\chi\in \mathcal{C}^{3}(\mathbb{R}, \mathbb{R})$ be a thrice differentiable function such that for some constants $L>0$ and $m\in \mathbb{Z}_{\geq 0}$,
\begin{equation}
  |\chi^{(\ell)}(x)| \le L \bigl(1 + |x|^m\bigr), \qquad \forall x\in\R,\ \ell=1,2,3.
\end{equation}
Let $w(z) = \chi(v(z))$. Then there exists a constant $c_m>0$, depending only on $m$, such that for all
$j\in [3]$, and all $z, z^*$ with $|z^*|\leq |z|$, it holds
\begin{align*}
    |w^{(j)} (z^*)|
    &\le  c_m\,L\,
      \sum_{\substack{j_1,j_2,j_3,j_4 \leq j+2m\\ j_2+j_3 \geq \lceil j/2 \rceil }} |z|^{j_1} |a|^{j_2} |b|^{j_3} |u|^{j_4}.
\end{align*}
\end{lemma}

\begin{proof}
Throughout this proof, we denote by $c_m$ different constants depending only on $m$. Using $|z^*|\le |z|$ and applying H\"{o}lder's inequality (this step requires $m\geq 1$, but the final bound holds trivially when $m=0$), we have for $\ell=1,2,3$,
\begin{align*}
    |\chi^{(\ell)}(v(z^*))| &\leq  L(1 + |u + az^* + b(z^*)^2|^m)\\
    &\leq L\l(1 + \l( |u| + |a||z| + |b||z|^2 \r)^m \r)\\
    &\le 3^{\max(m-1, 0)}L\bigl( 1 + |u|^m + |a|^m |z|^m + |b|^m |z|^{2m} \bigr)\\
    &\le c_m L \sum_{i_1,i_2,i_3,i_4\le 2m} |z|^{i_1} |a|^{i_2} |b|^{i_3} |u|^{i_4}.
\end{align*}
We now bound the derivatives of $w$. We have for $j=1$,
\begin{align*}
|w^{(1)}(z^*)|
&= \Bigl|\chi^{(1)}(v(z^*))\cdot (a+2bz^*)\Bigr| \\
&\le c_m L 
   \sum_{i_1,i_2,i_3,i_4\le 2m}
   |z|^{i_1} |a|^{i_2} |b|^{i_3} |u|^{i_4}\cdot \bigl(|a| + 2|b||z|\bigr) 
\\
&\le c_m L
   \sum_{\substack{j_1,j_2,j_3,j_4\le 1+2m\\ j_2+j_3\ge 1}}
   |z|^{j_1}|a|^{j_2}|b|^{j_3}|u|^{j_4},
\end{align*}
which yields the desired bound for $j=1$.  We have  similarly for $j=2$
\begin{align*}
|w^{(2)}(z^*)| &= \Bigl|\chi^{(2)}(v(z^*)) \cdot (a+2bz^*)^2 + 2\chi^{(1)}(v(z^*))\cdot b\Bigr| \\
&\le c_m L 
   \sum_{i_1,i_2,i_3,i_4\le 2m}
   |z|^{i_1} |a|^{i_2} |b|^{i_3} |u|^{i_4}
   \,\Bigl((|a|+2|b||z|)^2 + |b|\Bigr)
\\
&\le c_m L
   \sum_{\substack{j_1,j_2,j_3,j_4\le 2+2m\\ j_2+j_3\ge 1}}
   |z|^{j_1}|a|^{j_2}|b|^{j_3}|u|^{j_4},
\end{align*}
which yields the bound for $j=2$. Finally, we have for $j=3$
\begin{align*}
|w^{(3)}(z^*)|
&= \Bigl|\chi^{(3)}(v(z^*)) \cdot (a+2bz^*)^3
     + 6\chi^{(2)}(v(z^*))\cdot b(a+2bz^*)\Bigr| \\
&\le c_m L 
   \sum_{i_1,i_2,i_3,i_4\le 2m}
   |z|^{i_1} |a|^{i_2} |b|^{i_3} |u|^{i_4}
   \,\Bigl((|a|+2|b||z|)^3 + |b|(|a|+|b||z|)\Bigr)
\\
&\le c_m L
   \sum_{\substack{j_1,j_2,j_3,j_4\le 3+2m\\ j_2+j_3\ge 2}}
   |z|^{j_1}|a|^{j_2}|b|^{j_3}|u|^{j_4},
\end{align*}
which yields the bound for $j=3$. This ends the proof of the lemma.
\end{proof}

The next lemma provides a smooth approximation of pseudo-Lipschitz functions, together with quantitative bounds on the approximation error and the regularity of the approximant. This result will be used later to extend universality statements to non-differentiable test functions in Theorem~\ref{chapter2:lemma:state_ev}.
\begin{lemma}\label{chapter2:lemma:PL_approx}
Let $g:\R\to\R$ be pseudo-Lipschitz of order $k\in\mathbb{Z}_{\geq 1}$, i.e.,\ $|g(x)-g(y)|\le L(1+|x|^{k-1}+|y|^{k-1})|x-y|$ for all $x,y\in\R$, where $L>0$ is a constant independent of $x$ and $y$. Then, there exist positive constants $C_j, j\in \{0, 1, 2, 3\}$ depending only on $k$ and $L$, such that  for every $\sigma \in (0, 1]$, there exists an infinitely differentiable function $f_{\sigma}\in \mathcal{C}^{\infty}(\R, \R)$ satisfying:
\begin{enumerate}[label=(\roman*), ref=\roman*]
\item\label{chapter2:lemma:PL_approx.i}  $|f_{\sigma}(x)-g(x)|\le\sigma C_0(1+|x|^{k-1})$ for all $x\in \R$.
\item\label{chapter2:lemma:PL_approx.ii} $|f_{\sigma}^{(j)}(x)|\le C_j \sigma^{-j}(1+|x|^{k})$ for all $x\in \R$ and $j\in \{1, 2, 3\}$.
\end{enumerate}
\end{lemma}
\begin{proof}
For $\sigma>0$, set $\kappa_\sigma(u)=\frac{1}{\sigma\sqrt{2\pi}}e^{-u^2/(2\sigma^2)}$ and define 
\begin{align*}
    f_\sigma(x)=\int_{\R} g(y)\kappa_\sigma(x-y)\,dy=\E[g(x+\sigma Z)],
\end{align*}
where $Z$ is a random variable with distribution $\mathcal{N}(0,1)$. Then
\begin{align*}
    \l| f_\sigma(x)-g(x)\r| &\leq \int_{\R} \l| g(x+z)-g(x) \r| \kappa_\sigma(z)\,dz\\
    &\leq L\int_{\R} |z|\l(1+|x+z|^{k-1}+|x|^{k-1}\r) \kappa_\sigma(z)\,dz, \numberthis \label{chapter2:lemma:PL_approx.1}
\end{align*}
where we used the pseudo-Lipschitz bound in the last line. Using the weighted arithmetic-geometric mean inequality, it can be shown that  $|z|\l(1+|x+z|^{k-1}+|x|^{k-1}\r)$ is less than $ A\l(|z|+|z|^{k}+|z||x|^{k-1}\r)$ for some constant $A=A(k)$. Moreover, we have using Gaussian moment estimates  $\E|\sigma Z|^{\ell}\leq B\sigma^{\ell}$ for all $\ell \in [k]$, where $B=B(k)$ is a constant. Using the previous bounds and $\sigma\leq 1$ in (\ref{chapter2:lemma:PL_approx.1}) yields
\begin{align*}
    |f_\sigma(x)-g(x)|&\le LAB\l(\sigma + \sigma^{k} + \sigma |x|^{k-1}\r) \\
    &\leq C'\sigma(1+|x|^{k-1}),
\end{align*}
where $C'=C'(L, k)$ is a constant. This shows (\ref{chapter2:lemma:PL_approx.i}) with $C_0 = C'$. We next show  (\ref{chapter2:lemma:PL_approx.ii}). Differentiating under the integral gives
\begin{align*}
    f_\sigma^{(j)}(x)&=\int_{\R} g(y)\,\partial_x^j\kappa_\sigma(x-y)\,dy\\
    &=\sigma^{-j}\!\int_{\R} g(x-\sigma u)\,q_j(u)\frac{e^{-u^2/2}}{\sqrt{2\pi}}\,du, \numberthis \label{chapter2:lemma:PL_approx.2}
\end{align*}
where $q_j$ is a fixed polynomial. Since $|g(x)-g(0)|\le L(1+|x|^{k-1})|x|$, we have $|g(x)|\le A(1+|x|^{k})$ for some constant $A=A(L, k)$. Therefore
\begin{align*}
    |g(x-\sigma u)|&\le A(1+|x-\sigma u|^{k})\\
    &\le A(1+\l(|x|+ \sigma |u|\r)^{k}) \numberthis \label{chapter2:lemma:PL_approx.3}\\
    &\le A'(1+|x|^{k}+|u|^{k}),
\end{align*}
where we used $0< \sigma\leq 1$ in (\ref{chapter2:lemma:PL_approx.3}) and $A'=A'(L,k)$ is a constant. Using the above bound in (\ref{chapter2:lemma:PL_approx.2}) and Gaussian moments estimates, it follows that $|f_\sigma^{(j)}(x)|\le C_j\sigma^{-j}(1+|x|^{k})$ for $j=1,2,3,$ where $C_j=C_j(L, k)$ are constants. This shows (\ref{chapter2:lemma:PL_approx.ii}) and ends the proof of the lemma.
\end{proof}

\subsection{Algebraic Properties of LDPs}
In this section, we derive elementary algebraic properties of LDPs. The next lemma shows that the class of Low-Degree Polynomials is closed under products, and provides bounds on the coefficient norm of the resulting polynomial.

\begin{lemma}\label{chapter2:lemma:pre:1}
    Let $\X\in \R^{n\times n}$, $p \in \R[\X]_{\leq \Delta_1}$ and $q\in \R[\X]_{\leq \Delta_2}$, then
    \begin{align*}
        \|pq\|_{\rm coef} \leq 2^{\Delta_1+\Delta_2}\|p\|_{\rm coef} \|q\|_{\rm coef}.
    \end{align*}
\end{lemma}
\begin{proof}
    For $\pmb{\alpha} \in \U^{n\times n}$ with $c_{pq, \pmb{\alpha}}\neq 0$, the coefficient of the monomial $\X^{\pmb{\alpha}}$ in $pq$ is given by
    \begin{align*}
        c_{pq,\pmb{\alpha} } &= \sum_{\substack{\pmb{\alpha}_1, \pmb{\alpha}_2 \in \U^{n\times n} \\ \pmb{\alpha}_1+\pmb{\alpha}_2=\pmb{\alpha}}} c_{p, \pmb{\alpha}_1} c_{q, \pmb{\alpha}_2}.
    \end{align*}
    Each of the summands has absolute value bounded by $\|p\|_{\rm coef} \|q\|_{\rm coef}$. Since the degree of $pq$ is at most $\Delta_1+\Delta_2$ we have $\|\pmb{\alpha}\|_1 \leq \Delta_1 + \Delta_2$, therefore the number of pairs $(\pmb{\alpha}_1, \pmb{\alpha}_2)\in (\U^{n\times n})^2$ summing to $\pmb{\alpha}$ is bounded by $2^{\|\pmb{\alpha}\|_1}\leq 2^{\Delta_1 + \Delta_2}$. 
\end{proof}

The next lemma gives a  bound on the coefficient norm $\|\cdot\|_{\rm coef}$ of a sum of connected LDPs $p_i$ with zero constant coefficients, and containing subsets $S_i$ (i.e., $p_i \in \V_{S_i}$). The key point is that, under a growth condition on the cardinal of the unions of  $S_i$, a fixed monomial can only appear in a constant number of the LDPs $p_i$, since the size of its vertex set is limited by the degree constraint. This immediately yields a bound on the coefficient norm of $\sum_i p_i$. This result will be crucially useful when bounding sums of LDPs arising in the interpolation arguments, for example, when the LDPs are the Hamiltonian summands $p_i(\X)  \X_{ij} p_j(\X)$, the LDPs $(p_i,\ i\in[n])$ are obtained from the AMP approximation, and $S_{ij} = \{i, j\}$.

\begin{lemma}\label{chapter2:lemma:pre:2}
    Let $\X\in \R^{n\times n}$ and let $p_i \in \R[\X]_{\leq \Delta_i}, i=1,\ldots,m$. 
    Suppose $p_i$ are connected and $c_{p_i, \b{0}}=0$ (i.e. $p_i(\b{0})=0$) for all $i\in [m]$.
    Assume there exist subsets $S_i \subseteq[n], i=1,\ldots,m$ such that
    \begin{enumerate}
    \item $\forall i\in [m],\quad p_i \in \V_{S_i}$. 
    \item There exist two positive constants $c$ and $\gamma$ such that
    \begin{align*}
        \forall j\in [m], \ \forall 1\leq k_1<\ldots<k_j \leq m, \quad|S_{k_1}\cup \ldots \cup S_{k_j}| \geq c j^{\gamma}.
    \end{align*}
    \end{enumerate}
    Then 
    $$\l\|\sum_{i=1}^{m} p_i\r\|_{\rm coef} \leq \l(\frac{1+\max_{i\in [m]} \Delta_i}{c}\r)^{\frac{1}{\gamma}} \max_{i\in [m]} \|p_i\|_{\rm coef}.$$
\end{lemma}
\begin{proof}  
The idea of proof is as follows: If a monomial $\X^{\pmb{\alpha}}$ appears in many of the $p_i$, then all corresponding sets $S_i$ must be contained in the vertex set $V_{\pmb{\alpha}}$ of that monomial.   But the size of $V_{\pmb{\alpha}}$ is bounded, because  a connected monomial of degree at most $\Delta_i$ uses at most $1+\Delta_i$ vertices. Combining these two facts yields a bound on how many $p_i$ can contribute to the coefficient of $\X^{\pmb{\alpha}}$. We next make this argument formal. Fix $\pmb{\alpha} \in \U^{n\times n} \setminus \{\b{0}\}$, such that $c_{p_i, \pmb{\alpha}} \neq 0$ for at least one $i\in [n]$. The coefficient of the monomial $\X^{\pmb{\alpha}}$ in $s(\X) \triangleq \sum_{i=1}^{m} p_i(\X)$ is given by 
    \begin{align*}
        c_{s,\pmb{\alpha} } &= \sum_{i\in [m]} c_{p_i, \pmb{\alpha}}.\numberthis\label{chapter2:lemma:pre:2.1}
    \end{align*}
    Each of the summands above has absolute value bounded by $\max_{i\in[m]} \|p_i\|_{\rm coef}$.  Let $I(\pmb{\alpha})\triangleq\{i\in [m] \mid c_{p_i, \pmb{\alpha}}\neq 0\}$, and fix $i \in I(\pmb{\alpha})$. Next, we will use $|I(\pmb{\alpha})|$ to lower bound $|V_{\pmb{\alpha}}|$, and the connectivity of $\pmb{\alpha}$ to upper bound $|V_{\pmb{\alpha}}|$. Combining these bounds will yield the result of the lemma.   By construction, we have $c_{p_i, \pmb{\alpha}} \neq 0$. Recalling the conditions $p_i \in \V_{S_i}$ and $p_i(\b{0})=0$ in the statement of the lemma, and the definition of $V(p_i)$ from (\ref{chapter2:notation:Vp}), it follows that $S_i \subseteq V_{\pmb{\alpha}}$. Since $i$ was taken arbitrarily from $I(\pmb{\alpha})$, it follows that 
    \begin{align*}
        \bigcup_{i\in I(\pmb{\alpha})} S_i \subseteq V_{\pmb{\alpha}},
    \end{align*}
     which implies
    \begin{align*}
        |V_{{\pmb{\alpha}}}| \geq \l|\bigcup_{i\in I(\pmb{\alpha})} S_i\r| \geq c |I(\pmb{\alpha})|^{\gamma}, \numberthis\label{chapter2:lemma:pre:2.2}
    \end{align*}
    where the last inequality follows from the second assumption in the  lemma. 
    Since the number of nodes of a connected graph is at most one plus the number of edges, and $\pmb{\alpha}$ has a connected monomial graph (as all $p_i$ are connected, and $\pmb{\alpha}$ appears in at least one $p_i$ with nonzero coefficient $c_{p_i, \pmb{\alpha}}\neq 0$), it follows that 
    \begin{align*}
        |V_{{\pmb{\alpha}}}|&\leq 1 + |E_{\pmb{\alpha}}|\\
        &= 1 + \|\pmb{\alpha}\|_0 \\
        &\leq 1 + \|\pmb{\alpha}\|_1 \\
        &\leq  1+\max_{i\in[m]} \Delta_i,
    \end{align*}
    combining the above with (\ref{chapter2:lemma:pre:2.2}) yields
    \begin{align*}
        |I(\pmb{\alpha})| \leq \l( \frac{1+\max_{i\in[m]} \Delta_i}{c}\r)^{\frac{1}{\gamma}}.
    \end{align*}
    Using the above and (\ref{chapter2:lemma:pre:2.1}), we obtain
    \begin{align*}
        \forall \pmb{\alpha} \in \mathcal{U}^{n\times n}\setminus \{\b{0}\}, \quad |c_{s,\pmb{\alpha} }| &\leq  |I(\pmb{\alpha})| \max_{i\in[m]} \|p_i\|_{\rm coef} \leq \l( \frac{1+\max_{i\in[m]} \Delta_i}{c}\r)^{\frac{1}{\gamma}} \max_{i\in [m]} \|p_i\|_{\rm coef},
    \end{align*}
    where we note that while our previous analysis was restricted to $\pmb{\alpha}$ appearing with nonzero coefficient in at least one of the LDPs $p_i$, the above inequality still holds trivially when $c_{p_i, \pmb{\alpha}}=0, \forall i\in [n]$.   This concludes the proof as $\l\|\sum_{i=1}^{m} p_i\r\|_{\rm coef} = \max_{\pmb{\alpha} \in \mathcal{U}^{n\times n}\setminus \{\b{0}\}} |c_{s,\pmb{\alpha} }|$, since the constant monomial coefficient in $\sum_{i=1}^{m} p_i$ is zero.
\end{proof}

The next lemma establishes some useful properties of the projection operator that maps an LDP to its tree-structured component.
\begin{lemma}\label{chapter2:lemma:pre:3}
Let $\X\in \R^{n\times n}, p,q \in \R[\X]$. Then
\begin{itemize}
    \item $(p+q)^{\rm Tr} =  p^{\rm Tr} + q^{\rm Tr}$. 
    \item If  $r\in \{p,q\}$ is connected and satisfies $ r^{\rm Tr}=0$ then $(pq)^{\rm Tr}=0$.
\end{itemize}
\end{lemma}
\begin{proof}
    The first property is straightforward from the definition of projection into $\mathcal{T}_{n, 2}$. For the second, assume without loss of generality $r=p$, i.e., $p$ is connected and $p^{\rm Tr}=0$. Note in particular that the latter implies $c_{p,\b{0}}=0$, and thus $c_{pq, \b{0}}=0$ as well. Assume that $c_{(pq)^{\rm Tr}, \alphaa}\neq 0$ for some $\alphaa \in \mathcal{T}_{n, 2}\setminus \{\b{0}\}$. Since $c_{(pq)^{\rm Tr}, \alphaa} =\sum_{\substack{\alphaa_1,\alphaa_2\\ \alphaa_1+\alphaa_2=\alphaa}} c_{p, \alphaa_1}c_{q,\alphaa_2}$, there must exist nonzero $\alphaa_1, \alphaa_2$ such that $\alphaa_1+\alphaa_2=\alphaa, c_{p,\alphaa_1}\neq 0$ and, $c_{q,\alphaa_2}\neq 0$. Since $p^{\rm Tr}=0$, it follows that $\alphaa_1\not \in \mathcal{T}_{n, 2}$ which is equivalent to : $\max(\alphaa_1)>2$ or $\alphaa_1\not \in \T$. In the former case we have $\max(\alphaa_1)>2\implies \max(\alphaa)>2$ and thus $\alphaa \not \in \mathcal{T}_{n, 2}$ which contradicts the claim $c_{(pq)^{\rm Tr},\alphaa}\neq 0$, while in the latter case it follows that $G_{\alphaa_1}$ has a cycle (since $p$ is connected), and therefore $G_{\alphaa}=G_{\alphaa_1}\cup G_{\alphaa_2}$ has a cycle, which contradicts $\alphaa \in \mathcal{T}_{n, 2}$ and concludes the proof.
\end{proof}

\subsection{Properties of Random Graphs and Matrices}
In this section, we collect probabilistic properties of random graphs and random matrix ensembles that will be used throughout the paper. The next two lemmas derive operator norm bounds for the ${\rm Bernoulli}$ ensemble, $\G(n, \tau_n/n),\ \tau_n=\omega(\log n)$ as well as the Gaussian ensembles ${\rm GOE}(n)$ and ${\rm GOE}^0(n)$. These estimates will be useful in our discussion leading to the rounding argument in Section~\ref{chapter2:section:rounding}, as well as in the AMP approximation by LDP proofs in Section~\ref{chapter2:section:3}.

\begin{lemma}\label{chapter2:lemma:random:matrix:sparse}
Let $\Y \distrib \G(n, \tau_n/n)$ with $\omega(\log n)\leq \tau_n\leq n$. Then, there exists a constant $c>0$ such that $\|\Y\| \leq c$ with high probability as $n\to \infty$.
\end{lemma}
\begin{proof}
Using Theorem 2.7 in \cite{benaych2020spectral} (with $q=\sqrt{\tau_n}$) and the discussion thereafter, it follows that for every $\varepsilon>0$, if $\frac{\tau_n}{\log n} = \omega(1)$, then $\|\b{Y}\|\leq 2+\varepsilon$ w.h.p. as $n\to \infty$, which readily yields the result of the lemma.
\end{proof}

\begin{lemma}\label{chapter2:lemma:random:matrix:dense}
Let $\X$ be sampled from ${\rm GOE}(n)$ or ${\rm GOE}^0(n)$. Then, there exists a constant $c>0$ such that $\|\X\|\leq c$ with high probability as $n\to \infty$.
\end{lemma}
\begin{proof}
The result follows directly from applying Theorem 2.1.22 in~\cite{guionnet2010} for the maximal eigenvalue of Wigner random matrices, which captures both the ${\rm GOE}(n)$ and ${\rm GOE}^0(n)$ ensembles.
\end{proof}

The next  lemma derives a standard tail bound on the degree of a vertex in an \ER graph. 
\begin{lemma}\label{chapter2:lemma:degree_tail}
    Let $v$ be a vertex in $\mathbb{G}(n, p)$, where $p\in (0,1)$ may depend on $n$, and $\mu = np$. For any $t > 1$, the degree of $v$ satisfies the following tail bound,
    \[
        \p(\deg(v) > t \mu) \le \exp\left( -\mu (t \log t - t + 1) \right).
    \]
\end{lemma}

\begin{proof}
    The degree of vertex $v$ is a sum of independent Bernoulli variables, i.e.,  $X = \deg(v) \distrib \text{Binomial}(n-1, p)$.  Using stochastic domination and standard Chernoff bounds for the upper tail of a sum of independent random variables (see e.g., Theorem 2.3.1 in~\cite{VershyninBook2018}), we have for any  $k \ge \mu$,
    \begin{align*}
        \p(X \ge k) &\le \p({\rm Binomial}(n, p) \geq k)\leq \left( \frac{e \mu}{k} \right)^k e^{-\mu}.
    \end{align*}
    Substituting the threshold $k = t \mu$ 
    \begin{align*}
        \p(X \geq t \mu) &\le \left( \frac{e \mu}{t \mu} \right)^{t \mu} e^{-\mu} = \exp\left( -\mu \l(t \log t - t  + 1\r) \right).
    \end{align*}
    This concludes the proof.
\end{proof}

The next lemma derives bounds on the number of vertices with large degree in an \ER graph, as well as on the number of edges incident to them. These estimates will be used to show that removing such edges has a negligible effect on the graph structure, which is key for several subsequent lemmas related to the rounding argument in Section~\ref{chapter2:section:rounding}.
\begin{lemma}\label{chapter2:lemma:removed}
    Let $G$ be a graph sampled from $\mathbb{G}(n, p)$, where $p\in (0,1)$ may depend on $n$, and $\mu = np$. Fix a positive constant $C$, and let $G'$ be obtained from $G$ by removing all edges incident to a node with degree larger than $C\mu$. Let $L$ be the set of removed edges, and $S$ the set of nodes with degree above $C\mu$. If $\mu\geq 2$ and $C\geq 4$, then, there exist positive constants $\gamma_1, \gamma_2, \gamma_3$ depending on $C$  such that
    \begin{align}
        \quad \frac{\E[|L|]}{n\mu} &\leq \gamma_2  e^{-\gamma_1 \mu}, \label{chapter2:lemma:E_removed}\\
        \frac{\E[|S|]}{n} &\leq  e^{-\gamma_3 \mu}. \label{chapter2:lemma:S_removed}
    \end{align}
\end{lemma}
\begin{proof}
The bound on $\E[|S|]$ follows readily from Lemma \ref{chapter2:lemma:degree_tail} as
\begin{align*}
    \E[|S|] &\leq n \cdot \p\l({\rm Binomial}(n-1, p) > C\mu\r) \leq n e^{-\eta(C) \mu},
\end{align*}
where $\eta(t)=\mu(t\log t - t + 1)$. Thus, it suffices to take $\gamma_3 = \eta(C)$ in (\ref{chapter2:lemma:S_removed}), and noting that $\eta(C)\geq \eta(4)>0$ as $t\mapsto \eta(t)$ is increasing for $t\geq 1$. We now derive the bound on $\E[|L|]$. Note that $L$ is  at most the sum of the degrees of the removed vertices. Using Lemma \ref{chapter2:lemma:degree_tail}, the probability that a vertex has degree larger than $ t\mu$ for $t>1$ is bounded as:
\[
    \p(\deg(v) > t \mu) \le  e^{ -\mu \eta(t)},
\]
where $\eta(t) = t \log t - t + 1 >0$ for $t > 1$. We then have
\begin{align}
    \E[|L|] &\leq  n \cdot \E\l[ {\rm deg}(v)  1_{{\rm deg}(v) > C\mu} \r] \nonumber \\
    &\leq n \sum_{k = \lceil C\mu \rceil}^{n} k \cdot \p(\deg(v) = k) \nonumber \\
    &\leq n \sum_{k = \lceil C\mu \rceil}^{n} ke^{-\mu\eta\l(\frac{k}{\mu}\r) }.  \label{chapter2:lemma:removed.1}
\end{align}
The function $x\mapsto \mu e^{-\mu\eta\l(\frac{x}{\mu}\r) }$ is decreasing on $[C\mu, n]$. We then have from (\ref{chapter2:lemma:removed.1})
\begin{align*}
    \E[|L|] &\leq n \int_{ C\mu  -2}^{\infty} x e^{-\mu\eta\l(\frac{x}{\mu}\r) } \, dx \\
    &= n \mu^2 \int_{C - 2/\mu}^{\infty} t e^{- \mu \eta(t)} \, dt,
\end{align*}
where we used the change of variables $x =t\mu$. Let $t_0 = C - 2/\mu$. Using the convexity of $\eta(t)$, we lower bound the function $\eta(t)$ by the tangent at $t_0$: for $t \ge t_0$, $\eta(t) \ge \eta(t_0) + (t - t_0) \log t_0$. Therefore,
\begin{align}
    \E[|L|] &\leq n \mu^2 \int_{t_0}^{\infty} t e^{-\mu [\eta(t_0) + (t - t_0) \log t_0]} \, dt \nonumber \\
    &= n \mu^2 e^{-\mu \eta(t_0)} \int_{0}^{\infty} (u + t_0) e^{-u (\mu \log t_0)} \, du \quad (\text{substituting } u = t - t_0) \nonumber \\
    &= n \mu^2 e^{-\mu \eta(t_0)} \left( \frac{t_0}{\mu \log t_0} + \frac{1}{(\mu \log t_0)^2} \right) \nonumber \\
    &= n \mu \, e^{-\mu \eta(t_0)} \left( \frac{t_0}{\log t_0} + \frac{1}{\mu (\log t_0)^2} \right).
\end{align}
Note then that  $\eta(t_0) = \eta(C - 2/\mu) \geq \eta(3) > 0$, and $\frac{t_0}{\log t_0} + \frac{1}{\mu (\log t_0)^2} \leq t_0 + 1 \leq C+1$, which readily yields the desired bound on (\ref{chapter2:lemma:E_removed}), and ends the proof.
\end{proof}

The next lemma is a corollary of Theorem~2.1 in~\cite{LeLevinaVershynin2017} and provides a way to control the operator norm of centered adjacency matrices of \ER graphs after removing a set of edges. Crucially, the resulting bound depends on the maximum degree of the pruned graph, a feature we will exploit when removing edges incident to high-degree vertices.
\begin{lemma}[Theorem 2.1 in \cite{LeLevinaVershynin2017}]\label{chapter2:theorem:regularization}
Let $G$ be sampled from $\mathbb{G}(n,p)$, where $p\in (0,1)$ may depend on $n$, and let $\mu = np$. Consider any subset $S$ consisting of at most $10/p$ vertices, and let $G'$ be the graph obtained from $G$ by removing all edges incident to vertices in $S$. Let $\tau'$ be the maximal degree of $G'$, and let $\b{A}, \b{A}'$ be the adjacency matrices of $G$ and $G'$ respectively. Then, there exists a constant $\gamma>0$ such that the following holds with probability at least $1 - n^{-1}$
\begin{align}
    \|\mathbf{A}' - \mathbb{E}[\mathbf{A}]\| \le \gamma \bigl(\sqrt{np} + \sqrt{\tau'}\bigr).
\end{align}
\end{lemma}
\begin{proof}
    The result of the lemma follows immediately from applying Theorem 2.1 in \cite{LeLevinaVershynin2017} with $r=1$.
\end{proof}

The next lemma applies the general regularization result from Lemma~\ref{chapter2:theorem:regularization} to sparse graphs $\Gb(n,t/n)$, and equivalently restates the result for $\G(n,t/n)$. We will use this result to regularize adjacency matrices of sparse graphs.
\begin{lemma}\label{chapter2:lemma:pruned_norm}
Let $G$ be a graph sampled from $ \Gb(n, t/n)$ where $d\leq t\leq \tau_n \leq n/2$ and $\tau_n$ satisfies $\omega(1)\leq \tau_n \leq o(n)$. Denote by $\b{A}$ be the adjacency matrix of $G$, and let the matrices $\b{U}$ and  $\Y$ be similarly given by  
\begin{equation}
     \b{U}_{ij} = \b{A}_{ij} - \frac{t}{n}, \quad  \Y_{ij} = \frac{-\b{U}_{ij}}{\sqrt{t \l(1 - \frac{t}{n} \r)}} .
\end{equation}
In particular, note that $\Y \distrib \G(n, t/n)$. We define the set of ``high-degree'' vertices $S_{\rm bad}$ in $G$ as:
\[
    S_{\rm bad} = \{ i \in [n] \mid \deg(i) > C t \},
\]
where $C \geq 4$ is a  constant. Let $G^{\rm prune}$ be the graph obtained from $G$ by removing all edges incident to $S_{\rm bad}$. Denote by $\b{A}^{\rm prune}$ the adjacency matrix of $G^{\rm prune}$, and similarly
\begin{equation}
    \b{U}^{\rm prune}_{ij} = \b{A}^{\rm prune}_{ij} - \frac{t}{n}, \quad  \Y^{\rm prune}_{ij} = \frac{-\b{U}^{\rm prune}_{ij}}{\sqrt{t \l(1 - \frac{t}{n} \r)}} .
\end{equation}
Then, there exists a  constant $\theta$ (depending on $C$)  such that with probability $1 - o(1)$ as $n\to \infty$ followed by $d\to \infty$, the norm of the regularized matrix $\Y^{\rm prune}$ satisfies:
\begin{align}
    \|\Y^{\rm prune}\| \le \theta. \label{chapter2:lemma:pruned_norm.1}
\end{align}
\end{lemma}
\begin{proof}
    Let $\b{A}, \b{A}^{\rm prune}, \b{U}, \b{U}^{\rm prune}, \Y, \Y^{\rm prune}$ and $S_{\rm bad}$ be as in the lemma statement, and let $S_{\rm limit}\subseteq [n]$ be the subset of $\lfloor 10n/t \rfloor$ nodes with the highest degrees. Let $G'$ be the graph obtained from $G$ by removing the edges incident to $S = S_{\rm bad} \cap S_{\rm limit}$. Similarly, let $\b{A}'$ be the adjacency matrix of $G'$ and define
    \begin{align*}
        \b{U}'_{ij}=\b{A}'_{ij} - \frac{t}{n}, \quad \Y'_{ij} = \frac{-\b{U}'_{ij}}{\sqrt{t\l(1 -\frac{t}{n}\r)}}.
    \end{align*}
    We apply Lemma \ref{chapter2:theorem:regularization} on $G\distrib \Gb(n, t/n)$ and $G'$ with  $\tau'$ the maximum degree in the pruned graph $G'$, and $S$ the set of removed vertices (which satisfies the theorem condition $|S|\leq 10n/t$ by construction). Then, the following holds with probability at least $1-n^{-1}$
    \begin{align*}
        \|\mathbf{A}' - \E \mathbf{A}\| \leq \gamma(\sqrt{t} + \sqrt{\tau'}),
    \end{align*}
    where $\gamma$ is an explicit constant, and $\tau'$ is a random variable. In our notation, this also implies
    \begin{align*}
        \|\Y'\| \leq \gamma \frac{\sqrt{t} + \sqrt{\tau'}}{\sqrt{t\l(1 - \frac{t}{n}\r)}} \leq 2\gamma \l( 1 + \sqrt{\frac{\tau'}{t}}\r), \numberthis \label{chapter2:lemma:pruned_norm.2}
    \end{align*}
    where we used $ t \leq n/2$ in the last inequality. We next show that w.h.p. as $n\to \infty$ followed by $d \to \infty$, we have $\tau' \leq Ct$. Note that 
    \begin{align*}
       \{|S_{\rm bad}| \leq 10n/t\} = \{S_{\rm bad} \subseteq S\} \subseteq \{ \tau' \leq Ct \},  \numberthis \label{chapter2:lemma:pruned_norm.3}   
    \end{align*}
    and thus $\p(|S_{\rm bad}| \leq 10n /t)  \leq \p(\tau' \leq Ct)  $. It is thus sufficient to show that $\p(|S_{\rm bad}|\leq 10n/t )=1-o(1)$, which we do next.  We have by Lemma \ref{chapter2:lemma:removed} (in our notation, the set of high-degree vertices is $S_{\rm bad}$,  and $\mu = t$)
    \begin{align*}
        \E[|S_{\rm bad}|] \leq ne^{-\gamma t},
    \end{align*}  
    where $\gamma=\gamma(C)$ is a constant. Hence, we have by Markov's inequality in the regime $t\to\infty$,
    \begin{align*}
        \p\l(|S_{\rm bad}| \geq \lfloor 10n/t \rfloor\r) &\leq \frac{ne^{-\gamma t}}{\lfloor 10n /t  \rfloor} \leq O(t e^{-\gamma t}) = o(1).
    \end{align*}
    Combining the above with (\ref{chapter2:lemma:pruned_norm.3}), it follows that w.h.p. as $n\to\infty$ followed by  $d\to\infty$, we have $\tau' \leq Ct$. Moreover, note that conditionally on $|S_{\rm bad}|\leq 10n/t$, the graphs $G'$ and $G^{\rm prune}$ are identical. Using the previous observations and the fact that the bound $(\ref{chapter2:lemma:pruned_norm.2})$ holds w.h.p. as well, it follows from conditioning on the high probability event $\{|S_{\rm bad}| \leq 10n/t\}$ that w.h.p.
    \begin{align*}
        \|\Y^{\rm prune}\| &= \|\Y'\| \leq 2\gamma \l(1 + \sqrt{C} \r) = \theta(C),
    \end{align*}
    where $\theta(C)$ is a constant depending on $C$. This ends the proof of the lemma.
\end{proof}

\subsection{Probabilistic Bounds on LDPs}
In this section, we establish probabilistic bounds for LDPs evaluated on the random graph and matrix models considered in this paper. While some proofs are technically involved, they are elementary in nature and typically rely on a combination of asymptotic analysis and combinatorial counting. The main purpose of the results in this section is to support the Taylor expansion analyses used in our interpolation arguments. The next lemma collects some basic moment estimates that will be used later when evaluating expectations of LDP monomials.
\begin{lemma}\label{chapter2:lemma:pre:sparse}
Let  $\Y\distrib \G(n, t/n)$ where $d\leq t\leq \tau_n$ and $\tau_n$ satisfies $\omega(1)\leq \tau_n \leq o(n)$. Then 
\begin{enumerate}
    \item $\E[\Y_{ij}]=0,\ \E[\Y_{ij}^2]=\frac{1}{n},\ \forall 1\leq i<j \leq n$.
    \item $\E[\Y_{ij}^{k}] =\frac{ (-1)^k+o(1)}{n t^{\frac{k}{2}-1}},\  \E[|\Y_{ij}|^{k}] =\frac{ 1+o(1)}{n t^{\frac{k}{2}-1}},\ \forall k\geq 2,\ \forall 1\leq i<j \leq n$, where $k$ is constant (i.e., does not depend on $n, d$).
    \item  For all $ \alphaa \in \mathcal{U}^{n\times n}$ such that $ \forall 1\leq i<j\leq n,\ \alphaa_{ij}\neq 1$ and $\|\alphaa\|_1$ is bounded by a constant independent of $n$, it holds
    \begin{align*}
        \E[\Y^{\alphaa}] &= ((-1)^{\|\alphaa\|_1}+o(1)) n^{-\|\alphaa\|_0} t^{-\frac{\|\alphaa\|_1}{2} + \|\alphaa\|_0},\\
        \E[|\Y^{\alphaa}|] &= (1+o(1)) n^{-\|\alphaa\|_0} t^{-\frac{\|\alphaa\|_1}{2} + \|\alphaa\|_0},
    \end{align*}
    where $o(1)$ is a sequence $\varepsilon_{n, d}$ satisfying $\lim_{d\to \infty} \lim\sup_{n\to \infty} |\varepsilon_{n,d}|=0$.
\end{enumerate}
\end{lemma}
\begin{proof}
    The first property is immediate from the definition of the distributions $\G(n,t/n)$. For the second property,  we have for $k\geq 2$
    \begin{align*}
        \E[\Y_{ij}^k] &= \frac{(-1)^k}{(t(1-\frac{t}{n}))^{\frac{k}{2}}} \left( \l(-\frac{t}{n}\r)^{k}\l(1-\frac{t}{n}\r) + \l(1-\frac{t}{n}\r)^k \frac{t}{n} \right) \\
        &\sim  (-1)^k \frac{\frac{t}{n} (1-\frac{t}{n})^k}{(t(1-\frac{t}{n}))^{\frac{k}{2}}} \\
        &= \frac{(-1)^k+o(1)}{n t^{\frac{k}{2}-1}},
    \end{align*}
    where we used $t/n=o(1)$. Similar computations yield the  asymptotics for $\E[|\Y_{ij}|^k]$. Finally, the second property yields
    \begin{align*}
    \E[\Y^{\alphaa}] = \prod_{\alphaa_{ij}>0} \frac{(-1)^{\alphaa_{ij}}+o(1)}{n t^{\frac{\alphaa_{ij}}{2}-1}} = ((-1)^{\|\alphaa\|_1}+o(1)) n^{-\|\alphaa\|_0} t^{-\frac{\|\alphaa\|_1}{2}+\|\alphaa\|_0}.
    \end{align*}
     A similar computation yields the asymptotic bound for $\E[|\Y^{\alphaa}|]$. This concludes the proof of the Lemma.
\end{proof}

The next lemma shows that, for the distributions considered in this paper, tree-structured LDPs suffice to represent general LDPs. This justifies our focus on this class and allows us to systematically restrict our attention to tree-structured LDPs throughout this paper.

\begin{lemma}\label{chapter2:lemma:pre:6}
Let $\b{X}\distrib{\rm GOE}^0(n)$  and $\Y\distrib\G(n,t/n)$, where $d\leq t \leq \tau_n$ and $\tau_n$ satisfies $\omega(1) \leq \tau_n \leq o(n)$. Suppose $p\in \mathbb{R}[\b{X}]_{\leq \Delta}$ and $S\subseteq [n]$ satisfy
     \begin{enumerate}
         \item the LDP $p$ contains all nodes in $S$, i.e., $p\in \V_S$. 
        \item For each $\alphaa$ with $c_{p,\alphaa}\ne 0$ either of the following holds
         \begin{itemize}
             \item $\alphaa$ is connected. 
             \item There exists $\alphaa_1, \alphaa_2 \in \mathcal{U}^{n\times n} \setminus \mathcal{T}_{n, 2}$ with disjoint supports satisfying $\alphaa = \alphaa_1 + \alphaa_2$ and $\alphaa_1,\alphaa_2$ are connected.
         \end{itemize}
     \end{enumerate}
    Then, for every fixed $\Delta$, there exists $c=c(\Delta)$ independent of $n$ and $d$ such that
        \begin{align}
            \l| \E\l[  p (\b{X})-  p^{\rm Tr}(\b{X})\r] \r| & \leq \frac{c \|p\|_{\rm coef}}{n^{|S|}},  \label{chapter2:lemma.4.1.1}\\
            \l| \E\l[  p(\b{Y}) - p^{\rm Tr}(\b{Y})\r]\r| & \leq \frac{c \|p\|_{\rm coef}}{n^{|S|-2} t}.   \label{chapter2:lemma.4.1.2}
        \end{align}
    Furthermore, when $p$ is connected,  the following tighter concentration for $\Y $ holds
    $$\l| \E\l[  p(\b{Y}) - p^{\rm Tr}(\b{Y})\r]\r|  \leq \frac{c \|p\|_{\rm coef}}{n^{|S|-1} \sqrt{t}}. $$
\end{lemma}
We note that property (\ref{chapter2:lemma.4.1.1}) is known in the literature (see \cite{bayati2015universality} for a reference). We provide a  proof here for completeness.

\begin{proof}
We first show (\ref{chapter2:lemma.4.1.1}). Introduce 
\begin{align*}
\mathcal{K} &\triangleq \{\alphaa \in \mathcal{U}^{n\times n} \setminus \mathcal{T}_{n, 2} \mid \alphaa \text{ is connected}\},\\
\mathcal{W} &\triangleq \{(\alphaa_1, \alphaa_2)\in (\mathcal{U}^{n\times n}\setminus \mathcal{T}_{n, 2})^2 \mid  \alphaa_1, \alphaa_2 \text{ are connected},\ { \rm Supp}(\alphaa_1) \cap {\rm Supp}(\alphaa_2)=\emptyset \}.
\end{align*}
We have
\begin{align*}
    |\E[p(\X) - p^{\rm Tr}(\X)]| &\leq \l|\sum_{\alphaa \in \mathcal{K}} c_{p, \alphaa} \E[\X^{\alphaa}]\r| + \l|\sum_{(\alphaa_1, \alphaa_2) \in \mathcal{W}} c_{p, \alphaa_1+\alphaa_2} \E[\X^{\alphaa_1+\alphaa_2}]\r|.
\end{align*}
Note that $\forall \alphaa \in \U^{n\times n}, \E[\b{X}^{\alphaa}] = \gamma_{\alphaa} n^{-\frac{\|\alphaa\|_1}{2}} \b{1}_{2|\alphaa}$, where   $\gamma_{\alphaa}= \prod_{\substack{1\leq i<j \leq n \\2| \alphaa_{ij} , \alphaa_{ij}>0}} (\alphaa_{ij}-1)!!$. Moreover $\gamma_{\alphaa}\leq c_{\Delta}$ for some constant $c_{\Delta}$ depending only on $\Delta$. We then have
\begin{align*}
    |\E[p(\b{X}) - p^{\rm Tr}(\b{X})]| &\leq c_{\Delta}\|p\|_{\rm coef} \l(\sum_{\substack{\alphaa \in \mathcal{K} \\ 2|\alphaa }} n^{-\frac{\|\alphaa\|_1 }{2}} +  \sum_{\substack{(\alphaa_1, \alphaa_2) \in \mathcal{W}\\ 2|\alphaa_1, 2|\alphaa_2}} n^{-\frac{\|\alphaa_1\|_1 + \|\alphaa_2\|_1}{2}} \r).
\end{align*}
We claim that $\forall \alphaa' \in \{\alphaa, \alphaa_1, \alphaa_2\}$ appearing in the sums above, we have $|V_{\alphaa'}|\leq \frac{\|\alphaa'\|_1}{2}$.  There are two cases
\begin{itemize}
    \item  Case 1: $\alphaa' \not \in \mathcal{T}$.
    
    As $G_{\alphaa'}$ is connected but is not a tree, it follows that $|V_{\alphaa'}|\leq |E_{\alphaa'}|=\|\alphaa'\|_0$. Using $2|\alphaa'$, we have $\|\alphaa'\|_0\leq \frac{\|\alphaa'\|_1}{2}$ and thus $|V_{\alphaa'}|\leq \frac{\|\alphaa'\|_1}{2}$. 
    
    \item Case 2: $\alphaa'  \in \mathcal{T}$.
    
    We have $|V_{\alphaa'}|=|E_{\alphaa'}|+1=\|\alphaa'\|_0 +1$, moreover $2|\alphaa'$ and $ \alphaa'\not \in \mathcal{T}_{n, 2}, \alphaa'\in \mathcal{T}$ implies that  $\max(\alphaa')\geq 4$, therefore $\|\alphaa'\|_0 \leq \frac{\|\alphaa'\|_1-2}{2}= \frac{\|\alphaa'\|_1}{2}-1$. Hence $|V_{\alphaa'}|\leq \frac{\|\alphaa'\|_1}{2}$.
\end{itemize}
Thus the claim is verified. Given an integer $0\leq a\leq \Delta$ and a graph $H$, introduce the set
\begin{align*}
    \mathcal{F}_{a, H} &\triangleq \l\{ \alphaa \in \mathcal{K} \mid \|\alphaa\|_1=a,\  G_{\alphaa} \sim H ,\ 2|\alphaa\r\} \\
    &\cup \l\{ \alphaa \in \U^{n\times n} \mid\|\alphaa\|_1=a,\  G_{\alphaa} \sim H,\  \exists (\alphaa_1, \alphaa_2) \in \mathcal{W},\ \alphaa = \alphaa_1+\alphaa_2,\  2|\alphaa_1,\ 2|\alphaa_2\r\},
\end{align*} 
where $G_{\alphaa}\sim H$ means $G_{\alphaa}$ is isomorphic to $H$. Recalling $S\subseteq G_{\alphaa}$, it follows $|\mathcal{F}_{a,H}|\leq \beta_{\Delta }n^{|V_{H}|-|S|}$ where $\beta_{\Delta}$ is a constant depending on $\Delta$. From the previous analysis, it holds that $\forall \alphaa \in \mathcal{F}_{a,H},|V_H|=|V_{\alphaa}|\leq \frac{\|\alphaa\|_1}{2}$, therefore $|\mathcal{F}_{a,H}|\leq \beta_{\Delta }n^{\frac{\|\alphaa\|_1}{2}-|S|}$. Hence
\begin{align*}
    \l|\E\l[p(\X) - p^{\rm Tr}(\X)\r]\r| &\leq c_{\Delta}\|p\|_{\rm coef} \sum \sum_{\alphaa\in \mathcal{F}_{a,H}}  n^{-\frac{\|\alphaa\|_1}{2}}\\
    &\leq c_{\Delta}\|p\|_{\rm coef} \sum  \beta_{\Delta} n^{-|S|}\\
    &\leq c \|p\|_{\rm coef} n^{-|S|},
\end{align*}
where the first sum is taken over all pairs $(a,H)$ of (integer, graph) such that $0\leq a \leq \Delta$ and $H$ is a graph in the orbit generated by $G_{\alphaa}$ with $\alphaa\in  \U^{n\times n} \setminus \mathcal{T}_{n, 2}, c_{p, \alphaa}\neq 0$, and  $c$ is a constant depending on $\Delta$. This concludes the proof of the first part of the lemma.

We now prove the second part. We have $\U^{n\times n}= \U^{n\times n}_{1} \cup \U^{n\times n}_{*}$ where $\U^{n\times n}_{1}, \U^{n\times n}_{*}$ are given in (\ref{chapter2:U:notation}). Note in particular that $\forall \alphaa\in \U^{n\times n}_{1}, \E[\Y^{\alphaa}]=0$. Furthermore, we have by Lemma \ref{chapter2:lemma:pre:sparse} that $\forall \alphaa\in \U^{n\times n}_{*}, \E[\Y^{\alphaa}]=((-1)^{\|\alphaa\|_1}+o(1))  n^{-\|\alphaa\|_0} t^{-\frac{\|\alphaa\|_1}{2} + \|\alphaa\|_0 }$. Hence

\begin{align*}
    |\E[p(\Y) -  p^{\rm Tr}(\Y)]| &=\l|\sum_{\alphaa \in \U^{n\times n} \setminus \mathcal{T}_{n, 2}}  c_{p, \alphaa} \E [ \Y^{\alphaa}]\r|\\
    &=\l|\sum_{\alphaa \in \U_*^{n\times n} \setminus \mathcal{T}_{n, 2}}  c_{p, \alphaa} \E [ \Y^{\alphaa}]\r|\\
    &\leq \l|\sum_{\alphaa \in \mathcal{K} \cap \U_*^{n\times n}} c_{p, \alphaa} \E[\Y^{\alphaa}]\r| + \l|\sum_{\substack{(\alphaa_1, \alphaa_2) \in \mathcal{W}\\ \alphaa_1 + \alphaa_2 \in \U_*^{n\times n}}} c_{p, \alphaa_1+\alphaa_2} \E[\Y^{\alphaa_1+\alphaa_2}]\r| \\
    &\leq  \|p\|_{\rm coef} (1+o(1)) \l( \underbrace{\sum_{\alphaa \in \mathcal{K} \cap \U_*^{n\times n}} n^{-\|\alphaa\|_0} t^{-\frac{\|\alphaa\|_1}{2} + \|\alphaa\|_0}}_{A} \r.  \\&+ \l. \underbrace{\sum_{\substack{(\alphaa_1, \alphaa_2) \in \mathcal{W}\\ \alphaa_1 + \alphaa_2 \in \U_*^{n\times n}}} n^{-\|\alphaa_1\|_0 - \|\alphaa_2\|_0} t^{-\frac{\|\alphaa_1\|_1 + \|\alphaa_2\|_1}{2} + \|\alphaa_1\|_0+\|\alphaa_2\|_0}}_{B} \r).
\end{align*}
We will bound the above summation similarly to the Gaussian case. We first consider the summands in $A$ indexed by $\alphaa \in \mathcal{K}\cap\U_*^{n\times n} $. We distinguish between two cases 
\begin{itemize}
    \item Case 1: $\alphaa\in \mathcal{T}$. 
    
    Then $|V_{\alphaa}|= |E_{\alphaa}|+1=\|\alphaa\|_0 +1$.  Furthermore $\alphaa \in \U_*^{n\times n} \implies \|\alphaa\|_0 \leq \frac{\|\alphaa\|_1}{2}$, and the equality occurs if and only if every nonzero entry in $\alphaa$ equals $2$, but that would imply $\alphaa \in \mathcal{T}_{n, 2}$. Therefore the inequality is strict, i.e. $\|\alphaa\|_0 \leq \frac{\|\alphaa\|_1-1}{2}$. 
    
     \item Case 2: $\alphaa \not \in \mathcal{T}$.
     
     Then $|V_{\alphaa}|\leq |E_{\alphaa}|=\|\alphaa\|_0$. Furthermore $\alphaa \in \U_*^{n\times n} \implies \|\alphaa\|_0 \leq \frac{\|\alphaa\|_1}{2}$.
\end{itemize}
Given an integer $0\leq a\leq \Delta $ and a graph $H$, introduce the sets
\begin{equation}
    \begin{aligned}
        \mathcal{F}^{1}_{a, H} &\triangleq \{ \alphaa \in (\mathcal{K} \cap \U^{n\times n}_{*}  \cap \mathcal{T}) \setminus \mathcal{T}_{n, 2} \mid \|\alphaa\|_1 =a,\  G_{\alphaa} \sim H \},\\
        \mathcal{F}^{2}_{a, H} &\triangleq \{ \alphaa \in \mathcal{K} \cap \U^{n\times n}_{*} \setminus (\mathcal{T}_{n, 2}\cup \mathcal{T}) \mid \|\alphaa\|_1 =a,\  G_{\alphaa} \sim H \}.
    \end{aligned}
\end{equation}
Recalling $S\subseteq G_{\alphaa}$ it follows that there exists constants $\beta_{\Delta}^{1}, \beta_{\Delta}^2$  depending on $\Delta$ such that, $\forall \alphaa\in \mathcal{F}_{a,H}^{1}, |\mathcal{F}_{a,H}^{1}|\leq \beta_{\Delta}^1 n^{\|\alphaa\|_0+1 - |S|}$ and $\forall \alphaa\in \mathcal{F}_{a,H}^{2}, |\mathcal{F}_{a,H}^{2}|\leq \beta_{\Delta}^2 n^{\|\alphaa\|_0 - |S|}$.   We then have
\begin{align*}
    A &= \sum \sum_{\alphaa \in \mathcal{F}_{a,H}^1}n^{-\|\alphaa\|_0} t^{-\frac{\|\alphaa\|_1}{2} + \|\alphaa\|_0} +\sum \sum_{\alphaa \in  \mathcal{F}_{a,H}^2}n^{-\|\alphaa\|_0} t^{-\frac{\|\alphaa\|_1}{2} + \|\alphaa\|_0}\\
    &\leq \sum \sum_{\alphaa \in  \mathcal{F}_{a,H}^1}n^{-\|\alphaa\|_0} t^{-\frac{1}{2}} +\sum \sum_{\alphaa \in  \mathcal{F}_{a,H}^2}n^{-\|\alphaa\|_0}\\
    &\leq \sum \beta_{\Delta}^1 n^{-|S|+1}t^{-\frac{1}{2}} + \sum \beta_{\Delta}^2 n^{-|S|}\\
    &= O \left( n^{-|S|+1}t^{-\frac{1}{2}} \left(1 +  \frac{t^{\frac{1}{2}}}{n}\right) \right)\\
    &= O \left( n^{-|S|+1}t^{-\frac{1}{2}} \right),
\end{align*}
where the first sums are taken over all pairs $(a,H)$ of (integer, graph) such that $0\leq a\leq \Delta$ and $H$ is a graph in the orbit generated by $G_{\alphaa}$ with $\alphaa\in  \U_*^{n\times n} \setminus \mathcal{T}_{n, 2}, c_{p, \alphaa}\neq 0$. We now bound $B$ similarly. Let $\alphaa \triangleq  \alphaa_1 + \alphaa_2$ with $\alphaa_1, \alphaa_2$ as in $B$. We consider three cases

\begin{itemize}

    \item Case 1: $\alphaa_1, \alphaa_2 \in \mathcal{T}$.
    
    We have $|V_{\alphaa_1}|=|E_{\alphaa_1}|+1 = \|\alphaa_1\|_0+1$ and $|V_{\alphaa_2}|=|E_{\alphaa_2}|+1=\|\alphaa_2\|_0+1$. Therefore $|V_{\alphaa}| = \|\alphaa_1\|_0 + \|\alphaa_2\|_0 + 2$. Furthermore, similarly to Case 1 in the previous analysis of $A$, we obtain $\|\alphaa_1\|_0 \leq \frac{\|\alphaa_1\|_1-1}{2}, \|\alphaa_2\|_0 \leq \frac{\|\alphaa_2\|_1-1}{2}$.

    \item Case 2: $\alphaa_1, \alphaa_2 \not \in \mathcal{T}$.

    Since $G_{\alphaa_1}$ is connected, it follows that $|V_{\alphaa_1}|\leq |E_{\alphaa_1}|=\|\alphaa_1\|_0$. Similarly $|V_{\alphaa_2}|\leq \|\alphaa_2\|_0$. Therefore $|V_{\alphaa} |=|V_{\alphaa_1}|+|V_{\alphaa_2}| \leq \|\alphaa_1\|_0 + \|\alphaa_2\|_0$. Furthermore, similarly to Case 2 in the previous analysis of $A$, we obtain $\|\alphaa_1\|_0 \leq \frac{\|\alphaa_1\|_1}{2}, \|\alphaa_2\|_0\leq \frac{\|\alphaa_2\|_1}{2}$.

    \item Case 3: $\alphaa_1 \in \T, \alphaa_2 \not \in \T$.
    
    Combining the arguments of Cases 1, 2 above we obtain $|V_{\alphaa}|\leq \|\alphaa_1\|_0+\|\alphaa_2\|_0 +1$, and $\|\alphaa_1\|_0 \leq \frac{\|\alphaa_1\|_1-1}{2}, \|\alphaa_2\|_0 \leq \frac{\|\alphaa_2\|_1}{2}$.
\end{itemize}
Similarly to $A$, introduce the sets
\begin{equation}
    \begin{aligned}
        \mathcal{F}^{1}_{a, H} &\triangleq \l\{ (\alphaa_1, \alphaa_2)\in \mathcal{W} \mid \alphaa_1+\alphaa_2 \in  \U^{n\times n}_{*},\ \alphaa_1, \alphaa_2\in \T ,\ \|\alphaa\|_1 =a,\  G_{\alphaa} \sim H \r\},\\
        \mathcal{F}^{2}_{a, H} &\triangleq \l\{ (\alphaa_1, \alphaa_2)\in \mathcal{W} \mid \alphaa_1+\alphaa_2  \in  \U^{n\times n}_{*},\ \alphaa_1, \alphaa_2\not\in \T ,\ \|\alphaa\|_1 =a,\  G_{\alphaa} \sim H \r\},\\
        \mathcal{F}^{3}_{a, H} &\triangleq \l\{ (\alphaa_1, \alphaa_2)\in \mathcal{W} \mid \alphaa_1+\alphaa_2  \in  \U^{n\times n}_{*},\ \alphaa_1\in \T,\ \alphaa_2 \not\in \T ,\  \|\alphaa\|_1 =a,\  G_{\alphaa} \sim H \r\}.
    \end{aligned}
\end{equation}
There exists constants $\beta_{\Delta}^{1}, \beta_{\Delta}^2, \beta_{\Delta}^3$  depending on $\Delta$ such that, $\forall (\alphaa_1, \alphaa_2)\in \mathcal{F}_{a,H}^{1}, |\mathcal{F}_{a,H}^{1}|\leq \beta_{\Delta}^1 n^{\|\alphaa_1\|_0+\|\alphaa_2\|_0+2 - |S|}$, $\forall  (\alphaa_1, \alphaa_2)\in \mathcal{F}_{a,H}^{2}, |\mathcal{F}_{a,H}^{2}|\leq \beta_{\Delta}^2 n^{\|\alphaa_1\|_0+\|\alphaa_2\|_0 - |S|}$, and  $ \forall (\alphaa_1, \alphaa_2)\in \mathcal{F}_{a,H}^{3}, \allowbreak|\mathcal{F}_{a,H}^{3}|\leq \beta_{\Delta}^3 n^{\|\alphaa_1\|_0+\|\alphaa_2\|_0 +1 - |S|}$. We then have
\begin{align*}
    B &= \sum \sum_{(\alphaa_1, \alphaa_2) \in \mathcal{F}_{a,H}^1}n^{-\|\alphaa_1\|_0-\|\alphaa_2\|_0} t^{-\frac{\|\alphaa_1\|_1 +\|\alphaa_2\|_1}{2} + \|\alphaa_1\|_0 + \|\alphaa_2\|_0} \\
    &+\sum \sum_{(\alphaa_1, \alphaa_2) \in \mathcal{F}_{a,H}^2}n^{-\|\alphaa_1\|_0-\|\alphaa_2\|_0} t^{-\frac{\|\alphaa_1\|_1 +\|\alphaa_2\|_1}{2} + \|\alphaa_1\|_0 + \|\alphaa_2\|_0}\\
    &+\sum \sum_{(\alphaa_1, \alphaa_2) \in \mathcal{F}_{a,H}^3}n^{-\|\alphaa_1\|_0-\|\alphaa_2\|_0} t^{-\frac{\|\alphaa_1\|_1 +\|\alphaa_2\|_1}{2} + \|\alphaa_1\|_0 + \|\alphaa_2\|_0}\\
    &\leq \sum \beta_{\Delta}^1 n^{-|S|+2}t^{-1} + \sum \beta_{\Delta}^2 n^{-|S|}  + \sum \beta_{\Delta}^3 n^{-|S|+1} t^{-\frac{1}{2}}\\
    &= O \left( n^{-|S|+2}t^{-1} \left(1 + \frac{t}{n^2}+ \frac{t^{\frac{1}{2}}}{n}\right) \right)\\
    &= O \left( n^{-|S|+2}t^{-1} \right).
\end{align*}
Combining the bounds on $A$ and $B$ yields
\begin{align*}
    \l|\E\l[p(\Y) -  p^{\rm Tr}(\Y)\r]\r| \leq c \|p\|_{\rm coef} n^{-|S|+2} t^{-1},
\end{align*}
where $c$ is a constant. Finally, note that if we further assume $p$  connected, then $B$ is removed from our analysis. Therefore we obtain the tighter bound
\begin{align*}
    \l|\E\l[p(\Y) -  p^{\rm Tr}(\Y)\r]\r| \leq c \|p\|_{\rm coef} n^{-|S|+1} t^{-\frac{1}{2}},
\end{align*}
from the analysis of $A$ only. This concludes the proof of the second part of the lemma.

\end{proof}

The next lemma derives estimates on the scaling of connected LDPs. It complements Lemma~\ref{chapter2:lemma:pre:6} by showing the typical order of magnitude one should expect from LDPs evaluated on the distributions considered in this paper.
\begin{lemma}\label{chapter2:lemma:+1}
Suppose $1\leq d \leq t \leq \tau_n$ where $\tau_n$ satisfies $\omega(1)\leq \tau_n \leq o(n)$.  Let $\b{Y}\distrib \G(n, t/n)$ and $\b{X} \distrib {\rm GOE}^0(n)$. Let $\Delta>0$ be a constant independent of $n$ and $d$. Suppose $p$ is a connected LDP with ${\rm Deg}(p) \leq \Delta$ and $p$ contains all nodes in set $K \subseteq [n]$, i.e., $p\in \V_K$. Then, there exists a constant $c = c(\Delta)$ such that
\begin{align*}
    \max\l\{\l| \E\l[p(\b{X})\r] \r| \ \l| \E\l[p(\b{Y})\r] \r| \r\}  \leq |p(\b{0})|+ \frac{ c\|p\|_{\rm coef}}{n^{|K|-1}}. \numberthis \label{chapter2:lemma:+1.1}
\end{align*}
\end{lemma}    
\begin{proof}
Throughout this proof, we denote by $c$ different constants depending on $\Delta$.  We have
\begin{align*}
    \l| \E\l[p(\b{X}) \r] \r| &\leq |c_{p, \b{0} }| + \|p\|_{\rm coef} \l| \sum_{\pmb{\alpha}} \E\l[ \b{X}^{\pmb{\alpha}} \r] \r|\\
    &\leq|p(\b{0})| +  c \|p\|_{\rm coef}  \sum_{\pmb{\alpha}} 1_{2 | \pmb{\alpha}}n^{-\frac{\|\pmb{\alpha}\|_1}{2}}, \numberthis \label{chapter2:lemma:+1.2}
\end{align*}
where the sums are taken over $\{\pmb{\alpha}\in \mathcal{U}^{n\times n}\setminus \{\b{0}\} \mid c_{p, \pmb{\alpha}}\neq 0\}$, and we used the fact that $|\E\l[ \b{X}^{\pmb{\alpha}} \r]| \leq c 1_{2 | \pmb{\alpha}}n^{-\frac{\|\pmb{\alpha}\|_1}{2}}$, where $c=c(\Delta)$ is a constant. Let $\pmb{\alpha}$ as in the sum above with $2|\pmb{\alpha}$. We next show that $|V_{\pmb{\alpha}}|\leq \frac{\|\pmb{\alpha}\|_1}{2}+1$. As $\pmb{\alpha}$ is connected (because $p$ is connected), then $|V_{\pmb{\alpha}}|\leq |E_{\pmb{\alpha}}|+1 = \|\pmb{\alpha}\|_0 + 1$. Since $2|\pmb{\alpha}$, it follows that $\|\pmb{\alpha}\|_0 \leq \frac{\|\pmb{\alpha}\|_1}{2}$. Thus, $|V_{\pmb{\alpha}}|\leq \frac{\|\pmb{\alpha}\|_1}{2} + 1$.

For an integer $0\leq a \leq \Delta$ and a connected graph $H$, let  $\mathcal{F}_{a, H}$ be the set of all $\pmb{\alpha}$ with $c_{p, \pmb{\alpha}} \neq 0$, and satisfying $\|\pmb{\alpha}\|_1 = a$ and $G_{\pmb{\alpha}} \sim H$ (here, $\sim$ indicates graph isomorphism). Then, there exists a constant $c = c(\Delta)$ such that for all $\pmb{\alpha} \in \mathcal{F}_{a, H}$ we have $|\mathcal{F}_{a, H}| \leq c n^{\frac{\|\pmb{\alpha}\|_1}{2} +1 - |K|}$. Using this bound  in (\ref{chapter2:lemma:+1.2}) yields
\begin{align*}
    \l| \E\l[p(\b{X})  \r]\r| - |p(\b{0})|&\leq  c \|p\|_{\rm coef} \sum_{a, H} \sum_{\pmb{\alpha} \in \mathcal{F}_{a, H}} n^{-\frac{\|\pmb{\alpha}\|_1}{2}}\\
    &\leq c \|p\|_{\rm coef} \sum_{a, H} n^{1- |K|} \\
    &\leq c \|p\|_{\rm coef} n^{-|K|+1}. \numberthis \label{chapter2:lemma:+1.3}
\end{align*}
Similarly, we have
\begin{align*}
    \l| \E\l[p(\b{Y}) \r] \r| - |p(\b{0})|&\leq \|p\|_{\rm coef} \l| \sum_{\pmb{\alpha}} \E\l[ \b{Y}^{\pmb{\alpha}} \r] \r|\\
    &\leq  \|p\|_{\rm coef}\l| \sum_{\pmb{\alpha}} 1_{\pmb{\alpha}\in  \mathcal{U}^{n\times n}_*}(1+o(1)) n^{- \|\pmb{\alpha}\|_0} t^{-\frac{\|\pmb{\alpha}\|_1}{2} + \|\pmb{\alpha}\|_0}\r| \numberthis \label{chapter2:lemma:+1.4}\\
    &\leq   2\|p\|_{\rm coef}\sum_{\pmb{\alpha}} 1_{\pmb{\alpha} \in \mathcal{U}^{n\times n}_*}n^{- \|\pmb{\alpha}\|_0} t^{-\frac{\|\pmb{\alpha}\|_1}{2} + \|\pmb{\alpha}\|_0}, \numberthis\label{chapter2:lemma:+1.5}
\end{align*}
where we used Lemma \ref{chapter2:lemma:pre:sparse} in (\ref{chapter2:lemma:+1.4}). Let $\pmb{\alpha}$ as in the sum above with $\pmb{\alpha}\in \mathcal{U}^{n\times n}_*$. Since none of the entries of $\pmb{\alpha}$ equal $1$, it follows that $\|\pmb{\alpha}\|_0 \leq \frac{\|\pmb{\alpha}\|_1}{2}$.  Moreover, as $\pmb{\alpha}$ is connected, we have  $|V_{\pmb{\alpha}}|\leq |E_{\pmb{\alpha}}|+1 = \|\pmb{\alpha}\|_0+1$. Let $\mathcal{F}_{a, H}$ as defined previously.  Then, there exists  constant $c = c(\Delta)$ such that for all $\pmb{\alpha} \in \mathcal{F}_{a, H}$ we have $|\mathcal{F}_{a, H}| \leq c n^{\|\pmb{\alpha}\|_0 +1 - |K|}$. Using this bound  together with $\|\pmb{\alpha}\|_0 \leq \frac{\|\pmb{\alpha}\|_1}{2}$ in (\ref{chapter2:lemma:+1.5}), we obtain
\begin{align*}
    \l| \E\l[p(\b{Y})  \r]\r| - |p(\b{0})| &\leq c \|p\|_{\rm coef} \sum_{a, H} \sum_{\pmb{\alpha} \in \mathcal{F}_{a, H}} n^{-\|\pmb{\alpha}\|_0}\\
    &\leq c \|p\|_{\rm coef} \sum_{a, H} n^{1- |K|} \\
    &\leq c \|p\|_{\rm coef} n^{-|K|+1}, 
\end{align*}
combining the above with (\ref{chapter2:lemma:+1.3}) yields the result of the lemma and ends the proof.
\end{proof}

The next lemma establishes conditional moment bounds for the sparse distributions considered in this paper. These bounds will be used later to quantify the effect of edge removals on quantities given by LDPs.
\begin{lemma}\label{chapter2:lemma:1}
    Suppose $2\leq d\leq t \leq \tau_n$ where $\tau_n$ satisfies $\omega(1)\leq \tau_n\leq o(n)$, and let $G$ be a graph sampled from $ \Gb(n, t/n)$. Let $\b{A}$ be the adjacency matrix of $G$ and $\b{U} = \b{A} - \E\b{A}$. Let $S$ be the set of nodes with degrees larger than $Ct$ (where $C>1$ is a constant independent of $n, d$), and $L$ the set of edges incident to $S$. Let $\Delta>0$ be a constant independent of $n$ and $d$, and let $\pmb{\alpha} = (\pmb{\alpha}_{ij}, 1\leq i<j\leq n)$ be a vector of integers with $\|\pmb{\alpha}\|_1 \leq \Delta$. Then, there exists a  constant $c = c(\Delta, C)$ such that   
    \begin{align}
        \text{for all } \ 1\leq i<j\leq n, \quad \E\l[|\b{U}^{\pmb \alpha}| \mid (i,j) \in L\r] &\leq c \cdot \l(\frac{t}{n} \r)^{\|\pmb{\alpha}\|_0 -1_{\pmb{\alpha}_{ij}>0}    }. \label{chapter2:lemma:1.1}
    \end{align}
\end{lemma}
\begin{proof}
    Throughout this proof, we use $c$ to denote different constants depending on $\Delta$ and $C$. Fix a pair $(i,j)$ with $1\leq i<j \leq n$. We have
\begin{align}
    \E\l[|\b{U}^{\pmb{\alpha}}| \mid (i,j) \in L \r] &= \E\l[\prod_{1\leq u < v \leq n} |\b{U}_{u v}|^{\pmb{\alpha}_{uv}} \mid (i,j) \in L\r]. \label{chapter2:lemma:1.2}
\end{align}
Let
\begin{align*}
    E_{\rm ind} &\triangleq \{1\leq u<v \leq n \mid \{u, v\} \cap \{i, j\} = \emptyset, \ \pmb{\alpha}_{uv} > 0\},\\
    E_{\rm dep} & \triangleq \{1\leq u<v \leq n \mid |\{u, v\} \cap \{ i, j\}|=1, \ \pmb{\alpha}_{uv} > 0  \}.
\end{align*}
So that $\{1\leq u<v \leq n \mid \pmb{\alpha}_{uv} > 0\} \cup \{(i, j)\}$ is the disjoint union of $E_{\rm ind}$, $E_{\rm dep}$, and $\{(i, j)\}$. Note that the variables $(\b{U}_{uv})_{(u,v) \in E_{\rm ind}}$ are independent of $\{(i, j)\in L\}$ and $(\b{U}_{uv})_{(u,v) \in E_{\rm dep} \cup \{(i, j)\} }$. Moreover, conditional on $\{(i,j) \in L\}$, we have $\b{U}_{ij} = 1 - \frac{t}{n}$. Hence, decomposing the product in (\ref{chapter2:lemma:1.2}) over $E_{\rm ind} \cup E_{\rm dep} \cup \{(i, j)\}$ yields
\begin{align*}
        \E[|\b{U}^{\pmb{\alpha}}| \mid (i,j) \in L] &= \l(1 - \frac{t}{n}\r)^{\pmb{\alpha}_{ij}}\prod_{(u,v) \in E_{\rm ind}} \E\l[ |\b{U}_{uv}|^{\pmb{\alpha}_{uv}}\r] \cdot  \E\l[\prod_{(u,v) \in E_{\rm dep}} |\b{U}_{u v}|^{\pmb{\alpha}_{uv}} \mid (i,j) \in L\r]  \\
        &\leq c\prod_{(u, v) \in E_{\rm ind}} \l(\frac{t}{n}\r) \cdot  \underbrace{\E\l[\prod_{(u,v) \in E_{\rm dep}} |\b{U}_{u v}|^{\pmb{\alpha}_{uv}} \mid (i,j) \in L\r]}_{\Gamma_1}, \numberthis \label{chapter2:lemma:1.3}
\end{align*}
where we used the fact that $\E\l[ |\b{U}_{uv}|^k\r] \leq c t/n$, for all $1\leq k \leq \Delta$, combined with  $|E_{\rm ind}|$ $ \leq \Delta$. We next bound the term $\Gamma_1$ in the last line. We have
\begin{align*}
    \prod_{(u,v) \in E_{\rm dep}} |\b{U}_{u v}|^{\pmb{\alpha}_{uv}} &=  \prod_{(u,v) \in E_{\rm dep}} \l|\b{A}_{u v} - \frac{t}{n} \r|^{\pmb{\alpha}_{uv}} \\
    &\leq c \sum_{\pmb{\beta}, \pmb{\gamma}} \prod_{(u, v) \in E_{\rm dep}}\l| \b{A}_{u v}\r|^{\pmb{\beta}_{uv}}  \l(\frac{t}{n}\r)^{\pmb{\gamma}_{uv}}, \numberthis \label{chapter2:lemma:1.4}
\end{align*}
where the last line follows from expanding the products, and the first sum is over all $\pmb{\beta}, \pmb{\gamma} \in \mathcal{U}^{n\times n}$ supported on $E_{\rm dep}$ such that $\pmb{\beta}_{uv} + \pmb{\gamma}_{uv} = \pmb{\alpha}_{uv}$ for all $(u, v) \in E_{\rm dep}$. In particular, the number of summands is bounded by a constant in $\Delta$. Hence,
\begin{align*}
    \Gamma_1 &\leq c \max_{\pmb{\beta}, \pmb{\gamma}} \E\l[\prod_{(u, v): \pmb{\beta}_{uv} > 0}\l| \b{A}_{u v}\r|^{\pmb{\beta}_{uv}}  \l(\frac{t}{n}\r)^{\pmb{\gamma}_{uv}}\mid (i,j) \in L \r]\\
    &= c \max_{\pmb{\beta}, \pmb{\gamma}} \l(\frac{t}{n}\r)^{\|\pmb{\gamma}\|_1} \p\l(\bigcap_{(u, v): \pmb{\beta}_{uv} > 0} \{\b{A}_{u v} = 1\}   \mid (i,j) \in L \r). \numberthis \label{chapter2:lemma:1.5}
\end{align*}
Fix $\pmb{\beta}, \pmb{\gamma}$, and let $\mathcal{E}= \bigcap_{(u, v): \pmb{\beta}_{uv} > 0} \{\b{A}_{u v} = 1\}$. Introduce the events 
\begin{align*}
    H_i &= \{\b{A}_{ij} = 1,\ i\in S,\ j \not \in S\},\\
    H_j &= \{\b{A}_{ij} = 1,\ i\not \in S,\ j  \in S\},\\
    H_{ij} &= \{\b{A}_{ij} = 1,\ i\in S,\ j  \in S\},
\end{align*}
and note that $H_i, H_j$ and $H_{ij}$ are disjoint, and $\{(i, j) \in L\} = H_i \cup H_j \cup H_{ij}$. In particular, using the law of total probability, we have
\begin{align*}
    \p(\mathcal{E} \mid \{i, j\} \in L ) &\leq \max\l\{\p(\mathcal{E} \mid H_i ), \ \p(\mathcal{E} \mid H_j ), \ \p(\mathcal{E} \mid H_{ij} ) \r\}. \numberthis \label{chapter2:lemma:1.6}
\end{align*}
We next bound the conditional probabilities in the right hand side above. We first bound $\p(\mathcal{E} \mid H_i )$. Denote by $T$ the set of all edges $(u, v)$ appearing in $\mathcal{E}$. Recalling that $T\subseteq E_{\rm dep}$ and $E_{\rm dep}$ contains edges incident to exactly one of $\{i, j\}$, it follows that we can divide $T = T\setminus \{i, j\}$ into disjoint sets $T_i$ and $T_j$,  where $T_i$ contains all pairs  $(u, v)$ with $\{u, v\} \cap \{i, j\}=\{i\}$, and $T_j$ contains all pairs $(u, v)$ with $\{u, v\} \cap \{i, j\} =\{j\}$. Hence,
\begin{align*}
    \p\l(\mathcal{E} \mid H_i\r) &=  \underbrace{\p\l( \bigcap_{(u,v) \in T_i \cup T_j}\{\b{A}_{uv} = 1\}   \mid H_i\r)}_{\Gamma_2}. \numberthis \label{chapter2:lemma:1.7}
\end{align*}
Let
\begin{alignat}{2}
    &B_i = \{i \in S\}, \quad &&B_j = \{j\not \in S\}, \label{chapter2:lemma:1.8}\\
    &\mathcal{E}_i = \bigcap_{(u,v) \in T_i}\{\b{A}_{uv} = 1\}, \quad &&\mathcal{E}_j=\bigcap_{(u,v) \in T_j}\{\b{A}_{uv} = 1\}. \label{chapter2:lemma:1.9}   
\end{alignat}
We have
\begin{align*}
    \Gamma_2 &= \p(\mathcal{E}_i \cap \mathcal{E}_j \mid H_i)\\
    &=\p\l(\mathcal{E}_i \cap \mathcal{E}_j \mid B_i \cap B_j \cap \{\b{A}_{ij}=1\}\r)\\
    &= \frac{\p\l( \mathcal{E}_i \cap \mathcal{E}_j \cap B_i \cap B_j \mid \{\b{A}_{ij} = 1\}\r)}{\p(B_i \cap B_j \mid \{\b{A}_{ij} = 1\})}.
\end{align*}
Note that conditional on $\{\b{A}_{ij} = 1\}$, the variables $(\b{A}_{uv}, \ \{u, v\} \cap \{i, j\} \subseteq \{i\})$ and $(\b{A}_{uv}, \ \{u, v\} \cap \{i, j\} \subseteq \{j\})$ are independent. Therefore, conditional on $\{\b{A}_{ij} = 1\}$, the events $\mathcal{E}_i \cap B_i$ and $\mathcal{E}_j \cap B_j$ are independent. Hence
\begin{align*}
    \Gamma_2 &= \frac{\p\l( \mathcal{E}_i \cap B_i \mid \{\b{A}_{ij} = 1\}\r)\cdot\p\l( \mathcal{E}_j \cap B_j \mid \{\b{A}_{ij} = 1\}\r)}{\p(B_i \mid   \{\b{A}_{ij} = 1\})\cdot\p(B_j \mid   \{\b{A}_{ij} = 1\})}\\
    &= \p\l( \mathcal{E}_i \mid   B_i \cap \{\b{A}_{ij} = 1\}\r) \cdot \p\l( \mathcal{E}_j \mid   B_j \cap \{\b{A}_{ij} = 1\}\r). \numberthis \label{chapter2:lemma:1.10}
\end{align*}
Let $X \distrib {\rm Binomial}(n-2, t/n)$. Note that
\begin{align*}
    \binom{n-2}{|T_i|}\p\l( \mathcal{E}_i \mid B_i \cap \{\b{A}_{ij} = 1\}\r) \leq \E\l[ X^{|T_i|} \mid 1 + X > Ct\r]. \numberthis \label{chapter2:lemma:1.11}
\end{align*}
and similarly
\begin{align*}
    \binom{n-2}{|T_j|}\p\l( \mathcal{E}_j \mid B_j \cap \{\b{A}_{ij} = 1\}\r) \leq \E\l[ X^{|T_j|} \mid 1 + X \leq Ct\r]. \numberthis \label{chapter2:lemma:1.12}
\end{align*}
Since $\max(|T_i|,\ |T_j|) \leq \Delta$, there exists a constant $c$ depending on $\Delta$ and $C$ such that \begin{align*}
    \E\l[ X^{|T_i|} \mid 1 + X > Ct\r] &\leq c t^{|T_i|},\\
    \E\l[ X^{|T_j|} \mid 1 + X \leq Ct\r] &\leq c t^{|T_j|},\\
    \binom{n-2}{|T_i|} &\geq c n^{|T_i|},\\
    \binom{n-2}{|T_j|} &\geq c n^{|T_j|}.
\end{align*}
Using the above in (\ref{chapter2:lemma:1.11}) and (\ref{chapter2:lemma:1.12}), there exists a constant $c$ such that
\begin{align*}
    \p\l( \mathcal{E}_i \mid B_i \cap \{\b{A}_{ij} = 1\}\r) \leq c \l(\frac{t}{n}\r)^{|T_i|}, \quad 
    \p\l( \mathcal{E}_j \mid B_j \cap \{\b{A}_{ij} = 1\}\r) \leq c \l(\frac{t}{n}\r)^{|T_j|}.
\end{align*}
Combining these bounds with (\ref{chapter2:lemma:1.10}), we obtain
\begin{align*}
    \Gamma_2 &\leq c \l(\frac{t}{n}\r)^{|T_i| + |T_j|}\\
    &= c \l( \frac{t}{n}\r)^{|T|}\\
    &= c \l( \frac{t}{n}\r)^{\|\pmb{\beta}\|_0}.
\end{align*}
Together with (\ref{chapter2:lemma:1.7}), the previous bound readily gives 
\begin{align*}
    \p\l(\mathcal{E} \mid H_i\r) &\leq  c \l( \frac{t}{n}\r)^{\|\pmb{\beta}\|_0}.
\end{align*}
The same bound can be derived verbatim on $ \p\l(\mathcal{E} \mid H_j\r) $ and $ \p\l(\mathcal{E} \mid H_{ij}\r) $ by modifying the definitions of $B_i$ and $B_j$ in (\ref{chapter2:lemma:1.8}) as follows: $B_i=\{i\not \in S\},\ B_j=\{j \in S\}$ for $ \p\l(\mathcal{E} \mid H_j\r)$,  and $B_i=\{i\in S\},\ B_j = \{j\in S\}$ for $ \p\l(\mathcal{E} \mid H_{ij}\r)$. Therefore, we have in (\ref{chapter2:lemma:1.6})
\begin{align*}
    \p(\mathcal{E} \mid \{i, j\} \in L ) &\leq c \l( \frac{t}{n}\r)^{\|\pmb{\beta}\|_0 }.
\end{align*}
Combining the above with (\ref{chapter2:lemma:1.5}), it follows that there exists a constant $c= c(\Delta, C)$ such that
\begin{align*}
    \Gamma_1 &\leq c \max_{\pmb{\beta}, \pmb{\gamma}} \l(\frac{t}{n}\r)^{\|\pmb{\gamma}\|_1} \l(\frac{t}{n}\r)^{\|\pmb{\beta}\|_0 }\\
    &\leq c \l(\frac{t}{n} \r)^{|E_{\rm dep}|},
\end{align*}
where the last line follows from noting that for any $(u,v) \in E_{\rm dep}$, we have $|\pmb{\gamma}_{uv}|+ 1_{\pmb{\beta}_{uv}>0} \geq 1_{\pmb{\alpha}_{uv} > 0}$. Therefore, as $\pmb{\gamma}$ and $\pmb{\beta}$ are supported on $E_{\rm dep}$, we have $\|\pmb{\gamma}\|_1 + \|\pmb{\beta}\|_0 \geq |E_{\rm dep}|$. Using the above in  (\ref{chapter2:lemma:1.3}) yields
\begin{align*}
    \E\l[|\b{U}^{\pmb{\alpha}}| \mid (i,j) \in L \r]  &\leq c\l(\frac{t}{n}\r)^{|E_{\rm ind}| + |E_{\rm dep}|}\\
    &= c \l( \frac{t}{n} \r)^{\|\pmb{\alpha}\|_0 - 1_{\pmb{\alpha}_{ij}>0}}.
\end{align*}
This ends the proof of the lemma.
\end{proof}

The next lemma builds on Lemma~\ref{chapter2:lemma:1} to obtain conditional estimates for LDPs evaluated on sparse distributions,  where we condition similarly on edge removals. 
\begin{lemma}\label{chapter2:lemma:2}
Suppose  $2\leq d\leq  t \leq \tau_n \leq n/2$ where $\tau_n$ satisfies $\omega(1)\leq \tau_n \leq o(n)$, and let $G$ be a graph sampled from $ \Gb(n, t/n)$. Let $\b{A}$ be the adjacency matrix of $G$ and $\b{U} = \b{A} - \E\b{A}$. Let $S$ be the set of nodes with degrees larger than $Ct$ (where $C>1$ is a constant independent of $n, d$), and $L$ the set of edges incident to $S$. Let $\Delta>0$ be a constant independent of $n$ and $d$. Let $\Y = -\b{U} / \sqrt{t (1 - t/n)}$ so that $\Y \distrib \G(n, t/n)$. Suppose $p$ is a connected LDP with ${\rm Deg}(p) \leq \Delta$ and $p$ contains all nodes in set $K \subseteq [n]$ (i.e., $p\in \V_K$). 
 Then, there exists a constant $c = c(\Delta, C)$ such that for any $1\leq i<j \leq n$,
\begin{align*}
    \max\l\{\E\l[|p(\b{U})| \mid (i, j) \in L\r], \ \E\l[|p(\Y)| \mid (i, j) \in L\r] \r\} \leq |p(\b{0})|+ \frac{ c\|p\|_{\rm coef} t^{\Delta}}{n^{\min(|K\cup \{i, j\}|-2, \ |K|-1)}}. \numberthis \label{chapter2:lemma:2.1}
\end{align*}
\end{lemma}    
\begin{proof}
Throughout this proof, we denote by $c$ different constants depending on $\Delta$ and $C$. We first show the bound on $p(\b{U})$. We have
\begin{align*}
    \E\l[|p(\b{U})| \mid (i, j) \in L\r] &\leq |c_{p, \b{0} }| + \|p\|_{\rm coef} \sum_{\pmb{\alpha}} \E\l[ |\b{U}|^{\pmb{\alpha}} \mid (i,j) \in L\r]\\
    &\leq |p(\b{0})| +  c\|p\|_{\rm coef}  \sum_{\pmb{\alpha}} \l(\frac{t}{n}\r)^{\|\pmb{\alpha}\|_0 - 1_{\pmb{\alpha}_{ij} > 0}}\numberthis \label{chapter2:lemma:2.2}\\
    &\leq |p(\b{0})|+c\|p\|_{\rm coef}   t^{\Delta}\sum_{\pmb{\alpha}} \l(\frac{1}{n}\r)^{\|\pmb{\alpha}\|_0 - 1_{\pmb{\alpha}_{ij} > 0}}, \numberthis \label{chapter2:lemma:2.3}
\end{align*}
where the sums are over $\{\pmb{\alpha}\in \mathcal{U}^{n\times n} \setminus \{\b{0}\}\mid c_{p, \pmb{\alpha}} \neq 0\}$, we applied Lemma \ref{chapter2:lemma:1} in (\ref{chapter2:lemma:2.3}), and we used $\|\pmb{\alpha}\|_0 \leq \Delta$ together with $t\geq 1$ in the last line.  Let $\pmb{\alpha}$ as in the sum above with $ c_{p, \pmb{\alpha}} \neq 0$. Since $p$ is connected, $G_{\pmb{\alpha}}$ is connected. Therefore, we have $|V_{\pmb{\alpha}}| \leq |E_{\pmb{\alpha}}| + 1 = \|\pmb{\alpha}\|_0 + 1$. For an integer $0\leq a \leq \Delta$ and a connected graph $H$, introduce

\begin{align*}
    \mathcal{F}^{ij}_{a, H} &\triangleq \l\{ \pmb{\alpha} \in \mathcal{U}^{n\times n} \setminus \{\b{0}\}\mid c_{p, \pmb{\alpha}} \neq 0,\ \pmb{\alpha}_{ij} > 0,\  \|\pmb{\alpha}\|_1 = a,\ G_{\pmb{\alpha}} \sim H   \r\},\\
    \mathcal{F}^{0}_{a, H} &\triangleq \l\{ \pmb{\alpha} \in \mathcal{U}^{n\times n} \setminus \{\b{0}\}\mid c_{p, \pmb{\alpha}} \neq 0,\ \pmb{\alpha}_{ij} = 0,\  \|\pmb{\alpha}\|_1 = a,\ G_{\pmb{\alpha}} \sim H   \r\}.
\end{align*}
Note that for each $\pmb{\alpha} \in \mathcal{F}^{ij}_{a, H}$, we have $K \cup \{i, j\} \subseteq G_{\pmb{\alpha}}$. Thus, the size of $|\mathcal{F}^{ij}_{a, H}|$ is bounded by the remaining vertices labels count, which is at most $|V_{\pmb{\alpha}}| - |K \cup \{i, j\}| \leq \|\pmb{\alpha}\|_0 + 1 - |K \cup \{i, j\}|$, and we included $\{i, j\}$ with the set of contained nodes $K$ because $\pmb{\alpha}_{ij}=1$ forces $\{i,j\}\in G_{\pmb{\alpha}}$. Hence, there exists  a constant $c = c(\Delta)$ such that for all $\pmb{\alpha} \in \mathcal{F}^{ij}_{a, H}$ we have $|\mathcal{F}^{ij}_{a, H}| \leq c n^{\|\pmb{\alpha}\|_0+1-|K \cup \{i, j\}| }$. Similarly, for each $\pmb{\alpha} \in \mathcal{F}^{0}_{a, H}$, we have $K  \subseteq G_{\pmb{\alpha}}$, thus the size of $|\mathcal{F}^{0}_{a, H}|$ is bounded by the remaining vertices labels count, which is at most $|V_{\pmb{\alpha}}| - |K| \leq \|\pmb{\alpha}\|_0 + 1 - |K |$. Then, there exists a constant $c = c(\Delta)$ such that for all $\pmb{\alpha} \in \mathcal{F}^0_{a,H}$, we have $|\mathcal{F}^{0}_{a, H}| \leq  c n^{\|\pmb{\alpha}\|_0+1-|K|}$. Using the previous bounds  in (\ref{chapter2:lemma:2.3}) yields
\begin{align*}
    &\E\l[|p(\b{U})| \mid (i, j) \in L\r] - |p(\b{0})|\\
    &\leq c \|p\|_{\rm coef}t^{\Delta} \l( \sum_{a, H} \sum_{\pmb{\alpha} \in \mathcal{F}^{ij}_{a, H}} \l(\frac{1}{n}\r)^{\|\pmb{\alpha}\|_0 - 1} + \sum_{a, H} \sum_{\pmb{\alpha} \in \mathcal{F}^{0}_{a, H}} \l(\frac{1}{n}\r)^{\|\pmb{\alpha}\|_0} \r)\\
    &\leq c  \|p\|_{\rm coef}  t^{\Delta} \l( \sum_{a, H}  \l(\frac{1}{n}\r)^{|K \cup \{i, j\}| - 2} +  \sum_{a, H}  \l(\frac{1}{n}\r)^{|K| - 1} \r)\\
    &\leq 2c \|p\|_{\rm coef} t^{\Delta} \sum_{a, H} \frac{1}{n^{\min(|K\cup \{i, j\}|-2, \ |K|-1)}}\\
    & = c \| p\|_{\rm coef}\frac{t^{\Delta}}{n^{\min(|K\cup \{i, j\}|-2, \ |K|-1)}},
\end{align*}
where $c= c(\Delta, C)$ is a constant. This ends the proof of the bound on $p(\b{U})$. The bound on $p(\Y)$ is obtained readily by noticing that $|\Y|^{\pmb \alpha}= (1 + o(1))t^{-\|\pmb{\alpha}\|_1 /2 } |\b{U}|^{\pmb{\alpha}} \leq |\b{U}|^{\pmb{\alpha}}$ for $t\geq 1$ and $\|\pmb{\alpha}\|_1\leq \Delta$, which concludes the proof of the lemma.
\end{proof}

The next result synthesizes several of the preceding lemmas and serves as a key ingredient in the rounding proof in Section~\ref{chapter2:section:rounding}. Specifically, it shows that adjacency matrices of sparse distributions can be regularized via edge pruning to achieve uniformly bounded operator norms. Furthermore, LDP outputs $p(\Y)$ (and their projection into $[-1,1]^n$) evaluated on the unregularized matrix $\b{Y}$ attain approximately the same objective value when evaluated on the regularized matrix (i.e., $H(p(\Y), \Y)/2n \sim H(p(\Y), \Y^{\rm prune})/2n$).

\begin{lemma}\label{chapter2:lemma:3}
Suppose  $2\leq d\leq  t \leq \tau_n \leq n/2$ where $\tau_n$ satisfies $\omega(1)\leq \tau_n \leq o(n)$. Let $G$ be a graph sampled from $ \Gb(n, t/n)$, with adjacency matrix $\b{A}$. Introduce
\begin{align*}
    \b{U} = \b{A} - \E[\b{A}], \quad \b{Y} = - \frac{\b{U}}{\sqrt{t\l(1 - \frac{t}{n}\r)}},
\end{align*}
so that $\Y\distrib \G(n, t/n)$.  Let $S$ be the set of nodes with degrees larger than $4t$, and $L$ the set of edges incident to $S$. The graph $G^{\rm prune}$ is obtained from $G$ by removing all edges in $L$. Let $\b{A}^{\rm prune}$ be the adjacency matrix of $G^{\rm prune}$, and define similarly
\begin{align*}
    \b{U}^{\rm prune} = \b{A}^{\rm prune} - \E[\b{A}], \quad \Y^{\rm prune} = - \frac{\b{U}^{\rm prune}}{\sqrt{t\l(1 - \frac{t}{n}\r)}}.
\end{align*}
Suppose $p=(p_i,\ i \in [n])$ is a collection of connected LDPs with degrees bounded by a constant $\Delta$ independent of $n$ and $d$, and such that $p_i$ contains node $i$ for all $i\in [n]$ (i.e., $p_i\in \V_i$). Finally, let $\pi(\Y) \in [-1, 1]^n$ be the $\ell_2$ projection  of the vector $p(\Y)$ into the hypercube $[-1, 1]^n$, i.e.,
\begin{align*}
\pi(\b{Y}) \triangleq \underset{\b{z} \in [-1,1]^n}{\mathrm{argmin}} \|p(\b{Y}) - \b{z}\|_2.
\end{align*}
Then, there exist positive constants $c_1$ and $c_2$ depending on $\Delta$ such that the following holds 
\begin{align*}
     \E\l[\l| p(\Y)^\top \Y p(\Y) -  p(\Y)^\top \Y^{\rm prune} p(\Y) \r| \r] &\leq c_2 \max_{i\in [n]} \|p_i^2\|_{\rm coef} n t^{2\Delta + 1} e^{-c_1 t}, \numberthis \label{chapter2:lemma:3.1} \\
     \E\l[\l| \pi(\Y)^\top \Y \pi(\Y) -  \pi(\Y)^\top \Y^{\rm prune} \pi(\Y) \r| \r] &\leq c_2 nt e^{-c_1 t}. \numberthis \label{chapter2:lemma:3.2}
\end{align*}
\end{lemma}
\begin{proof}
    We first show (\ref{chapter2:lemma:3.2}). We have
    \begin{align*}
        \E\l[\l| \pi(\Y)^\top \l( \Y - \Y^{\rm prune}\r) \pi(\Y) \r| \r] &= \frac{2}{\sqrt{t \l(1 - \frac{t}{n}\r)}}  \E\l[ \l|\sum_{1\leq i<j \leq n} \pi_i(\Y) \pi_j(\Y) 1_{(i, j) \in L}\r| \r]\\
        &\leq 2\E\l[|L|\r], \numberthis \label{chapter2:lemma:3.3}
    \end{align*}
    where we used $|\pi_i(\Y)|,\ |\pi_j(\Y)|\leq 1$ and $t(1-t/n)\geq 1$ in the last line. Using Lemma \ref{chapter2:lemma:removed} with $C=4$, we have $\E[|L|]\leq c_2 nt e^{-c_1 t} $ for some positive constants $c_2, c_1$. Combining the latter with (\ref{chapter2:lemma:3.3}) readily yields the result of (\ref{chapter2:lemma:3.2}). We now show  (\ref{chapter2:lemma:3.1}) similarly.  We have
    \begin{align*}
         \E\l[ \l| p(\Y)^\top \l(\Y  -   \Y^{\rm prune}\r) p(\Y) \r| \r] &= \frac{2}{\sqrt{t \l(1 - \frac{t}{n} \r)}} \E\l[ \l|\sum_{1\leq i<j \leq n} p_i(\Y) p_j(\Y) 1_{(i, j) \in L}\r| \r]\\
        &\leq  \sum_{1\leq i<j \leq n} \E\l[ p_i(\Y)^2 1_{(i,j) \in L} \r] +\E\l[  p_j(\Y)^2 1_{(i, j) \in L} \r], \numberthis \label{chapter2:lemma:3.4}
    \end{align*}
where we used the arithmetic-geometric mean inequality in the last line. Note that $p_i^2$ is a connected LDP containing node $i$ (i.e., $(p_i)^2 \in \V_i$), and has degree at most $2\Delta$. Therefore, it follows from applying Lemma \ref{chapter2:lemma:2} to $p_i^2$ with $K=\{i\}$ and $C=4$ that there exists a constant $c=c(\Delta)$ such that
\begin{align*}
    \E\l[p_i(\Y)^2 1_{(i, j) \in L}\r] &=   \E\l[ p_i(\Y)^2 \mid (i, j) \in L\r] \cdot \p((i, j) \in L)\\
    &\leq \l( |p^2_i(\b{0})| + c\|p_i^2\|_{\rm coef} t^{2\Delta} \r)\cdot \p((i, j) \in L)\\
    &\leq \l(\|p^2_i\|_{\rm coef} + c\|p_i^2\|_{\rm coef} t^{2\Delta} \r)\cdot \p((i, j) \in L)\\
    &\leq c'\|p_i^2\|_{\rm coef} t^{2\Delta} \cdot \p((i, j) \in L),
\end{align*}
where $c'=\max(1,\ c)$ and we used $t\geq 1$ in the last line. Combining the above with (\ref{chapter2:lemma:3.4}), we have
\begin{align*}
    \E\l[\l| p(\Y)^\top \l(\Y  -   \Y^{\rm prune}\r) p(\Y) \r|\r] &\leq  \sum_{1\leq i<j \leq n}    c'\l(\|p_i^2\|_{\rm coef} + \|p_j^2\|_{\rm coef} \r)t^{2\Delta}  \cdot\p((i, j) \in L)\\
    &\leq 2c'\max_{i\in [n]} \|p^2_i\|_{\rm coef} t^{2\Delta} \cdot \E\l[\sum_{1\leq i< j \leq n} 1_{(i, j) \in L}\r]\\
    &= 2c'\max_{i\in [n]} \|p_i^2\|_{\rm coef}t^{2\Delta} \cdot \E\l[|L|\r].
\end{align*}
Using Lemma \ref{chapter2:lemma:removed}, it follows that there exist positive constants $c_1$ and $c_2$  such that $\E[|L|] \leq c_2 nt e^{-c_1 t}$. Therefore,
\begin{align*}
    \E\l[\l| p(\Y)^\top \Y p(\Y) -  p(\Y)^\top \Y^{\rm prune} p(\Y) \r|\r] &\leq 2c' c_2 \max_{i\in [n]} \|p_i^2\|_{\rm coef} nt^{2\Delta+1 } e^{-c_1 t},
\end{align*}
which shows (\ref{chapter2:lemma:3.1}) and ends the proof of the lemma.
\end{proof}

The next lemma establishes bounds on LDPs evaluated on distributions with mixed Gaussian and Bernoulli entries. These bounds will be used to control LDP quantities arising in the Lindeberg interpolation steps as we switch between Gaussian and Bernoulli disorder ensembles.
\begin{lemma}\label{chapter2:lemma:pre:7}
Let $\b{X}\distrib{\rm GOE}^0(n)$  and $\Y\distrib\G(n,t/n)$, where $d\leq t \leq \tau_n$ and $\tau_n$ satisfies $\omega(1)\leq \tau_n \leq o(n)$. Suppose $L_1 \cup L_2$ is a partition of the ordered pairs $1\leq i<j \leq n$. Consider the (symmetric)  matrix $\Z$ given by
\begin{equation}
    \begin{aligned}
        \Z_{ij} &= \X_{ij} \quad && \text{if } \quad (i,j) \in L_1,\\
        \Z_{ij} &= \Y_{ij} \quad && \text{if } \quad (i,j) \in L_2,
    \end{aligned}
\end{equation}
and  $\Z_{ii}=0$. Let $S\subseteq [n]$, $q\in \R[\X]_{\leq \Delta}$, 
and $e=(i,j)$ with $1\leq i<j\leq n$. Suppose
$\Z_e q(\Z)$ is connected and contains all nodes in $S$ (i.e., $\Z_e q(\Z) \in \V_S)$).
 For every $\Delta$, there exists $c=c(\Delta)$ independent of $n,d$ such that

\begin{equation}\label{chapter2:20}
    \begin{aligned}
        |\E[|\Z_e|^{\ell}q(\Z)]| &\leq \frac{c(\Delta)\|q\|_{\rm coef}}{n^{|S|+\frac{\ell}{2}-2}} \quad &&\text{if} \quad e\in L_1, \text{and} \quad  \ell\geq 2,\\
        |\E[|\Z_e|^\ell q(\Z)]| &\leq \frac{c(\Delta)\|q\|_{\rm coef}}{n^{|S|-1} t^{\frac{\ell}{2}-1}} \quad &&\text{if} \quad e\in L_2 ,\text{and} \quad \ell\geq 2.
    \end{aligned}
\end{equation}

\end{lemma}

\begin{proof}
For $\alphaa\in \U^{n\times n}$, we will write $\alphaa=\alphaa^1 + \alphaa^2$ where $\alphaa_{ij}^r = \alphaa_{ij}$ if $(i,j) \in L_r$, and $\alphaa_{ij}^r=0$ otherwise for $r=1,2$.  
Consider the case $e\in L_1$ in (\ref{chapter2:20}) and denote by $w(\Z)$ the LDP $\Z_e^{\ell} q(\Z)$. We will use the same notation introduced in the proof of Lemma \ref{chapter2:lemma:pre:6}. For $\alphaa\in \mathcal{U}^{n\times n}$, we let  $\bar{\alphaa} \in \mathcal{U}^{n\times n}$ where $\bar{\alphaa}_{ij} = \alphaa_{ij},\  \forall (i,j)\neq e$ and $\bar{\alphaa}_e=0$.  Note that $\E\l[\X^{\alphaa}\r]=\gamma_{\alphaa} n^{-\|\alphaa\|_1/2}$ where $\gamma_{\alphaa}=\prod_{\substack{1\leq i<j \leq n \\2| \alphaa_{ij} ,\  \alphaa_{ij}>0}} (\alphaa_{ij}-1)!!$ and $|\gamma_{\alphaa}|\leq c_{\Delta}$ for some constant $c_{\Delta}$, and similarly $\E\l[\l|\X\r|^{\alphaa}\r] \leq c'_{\Delta} n^{-\|\alphaa\|_1/2}$ for some constant $c'_{\Delta}$. Using the latter and Lemma \ref{chapter2:lemma:pre:sparse} we have
\begin{align*}
    &\l|\E\l[w(\Z)\r]\r|\\
    &=\l| \sum_{\alphaa  } c_{q,\alphaa} \E\l[|\Z_e|^{\ell} \Z^{\alphaa}\r] \r|\\
    &=\l| \sum_{\alphaa} c_{q,\alphaa} \E\l[|\X_{e}|^{\ell + \alphaa^1_e}\r]\E\l[\X^{\bar{\alphaa}^1}\r]\E\l[\Y^{\alphaa^2}\r]\r|\\
    &\leq 2\|q\|_{\rm coef} c_{\Delta}c'_{\Delta}  \sum_{\alphaa,c_{q,\alphaa}\neq 0} n^{-\frac{\ell + |\alphaa^1_e|  + \|\bar{\alphaa}^1\|_1}{2}} \b{1}_{2|\bar{\alphaa}^1 }\b{1}_{\alphaa^2\in \mathcal{U}^{n\times n}_*}  n^{-\|\alphaa^2\|_0} t^{- \frac{\|\alphaa^2\|_1}{2} + \|\alphaa^2\|_0}, \\
    &\leq 2\|q\|_{\rm coef} c_{\Delta}c'_{\Delta} \sum_{\alphaa,\ c_{q,\alphaa}\neq 0 } \b{1}_{2|\bar{\alphaa}^1 } \b{1}_{\alphaa^2\in \mathcal{U}^{n\times n}_*}n^{- \frac{\ell + \|\alphaa^1\|_1}{2} - \|\alphaa^2\|_0} t^{- \frac{\|\alphaa^2\|_1}{2} + \|\alphaa^2\|_0},
\end{align*}
where  we used $\E\l[\Y^{\alphaa^2}\r] = \b{1}_{\alphaa^2\in \mathcal{U}^{n\times n}_*} ((-1)^{\|\alphaa^2\|_1}+o(1)) n^{-\|\alphaa^2\|_0} t^{- \frac{\|\alphaa^2\|_1}{2} + \|\alphaa^2\|_0}$, and upper bounded $|((-1)^{\|\alphaa^2\|_1}+o(1))|$ by $2$ for large enough $n,d$.  Since $\alphaa^2\in  \mathcal{U}^{n\times n}_*$, we have $- \frac{\|\alphaa^2\|_1}{2} + \|\alphaa^2\|_0\leq 0$. Therefore 
\begin{align*}
    \l|\E\l[w(\Z)\r]\r| &\leq 2\|q\|_{\rm coef} c_{\Delta}c'_{\Delta} \sum_{\alphaa\in \mathcal{C}} \b{1}_{2|\bar{\alphaa}^1} \b{1}_{\alphaa^2\in \mathcal{U}^{n\times n}_*} n^{- \frac{\ell + \|\alphaa^1\|_1}{2} - \|\alphaa^2\|_0}.
\end{align*}
Let $\hat{\alphaa} = \ell[e] + \alphaa$ (so that $\Z^{\hat{\alphaa}} = \Z_e^{\ell} \cdot\Z^{\alphaa}$). Since $2|\bar{\alphaa}^1$, it follows that $\|\hat{\alphaa}^1\|_0 \leq \frac{\|\hat{\alphaa}^1\|_1- (\ell-2) }{2}$. Since the LDP $w$ is connected, it follows that $\hat{\alphaa}$ is connected, and furthermore  $|V_{\hat{\alphaa}}|\leq |E_{\hat{\alphaa}}|+1 = \|\hat{\alphaa}\|_0 + 1 = \|\hat{\alphaa}^1\|_0 +\|\alphaa^2\|_0 + 1\leq \frac{\|\hat{\alphaa}^1\|_1-(\ell-2)}{2} + \|\alphaa^2\|_0+1 =  \frac{\|\hat{\alphaa}^1\|_1}{2} + \|\alphaa^2\|_0+2 - \frac{\ell}{2} $. For an integer  $0\leq a \leq \Delta$ and a graph $H$, introduce the set
\begin{align}
    \mathcal{F}_{a, H} \triangleq \l\{ \alphaa \mid c_{q,\alphaa}\neq 0,\  \|\alphaa\|_1 =a,\  G_{\alphaa} \sim H ,\ 2|\bar{\alphaa}^1,\ \alphaa^2 \in \mathcal{U}^{n\times n}_*\r\}.
\end{align} 
Recalling that  $w\in \V_{S}$, it follows that 
$$\forall \alphaa \in \mathcal{F}_{a, H}, |\mathcal{F}_{a, H}|\leq \beta_{\Delta}n^{|V_{\hat{\alphaa}}|-|S|} \leq \beta_{\Delta}n^{\frac{\|\hat{\alphaa}^1\|_1}{2} + \|\alphaa^2\|_0+2-\frac{\ell}{2}  -|S|},$$ 
where $\beta_{\Delta}$ is a constant depending on $\Delta$. Hence
\begin{align*}
    \l|\E\l[w(\Z)\r]\r| &\leq 2\|q\|_{\rm coef} c_{\Delta}c'_{\Delta} \sum \sum_{\alphaa\in\mathcal{F}_{a, H} } \b{1}_{2|\bar{\alphaa}^1}\b{1}_{\alphaa^2\in \mathcal{U}^{n\times n}_*} n^{- \frac{\ell + \|\alphaa^1\|_1}{2} - \|\alphaa^2\|_0}\\
    &=2\|q\|_{\rm coef} c_{\Delta}c'_{\Delta} \sum \sum_{\alphaa\in\mathcal{F}_{a, H} } \b{1}_{2|\bar{\alphaa}^1}\b{1}_{\alphaa^2\in \mathcal{U}^{n\times n}_*} n^{- \frac{\|\hat{\alphaa}^1\|_1}{2} - \|\alphaa^2\|_0}\\
    &\leq 2\|q\|_{\rm coef} c_{\Delta}c'_{\Delta}\sum \beta_{\Delta} n^{2-\frac{\ell}{2}-|S|}\\
    &= \frac{O(\|q\|_{\rm coef})}{n^{|S|+\frac{\ell}{2}-2}},
\end{align*}
where the first sum is taken over  over all pairs $(a,H)$ of (integer, graph) such that $0\leq a \leq \Delta$ and $H$ is a graph in the orbit generated by $G_{\alphaa}$ with $c_{q, \alphaa}\neq 0$. Denoting
by $c(\Delta)$ the constant hidden in $O(\cdot)$, we obtain the result for the case $e\in L_1$.

Consider now the case $e\in L_2$ and denote again by $w(\Z)$ the LDP $\Z_e^\ell q(\Z)$. We have similarly to the previous case
\begin{align*}
    \l|\E\l[w(\Z)\r]\r| &\leq 2\|q\|_{\rm coef}c_{\Delta}c'_{\Delta} \sum_{\alphaa\in \mathcal{C}} \b{1}_{2|\alphaa^1} \b{1}_{\bar{\alphaa}^2\in \mathcal{U}^{n\times n}_*}n^{- \frac{\|\alphaa^1\|_1}{2} - \|\hat{\alphaa}^2\|_0} t^{- \frac{\|\hat{\alphaa}^2\|_1}{2} + \|\hat{\alphaa}^2\|_0}.
\end{align*}
As $\ell\geq 2$, it follows that $\bar{\alphaa}^2 \in\mathcal{U}^{n\times n}_*$ if and only if $\hat{\alphaa}^2\in \mathcal{U}^{n\times n}_*$. Let $\alphaa$ satisfy $2|\alphaa^1, \bar{\alphaa}^2 \in \mathcal{U}^{n\times n}_*$. Since $\hat{\alphaa}^2\in \mathcal{U}^{n\times n}_*$, all the nonzero values in $\hat{\alphaa}^2$ are at least $2$. It follows that $\|\hat{\alphaa}^2\|_0 \leq \frac{\|\hat{\alphaa}^2\|_1 - (\ell-2)}{2}$ and thus $-\frac{\|\hat{\alphaa}^2\|_1}{2}+\|\hat{\alphaa}^2\|_0 \leq -\frac{\ell}{2}+1$. Hence
\begin{align*}
    \l|\E\l[w(\Z)\r]\r| &\leq 2\|q\|_{\rm coef}c_{\Delta}c'_{\Delta} t^{-\frac{\ell}{2}+1} \sum_{\alphaa, c_{q,\alphaa}\neq 0} \b{1}_{2|\alphaa^1} \b{1}_{\bar{\alphaa}^2\in \mathcal{U}^{n\times n}_*} n^{- \frac{\|\alphaa^1\|_1}{2} - \|\hat{\alphaa}^2\|_0}.
\end{align*}
Since $w$ is connected, it follows that $\hat{\alphaa}$ is connected. Therefore, we have $|V_{\hat{\alphaa}}|\leq |E_{\hat{\alphaa}}|+1 = \|\hat{\alphaa}\|_0 + 1 = \|\alphaa^1\|_0 +\|\hat{\alphaa}^2\|_0 + 1\leq \frac{\|\alphaa^1\|_1}{2} + \|\hat{\alphaa}^2\|_0+1$, where the last inequality follows from $2|\alphaa^1$.
Recalling that $w\in \V_{S}$ and using the same notation for $\mathcal{F}_{a,H}$, it follows that $\forall \alphaa \in \mathcal{F}_{a, H}, |\mathcal{F}_{a, H}|\leq \beta_{\Delta}n^{|V_{\hat{\alphaa}}|-|S|} \leq \beta_{\Delta}n^{\frac{\|\alphaa^1\|_1}{2} + \|\hat{\alphaa}^2\|_0+1  -|S|}$. Hence
\begin{align*}
    |\E[w(\Z)]| &\leq 2\|q\|_{\rm coef}c_{\Delta}c'_{\Delta} t^{-\frac{\ell}{2}+1} \sum \sum_{\alphaa\in\mathcal{F}_{a, H} } \b{1}_{2|\alphaa^1} \b{1}_{\bar{\alphaa}^2\in \mathcal{U}^{n\times n}_*} n^{- \frac{\|\alphaa^1\|_1}{2} - \|\hat{\alphaa}^2\|_0}\\
    &\leq 2\|q\|_{\rm coef}c_{\Delta} c'_{\Delta}t^{-\frac{\ell}{2}+1} \sum \beta_{\Delta} n^{1-|S|}\\
    &= \frac{O(\|q\|_{\rm coef})}{n^{|S|-1}t^{\frac{\ell}{2}-1}},
\end{align*}
where the first sum is taken over all pairs $(a,H)$ of (integer, graph) such that $0\leq a \leq \Delta$ and $H$ is a graph in the orbit generated by $G_{\alphaa}$ with $\alphaa\in  \mathcal{C},\ c_{q, \alphaa}\neq 0$. Denoting
by $c(\Delta)$ the constant hidden in $O(\cdot)$, we obtain the result for the case $e\in L_2$. This concludes the proof.
\end{proof}

The next lemma restates the Lindeberg interpolation setup and notation used in the proof of Theorem~\ref{chapter2:lemma:state_ev}, and derives bounds on the third-order Taylor remainders arising in the interpolation analysis. It builds on several previous lemmas and crucially exploits the quadratic structure of tree-structured LDPs in $\mathcal{T}_{n,2,\Delta}$ to describe how their values change when a single disorder entry is switched from Gaussian to Bernoulli. Moreover, it uses Lemma~\ref{chapter2:lemma:pre:2} to control the coefficient scaling of various sums of LDPs constrained to contain specified vertices.

\begin{lemma}\label{chapter2:claim:5.13}
Let $\b{X}\distrib{\rm GOE}^0(n)$  and $\Y\distrib\G(n,t/n)$, where $d\leq t \leq \tau_n$ and $\tau_n$ satisfies $\omega(1)\leq \tau_n \leq o(n)$. Let $\ell\in [\binom{n}{2}]$ enumerate a pair $(i^*, j^*)$ with $(i^*, j^*)\in \binom{[n]}{2}$.
Suppose $L_1 \cup L_2$ is a partition of  pairs $1\leq i<j \leq n,\ (i,j)\ne (i^*, j^*)$ with $L_1\not = \emptyset$.  
Given  $i\in [n]$ let $u_{i,\ell},\ a_{i,\ell},\ b_{i,\ell}$ be  LDPs
in variables
$\b{x}=(\b{x}_{ij},\ 1\leq i<j \leq n)$, satisfying 
\begin{enumerate}
    \item $\forall i\in[n],\  u_{i, \ell}(\b{x}),\ \b{x}_\ell a_{i,\ell}(\b{x}),\ \b{x}_\ell b_{i,\ell}(\b{x})$ are connected and contain node $i$ (i.e., $u_{i, \ell}(\b{x})$, $  \b{x}_\ell a_{i,\ell}(\b{x}), \b{x}_\ell b_{i,\ell}(\b{x}) \in \V_i$).
    \item  $\forall i\in[n],\ u_{i,\ell}(\b{x}),\ a_{i,\ell}(\b{x}),\ b_{i,\ell}(\b{x})$ do not depend on the variable $\b{x}_{\ell}$ (i.e., $\b{x}_{i^* j^*}$).
    \item $\forall i\in[n],\ \|u_{i,\ell}\|_{\rm coef} + \|a_{i,\ell}\|_{\rm coef} + \|b_{i,\ell}\|_{\rm coef}\leq C$, where $C$ is a constant that does not depend on $n, d$ and $\ell$. 
\end{enumerate}
Let  $\Z^{\ell}$  be given by $\Z^{\ell}_{ii}=0,\ \forall i\in [n]$ and 
\begin{equation}
    \begin{aligned}
        \relax [\Z^{\ell}]_{ij} &= \b{X}_{ij} \quad && \text{if } \quad (i,j) \in L_1 \cup \{(i^*, j^*)\}, \\
        \relax [\Z^{\ell}]_{ij} &= \Y_{ij} \quad && \text{if } \quad (i,j) \in L_2.
    \end{aligned}
\end{equation}
Furthermore, let $\b{Z}^{\ell+1}$ be obtained from $\b{Z}^{\ell}$ by switching the disorder of the $\ell$-th edge $(i^*,j^*)$ from $\b{X}_\ell$ to $\b{Y}_\ell$. Let $\chi\in \mathcal{C}^{3}(\mathbb{R}, \mathbb{R})$ be a thrice differentiable function, and suppose that  there exist a constant $m\in \mathbb{Z}_{\geq 0}$ independent of $n$ and $d$, and a scalar $L_{n, d}>0$ that may depend on $n, d$ such that
\begin{equation}
  |\chi^{(r)}(x)| \le L_{n, d} \bigl(1 + |x|^m\bigr), \qquad \forall x\in\R,\ r=1,2,3.
\end{equation}
Finally, introduce
\begin{align*}
v_{i,\ell}(z) &= u_{i,\ell}(\b{Z}^{\ell}) + z a_{i,\ell}(\b{Z}^{\ell}) + z^2 b_{i,\ell}(\b{Z}^{\ell}),
\qquad \forall i\in [n],\\
    h(z)& =\sum_{i=1}^{n} \chi(v_{i, \ell}(z)).
\end{align*}
Then for every  $(s_1,m_1,s_2,m_2) \in \mathbb{Z}_{\geq 0}^4$ such that  $s_1 m_1 + s_2 m_2=3$, it holds
    \begin{align*}
        \E\l[\sup_{|x_{\ell}|\leq |\b{X}_{\ell}|}\l|\b{X}_\ell^3 \cdot \l(h^{(s_1)}(x_\ell)\r)^{m_1} \cdot \l(h^{(s_2)}(x_\ell)\r)^{m_2}\r|\r] &\leq  cL_{n,d}^{m_1 + m_2} n^{m_1+m_2-\frac{5}{2}},\numberthis \label{chapter2:claim.5.13.1}\\
        \E\l[\sup_{|y_{\ell}|\leq |\Y_{\ell}|}\l|\Y_\ell^3 \cdot  \l(h^{(s_1)}(y_\ell) \r)^{m_1} \cdot \l(h^{(s_2)}(y_\ell)\r)^{m_2}\r|\r] &\leq  cL_{n,d}^{m_1 + m_2} n^{m_1+m_2-2}t^{-\frac{1}{2}},  \numberthis \label{chapter2:claim.5.13.2}
    \end{align*}
    where $c$ is a constant independent of $n, d$ and $\ell$.
\end{lemma}
\begin{proof}
    We use the slight abuse of notation $u_{i,\ell},\ a_{i,\ell},\ b_{i,\ell}$ to denote $ u_{i,\ell}(\Z^\ell),\ a_{i,\ell}(\Z^\ell),$ $\ b_{i,\ell}(\Z^\ell)$. Moreover, throughout this proof, we use $c$ to denote different constants independent of $n,d$ and $\ell$.  We first consider the case $(s_1,m_1,s_2,m_2)= (2,1,1,1)$, the remaining cases will follow from similar arguments. Using Lemma~\ref{chapter2:claim:5.12}   
    \begin{align*}
        &|\b{X}_\ell^3 \cdot h^{(2)}(x_\ell) \cdot  h^{(1)}(x_\ell)|\\
        &\leq cL_{n,d}^2\sum  \sum |\b{X}_\ell|^{ 3+j_1^1+j_1^2} |a_{i_1,\ell}|^{j_2^1} |b_{i_1,\ell}|^{j_3^1} |u_{i_1,\ell}|^{j_4^1} |a_{i_2,\ell}|^{j_2^2} |b_{i_2,\ell}|^{j_3^2} |u_{i_2,\ell}|^{j_4^2}\\
          &\leq   c L_{n,d}^2 \sum \sum |\b{X}_\ell|^{3+ 2j_1^1} (a_{i_1,\ell}^{j_2^1} b_{i_1,\ell}^{j_3^1} u_{i_1,\ell}^{j_4^1})^2 + |\b{X}_\ell|^{3+2j_1^2} (a_{i_2,\ell}^{j_2^2} b_{i_2,\ell}^{j_3^2} u_{i_2,\ell}^{j_4^2})^2. \numberthis \label{chapter2:85.0}
    \end{align*}
Where the first summations are taken over the set $\mathcal{I} \triangleq \{(i_1, i_2), \ 1\leq i_1,i_2\leq n\}$, and the second summations are taken over the set
\begin{align*}
  \mathcal{J} \triangleq \l\{ j_s^t, t\in [2], \ s\in[4] \mid \max_{s\in[4]} (j_s^1) \leq 2+2m, \ \max_{s\in[4]} (j_s^2) \leq 1+2m, \ \min_{t\in[2]}(j^t_2+j^t_3)\geq 1\r\}.
\end{align*}
Furthermore, line (\ref{chapter2:85.0}) follows from  the arithmetic-geometric mean inequality. Using Lemma \ref{chapter2:lemma:pre:1}, it follows that $\b{X}_\ell^{ 3+ 2j_1^t}(a_{i_t,\ell}^{j_2^t} b_{i_t,\ell}^{j_3^t} u_{i_t,\ell}^{j_4^t})^2$ for $t=1,2$ are LDPs with coefficients bounded by a constant $c$ depending on $\Delta, m,$ and $C$. Moreover, these  LDPs are connected polynomials containing node $i_t$ (i.e., $\b{X}_\ell^{ 3+ 2j_1^t}(a_{i_t,\ell}^{j_2^t} b_{i_t,\ell}^{j_3^t} u_{i_t,\ell}^{j_4^t})^2 \in \V_{i_t}$). Indeed, $(u_{i_t,\ell}^{j_4^t})^2\in \V_{i_t}$ and is connected from the lemma assumptions, and $\b{X}_\ell^{ 3+ 2j_1^t}(a_{i_t,\ell}^{j_2^t} b_{i_t,\ell}^{j_3^t})^2$ is the product of powers of terms similar to $\b{X}_\ell a_{i_t,\ell},\  \b{X}_\ell b_{i_t,\ell}, \ \b{X}_\ell a_{i_t,\ell} b_{i_t,\ell}$ (as $j_2^t + j_3^t \geq 1)$, all of which are connected and contain $i_t$ by the lemma assumption. (Note that the connectivity of $\b{X}_\ell a_{i_t,\ell} b_{i_t,\ell}$ follows trivially from the connectivity of $\b{X}_\ell a_{i_t,\ell}, \ \b{X}_\ell b_{i_t,\ell}$ and the fact that the latter two LDPs share node $i_t$). Hence, $\b{X}_\ell^{ 3+ 2j_1^t}(a_{i_t,\ell}^{j_2^t} b_{i_t,\ell}^{j_3^t} u_{i_t,\ell}^{j_4^t})^2$ is a connected LDP containing node $i_t$ as a product of connected LDPs containing node $i_t$.

Since  the LDP $\b{X}_\ell^{ 3+ 2j_1^t}(a_{i_t,\ell}^{j_2^t} b_{i_t,\ell}^{j_3^t} u_{i_t,\ell}^{j_4^t})^2$ contains node $i_t$ for $t=1,2$, it follows by Lemma \ref{chapter2:lemma:pre:2} (using $S_1=\{i_1\}$ and $S_2=\{i_2\}$ respectively),   that the resulting LDP from summing the  LDPs above over $\mathcal{I},\mathcal{J}$
  has  coefficients bounded by $cn$, where the $n$ is due to double counting from the sum on $\mathcal{I}$. In particular, we have
\begin{align*}
    |\b{X}_\ell^3 \cdot h^{(2)}(x_\ell) \cdot  h^{(1)}(x_\ell)|  &\leq c n L_{n,d}^2 \b{X}_\ell^3 w(\Z^\ell), \\
    |\Y_\ell^3 \cdot h^{(2)}(y_\ell) \cdot  h^{(1)}(y_\ell)|  &\leq c n  L_{n,d}^2 \Y_\ell^3 w(\Z^{\ell+1}) ,
\end{align*}
where $w$ is a LDP with $\|w\|_{\rm coef}\leq c$  and $\b{X}_\ell^3 w(\Z^\ell), \ \Y_\ell^3 w(\Z^{\ell+1})$ are connected. Applying  (\ref{chapter2:20}) from Lemma \ref{chapter2:lemma:pre:7} (with $S$ the set of endpoints of edge $\ell$) yields
\begin{align*}
   \l| \E\l[ n \b{X}_\ell^3 w(\Z^\ell)\r] \r |&\leq c  L_{n,d}^2 n^{-\frac{1}{2}},\\
   \l| \E\l[ n \Y_\ell^3 w(\Z^{\ell+1})\r] \r |&\leq c  L_{n,d}^2 t^{-\frac{1}{2}}.
\end{align*}
Therefore
\begin{align*}
    \E\l[ \sup_{|x_{\ell}|\leq |\b{X}_{\ell}|}|\b{X}_\ell^3 \cdot h^{(2)}(x_\ell) \cdot  h^{(1)}(x_\ell)| \r] &\leq   L_{n,d}^2n^{-\frac{1}{2}} = c L_{n,d}^{m_1+m_2} n^{m_1+m_2-\frac{5}{2}},\\
    \E\l[ \sup_{|y_{\ell}|\leq |\Y_{\ell}|}|\Y_\ell^3 \cdot h^{(2)}(y_\ell) \cdot  h^{(1)}(y_\ell)| \r] &\leq  c L_{n,d}^2 t^{-\frac{1}{2}}  =  c L_{n,d}^{m_1 + m_2} n^{m_1+m_2-2}t^{-\frac{1}{2}},
\end{align*}
which ends the proof of (\ref{chapter2:claim.5.13.1}) and (\ref{chapter2:claim.5.13.2}) for $(s_1,m_1,s_2,m_2)=(2,1,1,1)$. The  case given by $(s_1,m_1,s_2,m_2)= (1,3,0,0)$ is identical in treatment, specifically, we can use Lemma \ref{chapter2:claim:5.12} to obtain the following bound,
\begin{align*}
|\b{X}_\ell^3 \cdot h^{(1)}(x_\ell)^3| &\leq  c L_{n,d}^3 \sum_{\mathcal{I}} \sum_{\mathcal{J}} \sum_{t=1}^{3} |\b{X}_\ell|^{3 + (4\b{1}_{t\leq 2} + 2 \b{1}_{t=3})j_1^t} (a_{i_t,\ell}^{j_2^t} b_{i_t,\ell}^{j_3^t} u_{i_t,\ell}^{j_4^t})^{4 \b{1}_{t\leq 2} + 2 \b{1}_{t=3}},
\end{align*}
where
\begin{align*}
    \mathcal{I} &\triangleq \{i_t, t\in [3] \mid  i_t \in[n]\},\\
    \mathcal{J} &\triangleq \{j_s^t, t\in [3],\ s\in [4] \mid \max_{s\in [4],\ t\in [3]}(j_s^t) \leq 1+2m,\ \min_{t\in [3]} (j_2^t+j_3^t)\geq 1\}.
\end{align*}
Moreover, the same bound holds for $\Y$. It suffices then to notice that each summand in the above sum is a connected LDP containing the corresponding $i_t$ node. Applying  Lemma \ref{chapter2:lemma:pre:2} yields that each summation can be written as $n^{m_1 + m_2-1} |\b{X}_\ell|^3 w(\Z^\ell)$ where $\b{X}_\ell^3 w(\Z^\ell)$ is a connected LDP with coefficients bounded by a constant $c$, and the $n^{m_1+m_2-1}$ accounts for double counting in the summation over all indices in $\{i_1,i_2,i_3\}$. The result of (\ref{chapter2:claim.5.13.1}) and (\ref{chapter2:claim.5.13.2}) then readily follows in the above cases by applying  (\ref{chapter2:20}) from Lemma \ref{chapter2:lemma:pre:7} (with $S$ the set of endpoints of edge $\ell$).

It remains to deal with the case $(s_1,m_1,s_2,m_2)=(3,1,0,0)$ which we do next. We have from Lemma \ref{chapter2:claim:5.12}
\begin{samepage}
\begin{align*}
    &|\b{X}_\ell^3 \cdot h^{(3)}(x_\ell)| \\
    &\leq c L_{n,d} |\b{X}_\ell|^3 \sum_{i=1}^{n} \sum_{\substack{j_1,j_2,j_3,j_4 \leq 3+2m\\ j_2+j_3\geq 2}} |\b{X}_\ell|^{j_1} |a_{i,\ell}|^{j_2} |b_{i,\ell}|^{j_3} |u_{i,\ell}|^{j_4}\\
    &\leq c L_{n,d} \sum_{i=1}^{n} \sum_{\substack{j_1,j_2,j_3,j_4 \leq 3+2m\\ j_2+j_3\geq 2}}  \bigg[|\b{X}_\ell|^{3+2j_1} \left( a_{i,\ell}^{2j_2} \b{1}_{j_2, j_3\geq 1} + b_{i,\ell}^{2} \b{1}_{j_2=0}+ a_{i,\ell}^{2j_2-2} \b{1}_{j_3=0} \right)|u_{i,\ell}|^{2j_4} \\&+  |\b{X}_\ell|^{3}  \left( b_{i,\ell}^{2j_3} \b{1}_{j_2, j_3\geq 1} + b_{i,\ell}^{2j_3-2} \b{1}_{j_2=0} + a_{i,\ell}^2 \b{1}_{j_3=0} \right) \bigg],
\end{align*}
\end{samepage}
where the last line follows from the arithmetic-geometric mean inequality. Note in particular that each term in $a_{i,\ell}, b_{i,\ell}$ in the summand has a positive even exponent. Moreover, all the terms in the summand are connected LDPs containing node $i$. Using  Lemma \ref{chapter2:lemma:pre:2}, it follows that the summation can then be written as $|\b{X}_\ell|^3 w(\Z^\ell)$ where $\b{X}_\ell^3 w(\Z^\ell)$ is a connected LDP with coefficients bounded by a constant $c$. We can then apply  (\ref{chapter2:20}) in Lemma \ref{chapter2:lemma:pre:7} to obtain

\begin{align*}
    \E\l[\sup_{|x_{\ell}|\leq |\b{X}_{\ell}|} \l|\b{X}_\ell^3 \cdot h^{(3)}(x_\ell) \r| \r] &\leq  cL_{n,d} n^{-\frac{3}{2}} = cL_{n,d}^{m_1 + m_2}  n^{m_1+m_2-\frac{5}{2}},\\
    \E\l[ \sup_{|y_{\ell}|\leq |\Y_{\ell}|} \l|\Y_\ell^3 \cdot h^{(3)}(y_\ell) \r| \r] &\leq   c L_{n,d} n^{-1}t^{-\frac{1}{2}} =  c L_{n,d}^{m_1+m_2} n^{m_1+m_2-2}t^{-\frac{1}{2}}.
\end{align*}
This concludes the proof of (\ref{chapter2:claim.5.13.1}) and (\ref{chapter2:claim.5.13.2}). 

\end{proof}

The next lemma is closely related to Lemma~\ref{chapter2:claim:5.13} and serves a similar purpose, but in the context of the interpolation analysis of the Hamiltonian value. Specifically, it restates the interpolation setup used in the proof of universality of the objective value in Theorem~\ref{chapter2:thm:main-formal} and derives bounds on the associated third-order Taylor remainders. Due to the structure of the Hamiltonian, the resulting LDP notation is lighter, for instance using $a_{\ell}$ instead of $a_{i,\ell}$.

\begin{lemma}\label{chapter2:lemma:5.10}
 Let $\b{X}\distrib{\rm GOE}^0(n)$  and $\Y\distrib\G(n,t/n)$, where $d\leq t \leq \tau_n $ and $\tau_n$ satisfies $\omega(1)\leq \tau_n \leq o(n)$. Let $\ell\in [\binom{n}{2}]$ enumerate a pair $(i^*, j^*)$ with $(i^*, j^*)\in \binom{[n]}{2}$. Suppose $L_1 \cup L_2$ is a partition of  pairs $1\leq i<j \leq n,\  (i,j)\ne (i^*, j^*)$ with $L_1\not = \emptyset$.   Given  $i\in [n]$ let $u_{\ell},\ a_{\ell},\ b_{\ell}$ be  LDPs in variables $\x=(\x_{ij},\ 1\leq i<j \leq n)$, satisfying
 \begin{enumerate}
    \item $u_{\ell}(\x),\ \x_\ell a_{\ell}(\x),\ \x_\ell b_{\ell}(\x)$ are connected, and $\ \x_\ell a_{\ell}(\x),\ \x_\ell b_{\ell}(\x)$ contain the endpoints of edge $\ell$ (i.e., $\b{x}_\ell a_{\ell}(\b{x}) ,\ \b{x}_\ell b_{\ell}(\b{x}) \in \V_{\{i^*, j^*\}}$).
    \item  $u_{\ell}(\x),\ a_{\ell}(\x),\ b_{\ell}(\x)$ do not depend on the variable $\x_{\ell}$ (i.e., $\b{x}_{i^* j^*}$).
    \item $\|u_{\ell}\|_{\rm coef} + \|a_{\ell}\|_{\rm coef} + \|b_{\ell}\|_{\rm coef}\leq C$, where $C$ is a constant independent of $n, d$ and $\ell$.
\end{enumerate}
 Let  $\Z^{\ell}$  be given by $\Z^{\ell}_{ii}=0,\ \forall i\in [n]$, and 
\begin{equation}
    \begin{aligned}
        \relax [\Z^{\ell}]_{ij} &= \X_{ij} \quad && \text{if } \quad (i,j) \in L_1 \cup \{(i^*, j^*)\}, \\
        \relax [\Z^{\ell}]_{ij} &= \Y_{ij} \quad && \text{if } \quad (i,j) \in L_2.
    \end{aligned}
\end{equation}
Furthermore, let $\b{Z}^{\ell+1}$ be obtained from $\b{Z}^{\ell}$ by switching the disorder of the $\ell$-th edge $(i^*,j^*)$ from $\b{X}_\ell$ to $\b{Y}_\ell$. Finally, let $\beta = \log n$, and introduce for $z\in \R$
\begin{align*}
v_{\ell}(z) &= u_{\ell}(\b{Z}^\ell) + z a_{\ell}(\b{Z}^\ell) + z^2 b_{\ell}(\b{Z}^\ell),
\qquad i\in [n],\\
    \xi_{\beta, c}(z)& =\psi_{\beta, c} (v_{\ell}(z)).
\end{align*}
Then, we have
\begin{align*}
    \l|\E\l[\sup_{|x_{\ell}|\leq |\b{X}_{\ell}|}\X_\ell^3 \cdot \xi^{(3)}(x_{\ell})\r]\r| &\leq c(\log n)^2 n^{-\frac{7}{2}},\\
    \l|\E\l[\sup_{|y_{\ell}|\leq |\b{Y}_{\ell}|}\Y_\ell^3 \cdot \xi^{(3)}(y_{\ell})\r]\r| &\leq  c(\log n)^2 n^{-3} t^{-\frac{1}{2}},
\end{align*}
where $c$ is a constant independent of $n, d$ and $\ell$.
\end{lemma}
\begin{proof}%[Proof of Lemma \ref{chapter2:lemma:5.10}]
We use the slight abuse of notation $u_{\ell},\ a_{\ell},\ b_{\ell}$ to denote $ u_{\ell}(\Z^\ell),\ a_{\ell}(\Z^\ell),\ b_{\ell}(\Z^\ell)$. Moreover, throughout this proof, we denote by $c$ different constants independent of $n, d$ and $\ell$. Let  $M_{\psi} \triangleq \|\psi^{(2)}\|_{\infty} + \|\psi^{(3)}\|_{\infty}$. Since $\beta= \log n$, it follows from item 2 of Lemma \ref{chapter2:lemma:pre:4}
\begin{align*}
    M_{\psi}  &\leq \left(\frac{\beta}{n}\right)^2 +\left(\frac{\beta}{n}\right)^3  \leq \frac{2(\log n)^2}{n^2}.
\end{align*}
Recall that for  $\ell\leq N-1$ we have $v_\ell(z) = u_\ell + z a_\ell + z^2 b_\ell$ and $\xi(z)= \psi(v_\ell(z))$. We have
\begin{align*}
    |\xi^{(3)}(z)|&=\l| \psi^{(3)}\bigl(v_\ell(z)\bigr)\cdot(a_\ell+2zb_\ell)^3+6\,\psi^{(2)}\bigl(v_\ell(z)\bigr)\cdot(a_\ell+2zb_\ell)\,b_\ell\r|\\
    &\leq c M_{\psi} \sum_{\substack{k_1+2k_2 = 3\\ k_1, k_2 \geq 0}} |a_\ell + 2z b_\ell|^{k_1} |b_\ell|^{k_2} \\
    &\leq c\frac{(\log n)^2}{n^2} \sum_{\substack{0\leq j_1,j_2,j_3 \leq 3}} |z|^{j_1} |a_\ell|^{j_2}|b_\ell|^{j_3}.
\end{align*}
Combining the above and $|x_{\ell}|\leq |\X_\ell|$, we can bound $|\xi^{(3)}(x_{\ell})|$ by
\begin{align*}
   |\xi^{(3)}(x_{\ell})| &\leq c\frac{(\log n)^2}{n^2} \sum_{\substack{0\leq j_1,j_2,j_3 \leq 3}} |\X_\ell|^{j_1} |a_\ell|^{j_2}|b_\ell|^{j_3}\\
   &\leq c\frac{(\log n)^2}{n^2} \sum_{\substack{0\leq j_1,j_2,j_3 \leq 3}} \l[(\X_\ell)^{2j_1} (a_\ell)^{2j_2}(b_\ell)^{2j_3} + 1\r],
\end{align*}
where we used the arithmetic-geometric mean inequality in the last line. Therefore
\begin{align*}
    \l|\E\l[\sup_{|x_{\ell}|\leq |\b{X}_{\ell}|}\X_\ell^3 \cdot \xi^{(3)}(x_{\ell})\r]\r| \leq c \frac{(\log n)^2}{n^2} \sum_{\substack{0\leq j_1,j_2,j_3 \leq 3}} \l(\E\l[|\X_\ell|^{3+2j_1} (a_\ell)^{2j_2}(b_\ell)^{2j_3} + |\X_\ell|^3\r]\r).
\end{align*}
Note that $\X_\ell^{3+2j_1} (a_\ell)^{2j_2}(b_\ell)^{2j_3}$ and $\X_\ell^3$  are connected LDPs containing edge $\ell$  as the product of connected LDPs sharing edge $\ell$ (i.e., $\X_\ell^{3+2j_1} (a_\ell)^{2j_2}(b_\ell)^{2j_3},\ \X_\ell^3 \in \V_{\{i^*, j^*\} }$, where $(i^*,j^*)$ is the edge enumerated by $\ell$). It follows from applying (\ref{chapter2:20}) in Lemma \ref{chapter2:lemma:pre:7} with $S$ the set of the endpoints of edge $\ell$
\begin{align*}
    \E\l[|\X_\ell|^{3}\r] &\leq cn^{-\frac{3}{2}}, \\
    \E\l[|\X_\ell|^{3+2j_1} (a_\ell)^{2j_2} (b_\ell)^{2j_3}  \r]  &\leq c n^{-(2 + \frac{3+2j_1}{2}-2)} = cn^{-\frac{3}{2}-j_1}\leq cn^{-\frac{3}{2}},
\end{align*}
where we used $j_1\geq 0$. It follows readily,
\begin{align}
    \l|\E\l[\sup_{|x_{\ell}|\leq |\b{X}_{\ell}|}\X_\ell^3 \cdot \xi^{(3)}(x_{\ell})\r]\r| &\leq c (\log n)^2 n^{-\frac{7}{2}} \label{chapter2:lemma.5.10.1}.
\end{align}
We bound $\l|\E\l[\sup_{|y_{\ell}|\leq |\b{Y}_{\ell}|}\Y_\ell^3 \cdot \xi^{(3)}(y_{\ell})\r]\r|$ identically: the only difference lies in the application of Lemma \ref{chapter2:lemma:pre:7}, which yields for the disorder matrix $\Y$,
\begin{align*}
    \E\l[ |\Y_\ell|^{3}\r] &\leq cn^{-1} t^{-\frac{1}{2}}, \\
    \E\l[|\Y_\ell|^{3+2j_1}  (a_\ell)^{2j_2}  (b_\ell)^{2 j_3}  \r]  &\leq  c n^{-1} t^{-\frac{3+2j_1}{2}+1} = c n^{-1}t^{-\frac{1}{2}-j_1}\leq c n^{-1}t^{-\frac{1}{2}},
\end{align*}
It follows  that
\begin{align}
    \l|\E\l[\sup_{|y_{\ell}|\leq |\b{Y}_{\ell}|}\Y_\ell^3 \cdot \xi^{(3)}(y_{\ell})\r]\r| &\leq c (\log n)^2 n^{-3}t^{-\frac{1}{2}} \label{chapter2:lemma.5.10.2}.
\end{align}
which ends the proof of Lemma \ref{chapter2:lemma:5.10}.
\end{proof}

\subsection{Ground State Energy in $\G(n, t/n)$}\label{chapter2:section:gound_state}

In this section, we provide a derivation of the ground state energy for the disorder $\G(n, t/n)$. As mentioned in the introduction, the result is known
when $t$ is a constant $d$ in the double limit sense $n\to \infty$ followed by $d\to\infty$~\cite{DemboMontanariSubhabrata, sen2018optimization}. It is not then surprising that it also holds in the intermediate regime $d\leq t\leq \tau_n$ where $\tau_n$ satisfies $\omega(1)\le \tau_n\leq n/2$. We provide the proof for completeness.

\begin{proposition}\label{chapter2:proposition:sk.sparse}
    Let $\Y \distrib \G(n, t/n)$ with $d\leq t\leq \tau_n$ where $\tau_n$ satisfies $\omega(1)\leq \tau_n\leq n/2$. Then, 
    \begin{align}
        \plimd \frac{1}{2n} \max_{\pmb{\sigma} \in \{\pm 1\}^n} \pmb{\sigma}^\top \Y \pmb{\sigma} = \Parisi^*. \label{chapter2:proposition:sk.sparse.1}
    \end{align}
\end{proposition}
\begin{proof}
Let $H_n(\pmb{\sigma}, \b{M}) \triangleq \pmb{\sigma}^\top \b{M} \pmb{\sigma}$. The proof proceeds in two steps. First, we show through an interpolation argument that $\E[\max_{\pmb{\sigma}\in \{\pm 1\}^n} H_n(\pmb{\sigma}, \b{Y})]/2n = \Parisi^* + o(1)$. Second, we use concentration results to obtain convergence in probability. Let $\b{X}\distrib {\rm GOE}^0(n)$, and introduce the following log-partition function with inverse temperature $\beta>0$
\begin{align*}
    F_n(\beta, \b{M}) = \frac{1}{2n\beta} \log\l(\sum_{\pmb{\sigma}\in\{\pm 1\}^n} \exp\l(\beta H_n(\pmb{\sigma}, \b{M})\r)\r).
\end{align*}
In the remainder of this proof, we set $\beta = \log t$. We next show the following through a Lindeberg interpolation
\begin{align}
    \l| \E[F_n(\beta, \b{Y})] - \E[F_n(\beta, \b{X})] \r| = o(1). \label{chapter2:proposition:sk.sparse.2}
\end{align}
Let the interpolation path disorders be given by $\b{Z}^0,\dots, \b{Z}^{N}$ with $N=\binom{n}{2}$ and $\b{Z}^0=\b{X}, \b{Z}^N=\b{Y}$. Fix $k\in[N]$, and suppose we switch $\b{X}_{ij}$ into $\b{Y}_{ij}$ to go from $\b{Z}^{k-1}$ to $\b{Z}^k$. Denote by $\b{Z}(u)$ the disorder obtained from $\b{Z}^{k-1}$ by replacing $\b{X}_{ij}$ with $u$. We have by Taylor's
\begin{align*}
    F_n(\beta, \b{Z}^k) - F_n(\beta, \b{Z}^{k-1}) &= F_n(\beta, \b{Z}(\b{Y}_{ij})) - F_n(\beta, \b{Z}(\b{X}_{ij}))\\
    &=\l(\b{Y}_{ij} - \b{X}_{ij}\r) \frac{\partial F_n(\beta, \b{Z}(0))}{\partial u} + \frac{\b{Y}_{ij}^2 - \b{X}_{ij}^2}{2} \frac{\partial^2 F_n(\beta, \b{Z}(0))}{\partial u^2} \\
    &+ \frac{\b{Y}_{ij}^3}{6} \frac{\partial^3 F_n(\beta, \b{Z}(\hat{u}))}{\partial u^3}-\frac{\b{X}_{ij}^3}{6} \frac{\partial^3 F_n(\beta, \b{Z}(\tilde{u}))}{\partial u^3},
\end{align*}
where $\hat{u}, \tilde{u} \in (\min(\b{X}_{ij},\b{Y}_{ij}), \max(\b{X}_{ij}, \b{Y}_{ij}))$. Taking the expectation over the disorders $\b{X}, \b{Y}$ yields
\begin{align}
    \l|\E\l[F_n(\beta, \b{Z}^k)\r] - \E\l[F_n(\beta, \b{Z}^{k-1})\r]\r| &\leq \frac{\sup_{u\in \R} \l|\frac{\partial^3 F_n(\beta, \b{Z}(u))}{\partial u^3}\r|}{6}\l(\E[|\b{X}_{ij}|^3] + \E[|\b{Y}_{ij}|^3] \r) \label{chapter2:proposition:sk.sparse.3}.
\end{align}
Straightforward computations of the derivatives of $u\mapsto F_n(\beta, \b{Z}(u))$ yield
\begin{align*}
    \frac{\partial F_n(\beta, \b{Z}(u))}{\partial u} &= \frac{1}{n} \langle \pmb{\sigma}_i \pmb{\sigma}_j \rangle_G,\\
    \frac{\partial^2 F_n(\beta, \b{Z}(u))}{\partial u^2} &= \frac{2\beta}{n} \l( 1 - \langle \pmb{\sigma}_i \pmb{\sigma}_j \rangle_G^2 \r)\\
    \frac{\partial^3 F_n(\beta, \b{Z}(u))}{\partial u^3} &= -\frac{8\beta^2}{n} \langle \pmb{\sigma}_i \pmb{\sigma}_j\rangle_G \l(1 - \langle \pmb{\sigma}_i \pmb{\sigma}_j \rangle^2_G  \r),
\end{align*}
where we use the notation $\langle f \rangle_G$ for a function $f=f(\pmb{\sigma})$ to denote the Gibbs average. Namely, $\langle f \rangle_G = \sum_{\pmb{\sigma}} f(\pmb{\sigma}) e^{\beta H_n(\pmb{\sigma}, \b{Z}(u))} /Z$ and $Z = \sum_{\pmb{\sigma}} e^{\beta H_n(\pmb{\sigma}, \b{Z}(u))} $. Therefore $|\langle \pmb{\sigma}_i \pmb{\sigma}_j \rangle_G|\leq 1$, which implies the bound $\sup_{u\in \R} \l| \frac{\partial^3 F_n(\beta, \b{Z}(u))}{\partial u^3}\r| \leq 8\beta^2 n^{-1}$.  Using Lemma \ref{chapter2:lemma:pre:sparse}, we have $\E[|\b{Y}_{ij}|^3]\leq 2n^{-1}t^{-\frac{1}{2}}$ for large enough $n$. Furthermore, we have $\E[|\b{X}_{ij}|^3] = cn^{-\frac{3}{2}}\leq 2n^{-\frac{3}{2}}$ where $c = 2\sqrt{2}/\sqrt{\pi} \approx 1.59$. It follows that
\begin{align*}
    \l|\E\l[F_n(\beta, \b{Z}^k)\r] - \E\l[F_n(\beta, \b{Z}^{k-1})\r]\r| &\leq \frac{16 \beta^2 n^{-1}\l(n^{-1} t^{-\frac{1}{2}} + n^{-\frac{3}{2}}\r)}{6}\\
    &\leq 6\beta^2n^{-2}t^{-\frac{1}{2}},
\end{align*}
where we used $t\leq n$ in the last line. Using the above in (\ref{chapter2:proposition:sk.sparse.3}), we have
\begin{align*}
    \l| \E[F_n(\beta, \b{Y})] - \E[F_n(\beta, \b{X})] \r| &\leq \sum_{k=1}^{N} \l|\E\l[F_n(\beta, \b{Z}^k)\r] - \E\l[F_n(\beta, \b{Z}^{k-1})\r] \r|\\
    &\leq \binom{n}{2} 8\beta^2n^{-2}t^{-\frac{1}{2}}\\
    &\leq 4\beta^2 t^{-\frac{1}{2}}.
\end{align*}
Recalling that $\beta=\log t$, it follows  from the above that $\l| \E[F_n(\beta, \b{Y})] - \E[F_n(\beta, \b{X})] \r| = o(1)$ as $n\to\infty$ followed by $d\to\infty$, which yields (\ref{chapter2:proposition:sk.sparse.2}). We next use (\ref{chapter2:proposition:sk.sparse.2}) and concentration arguments to show convergence in probability of $\max_{\pmb{\sigma} \in \{\pm 1\}^n} \frac{H_n(\pmb{\sigma}, \b{Y})}{2n}$. Note that
\begin{align*}
   \forall \b{M}\in\R^{n\times n}, \quad \frac{1}{2n}\max_{\pmb{\sigma}\in \{\pm 1\}^n} H_n(\pmb{\sigma}, \b{M}) \leq F_n(\beta, \b{M}) \leq \frac{\log 2}{2\beta} + \frac{1}{2n}\max_{\pmb{\sigma}\in \{\pm 1\}^n} H_n(\pmb{\sigma}, \b{M}).
\end{align*}
Combining the above with (\ref{chapter2:proposition:sk.sparse.2}) and using $\log (2) /\beta=o(1)$ yields
\begin{align*}
    \E\l[ \frac{1}{2n}\max_{\pmb{\sigma}\in \{\pm 1\}^n} H_n(\pmb{\sigma}, \b{Y})\r] &= \E\l[ \frac{1}{2n}\max_{\pmb{\sigma}\in \{\pm 1\}^n} H_n(\pmb{\sigma}, \b{X})\r] + o(1) \numberthis \label{chapter2:proposition:sk.sparse.lindeberg}.
\end{align*}
The function $\b{X} \mapsto \frac{1}{2n} \max_{\pmb{\sigma}\in \{\pm 1\}^n} H_n(\pmb{\sigma}, \b{X})$ is $1/\sqrt{2}$-Lipschitz in the centered Gaussian variables $(\b{X}_{ij})_{1\leq i<j \leq n}$ with variance $1/n$. Therefore, we have by concentration of measure
\begin{align*}
    \forall u>0, \quad \p\l( \l| \frac{1}{2n} \max_{\pmb{\sigma}\in \{\pm 1\}^n} H_n(\pmb{\sigma}, \b{X}) - \E\l[\frac{1}{2n} \max_{\pmb{\sigma}\in \{\pm 1 \}^n} H_n(\pmb{\sigma}, \b{X})\r] \r| \geq u\r) \leq 2e^{-cnu^2},
\end{align*}
where $c$ is some explicit constant. This yields
\begin{align}
    \plim \frac{1}{2n} \max_{\pmb{\sigma}\in \{\pm 1\}^n} H_n(\pmb{\sigma}, \b{X})  =\mathcal{P}^* = \lim_{n\to \infty} \E\l[\frac{1}{2n} \max_{\pmb{\sigma}\in \{\pm 1\}^n} H_n(\pmb{\sigma}, \b{X})\r], \label{chapter2:proposition:sk.sparse.4}
\end{align}
where we used the known fact (e.g. Theorem 1.1 in \cite{talagrand2006parisi})  that $\plim \frac{1}{2n} \underset{\pmb{\sigma}\in \{\pm 1\}^n}{\max} H_n(\pmb{\sigma}, \b{X})  = \Parisi^*$. Combining the above with (\ref{chapter2:proposition:sk.sparse.lindeberg}), we obtain
\begin{align*}
    \lim_{\substack{n\to \infty \\ \text{then } d\to\infty}} \E\l[\frac{1}{2n} \max_{\pmb{\sigma}\in \{\pm 1\}^n} H_n(\pmb{\sigma}, \b{Y})\r] = \mathcal{P}^*. \numberthis \label{chapter2:proposition:sk.sparse.mean}
\end{align*}
Let $\b{Y}'$ be the matrix obtained from $\b{Y}$ by flipping the value of the $(i,j)$-th Bernoulli variable in $\G(n, t/n)$. Note that $\Delta_{ij}\triangleq |\b{Y}_{ij} - \b{Y}'_{ij}| =\frac{1}{\sqrt{t\l(1-\frac{t}{n}\r)}} \leq \sqrt{2t^{-1}}$, where we used $t\leq n/2$ in the last inequality. We then have 
\begin{align*}
    \l|\max_{\pmb{\sigma} \in \{\pm 1\}^n} H_n(\pmb{\sigma}, \b{Y}) - \max_{\pmb{\sigma} \in \{\pm 1\}^n} H_n(\pmb{\sigma}, \b{Y}') \r| &\leq \max_{\pmb{\sigma} \in \{\pm 1\}^n}|\pmb{\sigma}^\top (\b{Y} - \b{Y}') \pmb{\sigma}|\\
    &=2\Delta_{ij}.
\end{align*}
Using McDiarmid's inequality and noting that $\sum_{1\leq i<j \leq n} \Delta_{ij}^2 \leq n^2 t^{-1}$, it follows that
\begin{align*}
    \forall u>0, \quad \p\l( \l| \frac{1}{2n} \max_{\pmb{\sigma}\in \{\pm 1\}^n} H_n(\pmb{\sigma}, \b{Y}) - \E\l[\frac{1}{2n} \max_{\pmb{\sigma}\in \{\pm 1\}^n} H_n(\pmb{\sigma}, \b{Y})\r] \r| \geq u\r) &\leq \exp\l(-ct u^2\r), 
\end{align*}
where $c$ is some explicit constant. The above  together with (\ref{chapter2:proposition:sk.sparse.mean}) and $t=\omega(1)$ implies
\begin{align*}
    \plimd \frac{1}{2n} \max_{\pmb{\sigma}\in \{\pm 1\}^n} H_n(\pmb{\sigma}, \b{Y}) = \mathcal{P}^*.
\end{align*}
This ends the proof.
\end{proof}

\section{Proof of Theorem~\ref{chapter2:lemma:state_ev}}\label{chapter2:section:state_ev}
In this section, we prove our first main result, Theorem~\ref{chapter2:lemma:state_ev}. We first show the result for sequences of smooth test functions as stated in the following intermediary proposition.

\begin{proposition}\label{chapter2:lemma:state_ev_smooth}
    Let $\b{X}\distrib {\rm GOE}^0(n)$ and $\b{Y}\distrib \G(n, t/n)$ with $d\leq t \leq \tau_n$ where $\tau_n$ satisfies $\omega(1)\leq \tau_n \leq o(n)$.
    Suppose $\chi_{n,d}:\R\to \R$ are functions satisfying  
    \begin{enumerate}
        \item $\chi_{n, d}$ is three times differentiable.
        \item The first three derivatives of $\chi_{n, d}$  satisfy the growth condition $|\chi_{n,d}^{(\ell)}(x)|\leq \log(d)^{q}(1+|x|^m), \ell=1,2,3,$ for some integers $q, \ m\in\mathbb{Z}_{\geq 0}$ that do not depend on $n$,  $d$.  
    \end{enumerate}
    Let $(p_i,\ i\in [n])$ be a collection of LDPs satisfying
    \begin{itemize}
        \item There exists a constant $C>0$ independent of $n$ and $d$ such that $\forall i\in [n], \ \|p_i\|_{\rm coef} \leq C$.
        \item There exists a constant $\Delta>0$ independent of $n$ and $d$ such that $\forall i \in [n], \ p_i \in \mathcal{T}_{n, 2, \Delta}$.
        \item For all $i\in [n]$, the LDP $p_i$ contains node $i$ (i.e., $p_i\in \V_i$).
    \end{itemize}
    Finally, let $\kappa\in \mathbb{R}$ be a scalar. Then, the following two statements are equivalent,
    \begin{align*}
        \plimnd \frac{1}{n} \sum_{i=1}^{n} \chi_{n, d}(p_i(\b{X})) &= \kappa, \numberthis \label{chapter2:lemma:state_ev_smooth.1}\\
        \plimnd \frac{1}{n} \sum_{i=1}^{n} \chi_{n, d}(p_i(\b{Y})) &= \kappa. \numberthis \label{chapter2:lemma:state_ev_smooth.2}
    \end{align*}
\end{proposition}
\begin{proof}
We use an interpolation argument from the Gaussian disorder $\b{X}\distrib {\rm GOE}^0(n)$ to the  disorder $\Y\distrib\G(n, t/n)$.  For $c, \beta \in \R$ recall $\phi$ and $\psi_{\beta,c}$ from Lemma \ref{chapter2:lemma:pre:4}. For convenience, we  rewrite $\psi_{\beta, c}$  as
\begin{align*}
    \psi_{\beta,c}(z)=   \phi\l(\beta \l(\frac{z}{n} - c\r)\r).
\end{align*}
Throughout this proof, we will assume that $\beta = \pm \log d $. Introduce the following functions of $\b{x} = (\b{x}_{ij},\ 1\leq i<j \leq n)$
\begin{align*}
     s(\b{x}) &= \sum_{i=1}^{n} \chi_{n, d}(p_i(\b{x})), \\
     f(\b{x}) &= \psi_{\beta,c} (s(\b{x})).
\end{align*}
We first show the following,
    \begin{align}
    \l| \E[f(\Y)]  - \E[f(\b{X})] \r| = o(1). \label{chapter2:lemma.5.11}
    \end{align}
We consider the interpolation path from $\b{X}$ to $\Y$ given by $\Z^0,\ldots,\Z^N$ where $\Z^0=\b{X},\ \Z^N=\Y$ and $N\triangleq \binom{n}{2}$.  For $\ell\in[N]$ we will identify $\ell$ with its corresponding (enumerated) edge $(ij)$, and thus write $\b{X}_\ell,\Y_\ell$ instead of $\b{X}_{ij}, \Y_{ij}$. We have
\begin{align*}
    \E[f(\Y)]  - \E[f(\b{X})] &= \sum_{\ell=0}^{N-1} \E[f(\Z^{\ell+1})] - \E[f(\Z^\ell)].
\end{align*}
Since $\forall i\in[n],\ p_i \in \mathcal{T}_{n, 2, \Delta}$, it follows that the monomials $\Z^{\pmb{\alpha}}$ in $ p_i(\Z)$ satisfy $\max(\pmb{\alpha}) \leq 2$. In particular, we can write for $\ell\leq N-1,\ i\in[ n]$
\begin{align}
     p_i(\Z^\ell) &= u_{i,\ell}(\Z^\ell) + \b{X}_{\ell} a_{i,\ell}(\Z^{\ell}) + \b{X}_{\ell}^2 b_{i,\ell}(\Z^\ell) \label{chapter2:lemma.5.11.1},
\end{align}
where $u_{i,\ell}(\b{x}),\ a_{i,\ell}(\b{x}),\ b_{i,\ell}(\b{x})$ are LDPs in the variables $\b{x}=(\b{x}_{ij},\ 1\leq i<j\leq n)$ which do not depend on $\b{x}_\ell$. Furthermore, since we switch the distribution of $[\Z^\ell]_\ell$ from $\b{X}_\ell$ to $\Y_\ell$ to obtain $\Z^{\ell+1}$ (and keep everything else unchanged), it follows that
\begin{align}
     p_i(\Z^{\ell+1})  &= u_{i,\ell}(\Z^\ell) + \Y_{\ell} a_{i,\ell}(\Z^{\ell}) + \Y_{\ell}^2 b_{i,\ell}(\Z^\ell) \label{chapter2:lemma.5.11.2}.
\end{align}
 Two important elementary properties of $u_{i,\ell},\ a_{i,\ell},\ b_{i,\ell}$ are
\begin{align}
    &\forall i\in[n],\ \ell\in [0,N-1],\ u_{i,\ell}(\b{x}),\ \b{x}_\ell a_{i,\ell}(\b{x}),\  \b{x}_\ell b_{i,\ell}(\b{x}) \text{ are connected and contain node } i.\label{chapter2:uab:3}\\
    &\forall i\in[n],\ \ell\in [0,N-1],\  \|u_{i,\ell}\|_{\rm coef} + \|a_{i,\ell}\|_{\rm coef} + \|b_{i,\ell}\|_{\rm coef}\leq 3C. \label{chapter2:uab:4}
\end{align}
The connectedness of $u_\ell(\b{x}),\ \b{x}_\ell a_\ell(\b{x}),\ \b{x}_\ell b_\ell(\b{x})$ in (\ref{chapter2:uab:3}) follows from first noting  that $u_{\ell}(\b{x}),\ \b{x}_\ell^2 a_\ell(\b{x}),\ \b{x}_\ell^2 b_\ell(\b{x})$ are connected (as $p_i(\b{x})$ is connected) and using the fact that $\b{x}_\ell w (\b{x}) $ is connected if and only if $(\b{x}_\ell)^k w(\b{x})$ is connected for all $k\in \mathbb{Z}_{\geq 1}$, where $w$ is any LDP.  The second part of property (\ref{chapter2:uab:3}) means that $u_{i, \ell}(\b{x}),\  \b{x}_\ell a_{i, \ell}(\b{x}),$ $\ \b{x}_\ell b_{i, \ell}(\b{x}) \in \V_i$, and follows from $p_i \in \V_i$. Finally, property (\ref{chapter2:uab:4}) follows immediately from $\forall i\in [n], \  \|p_i\|_{\rm coef}\leq C$ in the proposition statement.  For brevity, we use the slight abuse of notation $u_{i,\ell},\  a_{i,\ell},\ b_{i,\ell}$ to denote $ u_{i,\ell}(\Z^\ell),\  a_{i,\ell}(\Z^\ell),\ b_{i,\ell}(\Z^\ell)$. Introduce the following  functions 
\begin{align*}
    v_{i,\ell}(z) &=  u_{i,\ell} + z a_{i,\ell} + z^2 b_{i,\ell} ,\\
    h_{\ell}(z) &= \sum_{i=1}^{n} \chi_{n, d}(v_{i,\ell}(z)),\\
    g_{\ell}(z)&= \psi_{\beta, c} \left(h_{\ell}(z)\right). 
\end{align*}
In particular, note that for $\ell\leq N-1$
\begin{align*}
    f(\Z^{\ell+1}) - f(\Z^\ell) = g_\ell(\Y_{\ell}) - g_\ell(\b{X}_\ell) \numberthis \label{chapter2:interpol:step}.
\end{align*}
Using a third-order Taylor expansion, there exists $\b{X}_\ell^* \in [- |\b{X}_\ell|, |\b{X}_\ell|] $ and $\Y_\ell^* \in [-|\Y_\ell|, |\Y_\ell|]$ such that
    \begin{align*}
        g_\ell(\b{X}_\ell) &= g_\ell(0) + \b{X}_\ell g_\ell^{(1)}(0) + \frac{\b{X}_\ell^2}{2} g^{(2)}_\ell(0)   + \frac{\b{X}_\ell^3}{6} g_\ell^{(3)}(\b{X}_\ell^*),\\
        g_\ell(\Y_\ell) &= g_\ell(0) + \Y_\ell g^{(1)}_\ell(0) + \frac{\Y_\ell^2}{2} g^{(2)}_\ell(0)  + \frac{\Y_\ell^3}{6} g_\ell^{(3)}(\Y_\ell^*).
    \end{align*}
Taking the difference of the above and the expectation yields
\begin{align*}
    \E[g_\ell(\Y_\ell)] - \E[g_\ell(\b{X}_\ell)] = \frac{\E\l[\Y_\ell^3 g^{(3)}_\ell(\Y_\ell^*)\r] - \E\l[\b{X}_\ell^3 g^{(3)}_\ell(\b{X}_\ell^*)\r]}{6} \numberthis \label{chapter2:lemma.5.11.3}.
\end{align*}
Differentiating $g_\ell$, we have
\begin{align*}
    g_{\ell}^{(3)}(z)&= \psi_{\beta,c}^{(3)} \left(h_{\ell}(z)\right) \cdot \left(h_{\ell}^{(1)}(z)\right)^{3} + 3\,\psi_{\beta,c}^{(2)}\!\left(h_{\ell}(z)\right)  \cdot  h_{\ell}^{(1)}(z) \cdot h_{\ell}^{(2)}(z) \\
    &+ \psi_{\beta,c}^{(1)}\!\left(h_{\ell}(z)\right) \cdot  h_{\ell}^{(3)}(z).
\end{align*}
Combining the above with $\|\psi_{\beta,c}^{(j)}\|_{\infty} \leq (|\beta|/n)^j = (\log(d)/ n)^j$ for $j=1,2,3$ from Lemma \ref{chapter2:lemma:pre:4}, it follows that
\begin{align*}
    |g_{\ell}^{(3)}(z)| &\leq  \sum_{\substack{s_1,m_1,s_2,m_2 \in \mathbb{Z}_{\geq 0}\\s_1 m_1 + s_2 m_2 = 3}}  \l( \frac{\log d}{n}\r)^{m_1 + m_2} \l|h_{\ell}^{(s_1)}(z)\r|^{m_1} \l|h_{\ell}^{(s_2)}(z)\r|^{m_2}.
\end{align*}
Therefore
\begin{align*}
     \l| \E\l[\b{X}_\ell^3 g_\ell^{(3)}(\b{X}_\ell^*)\r]\r| &\leq   \sum_{\mathcal{I}}  \l( \frac{\log d}{n}\r)^{m_1 + m_2} \E \l[\sup_{|z|\leq |\b{X}_{\ell}|} |\b{X}_\ell|^3 \l|h_{\ell}^{(s_1)}(z)\r|^{m_1} \l|h_{\ell}^{(s_2)}(z)\r|^{m_2}\r],
\end{align*}
where the sum is over the set $\mathcal{I} = \{s_1,m_1,s_2,m_2 \in \mathbb{Z}_{\geq 0} \mid s_1 m_1 + s_2 m_2 = 3\}$.
Combining the latter bound with (\ref{chapter2:claim.5.13.1}) from applying Lemma \ref{chapter2:claim:5.13} on $\chi_{n, d}$ with $L_{n,d}=\log(d)^q$, we have
\begin{align*}
     \l| \E[\b{X}_\ell^3 g_\ell^{(3)}(\b{X}_\ell^*)]\r| &\leq   \sum_{\mathcal{I}}  c\l( \frac{\log d}{n}\r)^{m_1 + m_2}  \cdot  \l(\log(d)^{q(m_1 + m_2)} n^{m_1 + m_2 - \frac{5}{2}}\r)\\
     &\leq c\log(d)^{3(1+q)}n^{-\frac{5}{2}}, \numberthis \label{chapter2:lemma.5.11.7}
\end{align*}
where $c$ denotes different constants independent of $n,d$ and $\ell$. Following the same arguments above, we have with (\ref{chapter2:claim.5.13.2}) from Lemma \ref{chapter2:claim:5.13}
\begin{align*}
    \l| \E[\Y_\ell^3 g_\ell^{(3)}(\Y_\ell^*)]\r| &\leq c\log(d)^{3(1+q)}n^{-2} t^{-\frac{1}{2}} \numberthis \label{chapter2:lemma.5.11.8}.
\end{align*}
Combining (\ref{chapter2:lemma.5.11.7}) and (\ref{chapter2:lemma.5.11.8}) in (\ref{chapter2:lemma.5.11.3}), we have
\begin{align*}
    \l|\E[g_\ell(\Y_\ell)] - \E[g_\ell(\b{X}_\ell)]\r| &\leq  c\log(d)^{3(1+q)}\l(n^{-\frac{5}{2}} + n^{-2} t^{-\frac{1}{2}} \r)\leq   2c \log(d)^{3(1+q)} n^{-2} t^{-\frac{1}{2}}.
\end{align*}
Summing the above over $\ell\leq N-1 \sim n^2/2$ and using (\ref{chapter2:interpol:step}) yields
\begin{align*}
    |\E[f(\Y)] - \E[f(\b{X})]| &\leq  O\l(\log(d)^{3(1+q)} t^{-\frac{1}{2}}  \r) =o(1) \numberthis \label{chapter2:interpol},
\end{align*}
which ends the proof of (\ref{chapter2:lemma.5.11}). We now show the implication (\ref{chapter2:lemma:state_ev_smooth.1}) $\implies$ (\ref{chapter2:lemma:state_ev_smooth.2}) with limit $\kappa \in \R$. The other implication will follow from identical arguments, and thus we skip its proof for brevity. Fix $\varepsilon>0$ and set 
\begin{alignat}{3}
    c^+ &= \kappa + \varepsilon, \quad &&c^- &&= \kappa - \varepsilon,\nonumber \\
    \beta^+ &= \log d, \quad &&\beta^- &&= -\log d, \nonumber \\
    f^+ &= \psi_{\beta^+, c^+}, \quad &&f^- &&= \psi_{\beta^-, c^-}. \nonumber
\end{alignat}
Since $0\leq f^+ \leq 1$, and $f^+$ is increasing, we have
\begin{align*}
    \E[f^+(\b{X})] &= \E\l[\psi_{\beta^+, c^+}(s(\b{X}))\r]\\
    &= \E\l[\psi_{\beta^+, c^+}(s(\b{X})) \cdot 1_{\frac{s(\b{X})}{n} \geq \kappa + \frac{\varepsilon}{2}}\r] + \E\l[\psi_{\beta^+, c^+}(s(\b{X})) \cdot 1_{\frac{s(\b{X})}{n} < \kappa + \frac{\varepsilon}{2}}\r]\\
    &\leq \p\l( \frac{s(\b{X})}{n}\geq \kappa + \frac{\varepsilon}{2}\r) +  \psi_{\beta^+, c^+}\l(\kappa + \frac{\varepsilon}{2}\r)\\
    &= o(1) + \phi\l(-\frac{\varepsilon \log d}{2}\r) \numberthis \label{chapter2:lemma:state_ev_smooth.3}\\
    &= o(1) \numberthis \label{chapter2:lemma:state_ev_smooth.4},
\end{align*}
where we used $\plimnd s(\b{X})/n = \kappa $ in (\ref{chapter2:lemma:state_ev_smooth.3}), and $\varepsilon \log d=\omega(1)$ in (\ref{chapter2:lemma:state_ev_smooth.4}). Similarly, since $0\leq f^-\leq 1$ and $f^-$ is decreasing, we have
\begin{align*}
    \E[f^-(\b{X})] &= \E\l[\psi_{\beta^-, c^-}(s(\b{X}))\r]\\
    &= \E\l[\psi_{\beta^-, c^-}(s(\b{X})) \cdot 1_{\frac{s(\b{X})}{n} \leq \kappa - \frac{\varepsilon}{2}}\r] + \E\l[\psi_{\beta^-, c^-}(s(\b{X})) \cdot 1_{\frac{s(\b{X})}{n} > \kappa - \frac{\varepsilon}{2}}\r]\\
    &\leq \p\l( \frac{s(\b{X})}{n}\leq \kappa - \frac{\varepsilon}{2}\r) +  \psi_{\beta^-, c^-}\l(\kappa - \frac{\varepsilon}{2}\r)\\
    &= o(1) + \phi\l(-\frac{\varepsilon \log d}{2}\r)\\
    &= o(1).
\end{align*}
Therefore $\E[f^+(\b{X})]=o(1)$ and $\E[f^-(\b{X})]=o(1)$. Combining the latter with (\ref{chapter2:interpol}), we have 
\begin{align*}
    \E[f^+(\b{Y})]&=o(1), \numberthis \label{chapter2:lemma:state_ev_smooth.5}\\
    \E[f^-(\b{Y})]&=o(1). \numberthis \label{chapter2:lemma:state_ev_smooth.6}
\end{align*}
Using (\ref{chapter2:lemma:state_ev_smooth.5}) and $f^+\geq 0$, we have
\begin{align*}
     o(1) &= \E\l[f^+(\b{Y})\r]\\
     &\geq \E\l[\psi_{\beta^+, c^+}(s(\b{Y}))  \mid\frac{s(\b{Y})}{n} \geq \kappa + 2\varepsilon \r]  \cdot \p\l(\frac{s(\b{Y})}{n} \geq \kappa + 2\varepsilon\r)\\
     &\geq \phi\l(\varepsilon \log d \r)  \cdot \p\l(\frac{s(\b{Y})}{n} \geq \kappa + 2\varepsilon\r)\\
     &= (1 + o(1)) \cdot \p\l(\frac{s(\b{Y})}{n} \geq \kappa + 2\varepsilon\r),
\end{align*}
where we used $\varepsilon \log d = \omega(1)$ in the last line. The above rearranges to 
\begin{align*}
    \p\l(\frac{s(\b{Y})}{n} \geq \kappa + 2\varepsilon\r) = o(1). \numberthis \label{chapter2:lemma:state_ev_smooth.7}
\end{align*}
Similarly
\begin{align*}
     o(1) &= \E\l[f^-(\b{Y})\r]\\
     &\geq \E\l[ \psi_{\beta^-, c^-}(s(\b{Y}))  \mid \frac{s(\b{Y})}{n} \leq \kappa - 2\varepsilon \r] \cdot  \p\l(\frac{s(\b{Y})}{n} \leq \kappa - 2\varepsilon\r)\\
     &\geq \phi\l(\varepsilon \log d \r) \cdot \p\l(\frac{s(\b{Y})}{n} \leq \kappa - 2\varepsilon\r)\\
     &= (1 + o(1))\cdot \p\l(\frac{s(\b{Y})}{n} \leq \kappa - 2\varepsilon\r),
\end{align*}
which rearranges to 
\begin{align*}
    \p\l(\frac{s(\b{Y})}{n} \leq \kappa - 2\varepsilon\r) = o(1). \numberthis \label{chapter2:lemma:state_ev_smooth.8}
\end{align*}
Combining (\ref{chapter2:lemma:state_ev_smooth.7}) and (\ref{chapter2:lemma:state_ev_smooth.8}) yields
\begin{align*}
    \plimnd \frac{s(\b{Y})}{n} = \kappa,
\end{align*}
which shows (\ref{chapter2:lemma:state_ev_smooth.1}) $\implies$ (\ref{chapter2:lemma:state_ev_smooth.2}) and ends the proof of the Proposition.
\end{proof}

We now prove Theorem~\ref{chapter2:lemma:state_ev}
\begin{proof}
    Let $\X \distrib {\rm GOE}^0(n)$ and $\Y\distrib\G(n, t/n)$ with $t$ as in the statement of the theorem. We next prove the implication (\ref{chapter2:lemma:state_ev.1}) $\implies$ (\ref{chapter2:lemma:state_ev.2}). The other implication will follow from identical arguments, and we omit its details. Fix $\kappa\in \R$ and suppose (\ref{chapter2:lemma:state_ev.1}) holds.     Applying Lemma \ref{chapter2:lemma:PL_approx} to $\chi$ with $\sigma=1/\log d$, it follows that there exists a thrice differentiable scalar function $f_d = f_{\sigma}$, and a constant $c$ depending only on $k$ and $L$ such that
    \begin{align*}
        \forall x\in \R, \quad |\chi(x) - f_{d}(x)| &\leq  \frac{c}{\log d}(1 + |x|^{k-1}),\numberthis \label{chapter2:lemma:state_ev.3}\\
        \forall x\in \R, \ j\in[3], \quad |f^{(j)}(x)| &\leq c\log(d)^{j}(1 + |x|^{k}). \numberthis \label{chapter2:lemma:state_ev.4}
    \end{align*}
    We have
    \begin{align*}
        &\frac{1}{n} \sum_{i=1}^{n} \chi(p_i(\b{Y}))- \frac{1}{n} \sum_{i=1}^{n} \chi(p_i(\b{X})) \\
        &= \underbrace{\frac{1}{n}\sum_{i=1}^{n} \l[\chi(p_i(\b{Y})) - f_{d}(p_i(\b{Y}))\r]}_{\Gamma_1} + \underbrace{\frac{1}{n}\sum_{i=1}^{n} \l[f_{d}(p_i(\b{Y})) - f_{d}(p_i(\b{X}))\r]}_{\Gamma_2}\\
        &+ \underbrace{\frac{1}{n}\sum_{i=1}^{n} \l[f_{d}(p_i(\b{X})) - \chi(p_i(\b{X}))\r]}_{\Gamma_3}.
    \end{align*}
    Using (\ref{chapter2:lemma:state_ev.3}), we have
    \begin{align*}
        |\Gamma_1| &\leq \frac{1}{n} \sum_{i=1}^{n} \frac{c}{\log d} (1 + |p_i(\b{Y})|^{k-1})\\
        &\leq \frac{c}{n\log d} \sum_{i=1}^{n} (2 + (p_i(\b{Y}))^{2k -2}) \numberthis \label{chapter2:lemma:state_ev.5}\\
        &= \frac{2c}{\log d} + \frac{c}{n\log d} \sum_{i=1}^{n} (p_i(\b{Y}))^{2k-2},\numberthis \label{chapter2:lemma:state_ev.6}
    \end{align*}
    where we used the arithmetic-geometric mean inequality in (\ref{chapter2:lemma:state_ev.5}).
    Note that since $p_i\in \mathcal{T}_{n, 2, \Delta}$, we have that $(p_i)^{2k-2}$ is a connected LDP containing node $i$ (i.e., $(p_i)^{2k-2} \in \V_i$), and with degree bounded as ${\rm Deg}\l((p_i)^{2k-2}\r) \leq (2k-2)\Delta$. Moreover, we have from Lemma \ref{chapter2:lemma:pre:1} that $\l\|(p_i)^{2k-2} \r\| \leq 2^{(2k-2)^2\Delta} \|p_i\|^{2k-2} \leq 2^{(2k-2)^2\Delta}C^{2k-2}$. Henceforth, using Lemma \ref{chapter2:lemma:+1} on $(p_i)^{2k-2}$ with node set $K=\{i\}$, it follows that there exists a constant $\gamma=\gamma(\Delta, k, C)$ depending on $\Delta$, $k$ and $C$ such that
    \begin{align*}
        \E\l[\l(p_i(\b{Y})\r)^{2k-2}\r] &\leq \gamma.
    \end{align*}
    Using the above and taking the expectation in (\ref{chapter2:lemma:state_ev.6}) yields
    \begin{align*}
        \E\l[|\Gamma_1|\r] &\leq \frac{(2+\gamma)c}{\log d} = O\l(\frac{1}{\log d}\r). 
    \end{align*}
    Thus, we have by Markov's inequality
    \begin{align*}
        \p\l( |\Gamma_1| \geq \frac{1}{\sqrt{\log d}}\r) \leq O\l(\frac{1}{\sqrt{\log d}}\r) = o(1),
    \end{align*}
    Henceforth
    \begin{align*}
        \plimnd \Gamma_1 = 0.\numberthis \label{chapter2:lemma:state_ev.7}
    \end{align*}
    Using the same arguments above together with $\E\l[(p_i(\b{X}))^{2k-2}\r]\leq \gamma$ from Lemma \ref{chapter2:lemma:+1}, it also holds
    \begin{align*}
        \plimnd \Gamma_3 = 0, \numberthis \label{chapter2:lemma:state_ev.8}
    \end{align*}
    In particular, combining (\ref{chapter2:lemma:state_ev.1}) and (\ref{chapter2:lemma:state_ev.8}), we have
    \begin{align*}
        \plimnd \frac{1}{n}\sum_{i=1}^{n} f_{d}(p_i(\b{X}))  &= \kappa. \numberthis \label{chapter2:lemma:state_ev.9}
    \end{align*}
    We now deal with $\Gamma_2$. Note that from (\ref{chapter2:lemma:state_ev.3}) and (\ref{chapter2:lemma:state_ev.4}), it follows that the sequence of functions $f_{n, d} = f_{d}$ satisfies the conditions of Proposition \ref{chapter2:lemma:state_ev_smooth}. Hence, applying the previous lemma and using the limit (\ref{chapter2:lemma:state_ev.9}), it follows that
   \begin{align*}
        \plimnd \frac{1}{n}\sum_{i=1}^{n} f_{d}(p_i(\b{Y}))  &= \kappa, \numberthis \label{chapter2:lemma:state_ev.10}
    \end{align*}
    and thus $\plimnd \Gamma_2 = 0$. Combining the latter with (\ref{chapter2:lemma:state_ev.8}) and (\ref{chapter2:lemma:state_ev.7}),  we have
    \begin{align*}
        \plimnd \sum_{i=1}^{n} \chi(p_i(\b{Y})) &= \plimnd \l( \sum_{i=1}^{n} \chi(p_i(\b{Y})) +  \Gamma_1 + \Gamma_2 + \Gamma_3 \r)\\
        &= \kappa,
    \end{align*}
    which shows (\ref{chapter2:lemma:state_ev.2}) and ends the proof of the implication (\ref{chapter2:lemma:state_ev.1}) $\implies$ (\ref{chapter2:lemma:state_ev.2}). This concludes the proof of the Theorem.
\end{proof}

\section{Proof of Theorem~\ref{chapter2:thm:main-formal}}\label{chapter2:section:6}
%\subsection{Performance of $\A_{\rm LD}^{\rm Tree}$ on Sparse Disorder}

In this section, we complete the proof of our second main result, Theorem~\ref{chapter2:thm:main-formal}.
Fix a sequence of polynomials $p=(p_i,\ i\in [n])$ satisfying the assumptions of the theorem. 
We first show \ref{chapter2:thm:main-formal.a}.  Specifically, we will show $\ref{chapter2:thm:main-formal.1} \implies  \ref{chapter2:thm:main-formal.2}$ through interpolation arguments from the Gaussian disorder $\X$ to the disorder $\Y$. The exact same proof of the latter can be used verbatim to show $\ref{chapter2:thm:main-formal.2}\implies \ref{chapter2:thm:main-formal.1}$  by reversing the interpolation order (from the disorder $\Y$ to the Gaussian disorder $\X$). Suppose \ref{chapter2:thm:main-formal.1} holds and let $d\leq t \leq \tau_n $.
\subsection{Proof of $(\ref{chapter2:eq:P-limit}) \implies (\ref{chapter2:eq:P-limit-Y.1})$}\label{chapter2:section:proof.P-limit}
\begin{proof}%[Proof of~(\ref{chapter2:eq:P-limit-Y})]
  For $1\leq i<j \leq n$, introduce the LDP 
$$
r_{ij}(\X) \triangleq p_i(\X) \X_{ij}  p_j(\X).
$$ Since $p_i,\ p_j$ contain nodes $i,\ j$ respectively (i.e., $p_i\in\V_i $ and $p_j\in \V_j$), it  follows that $r_{ij}$ is connected and contains both nodes $i$ and $j$ (i.e., $r_{ij} \in \V_{\{i, j\}}$). Indeed, the factor graphs of the monomials in $r_{ij}$ are the union of a tree rooted at $i$, a tree rooted at $j$, and the edge-tree $[ij]$. Furthermore, using Lemma \ref{chapter2:lemma:pre:1}, it follows that
\begin{align*}
    \|r_{ij}\|_{\rm coef} \leq 2^{2\Delta+1} \|p_i\|_{\rm coef} \|p_j\|_{\rm coef}\leq 2^{2\Delta + 1} c^2,
\end{align*}
where the last part follows from the assumption $\|p_i\|_{\rm coef}\leq c$.
Thus $r_{ij}$ is a connected LDP with constant coefficients. 
In fact the following stronger property holds, as an implication of which the contribution
of the non-tree
parts of $r_{ij}$ is asymptotically negligible.
\begin{lemma} \label{chapter2:lemma:5.8}
The following holds
\begin{align}
    \l\|  \sum_{1\leq i<j\leq n} r_{ij} \r\|_{\rm coef} = O(1) \label{chapter2:lemma.5.8.5}.
\end{align}
Furthermore, for every $\varepsilon>0$ 
   \begin{align}
       \frac{1}{n} \sum_{1\leq i<j \leq n}  r^{\rm Tr}_{ij}(\X) \geq \mathcal{P}
       -\varepsilon, \label{chapter2:lemma.5.8.1}
   \end{align} 
   w.h.p. as $n\to\infty$.
\end{lemma}
\begin{proof}[Proof of Lemma \ref{chapter2:lemma:5.8}]
 Let $q\triangleq \sum_{1\leq i<j\leq n} (r_{ij} -  r^{\rm Tr}_{ij})$. Since $ q^{\rm Tr}=0$ and $q$ is connected, it follows that $(q^2)^{\rm Tr}=0$ by Lemma \ref{chapter2:lemma:pre:3}. Applying (\ref{chapter2:lemma.4.1.1}) from Lemma \ref{chapter2:lemma:pre:6} to $q^2$ with $S\triangleq \emptyset$, it follows that
\begin{align}
    \E \l[ q^2(\X) \r]  &= \E \l[ q^2(\X) - (q^2)^{\rm Tr}(\X)\r] =O(\|q^2\|_{\rm coef}) \label{chapter2:lemma.5.8.2}.
\end{align}
Since the degree of each $ p_i$ is at most $\Delta$, it follows that the degree of $r_{ij}$ is at most $2\Delta+1$, and hence the degree of $q$ is at most $2\Delta+1$. Using Lemma \ref{chapter2:lemma:pre:1}, it follows
\begin{align}
    \|q^2\|_{\rm coef} &\leq 2^{4\Delta +2} \|q\|_{\rm coef}^2. \label{chapter2:lemma.5.8.3}
\end{align}
By definition of $q$
\begin{align}
    \|q\|_{\rm coef} &\leq \left \| \sum_{1\leq i<j\leq n} r_{ij}\right\|_{\rm coef} +  \left\| \sum_{1\leq i<j\leq n}  r^{\rm Tr}_{ij} \right\|_{\rm coef} \leq 2 \left\| \sum_{1\leq i<j\leq n} r_{ij}\right\|_{\rm coef}. \label{chapter2:lemma.5.8.4}
\end{align}
Using Lemma \ref{chapter2:lemma:pre:2} with sets $S_{i,j}\triangleq \{i,j\}$ for $1\leq i<j\leq n$ and $\gamma=\frac{1}{2}, c=\sqrt{2}$ (any set of $s$ edges touches at least $\sqrt{2s}$ nodes), it follows that
\begin{align}
    \l\|  \sum_{1\leq i<j\leq n} r_{ij} \r\|_{\rm coef} &\leq \l(\frac{1+\Delta}{\sqrt{2}}\r)^{2} \max_{1\leq i<j \leq n}   \|r_{ij}\|_{\rm coef} \leq \l(\frac{1+\Delta}{\sqrt{2}}\r)^{2} 2^{2\Delta +1}c^2= O(1) ,
\end{align}
and (\ref{chapter2:lemma.5.8.5}) is established.
Combining (\ref{chapter2:lemma.5.8.5}) and (\ref{chapter2:lemma.5.8.4}) in (\ref{chapter2:lemma.5.8.3}) yields $\l\|q^2\r\|_{\rm coef} =O(1)$. Plugging the latter in (\ref{chapter2:lemma.5.8.2})
\begin{align*}
    \E \l[q^2(\X) \r]  = O(1).
\end{align*}
By Chebyshev's inequality it follows that
\begin{align*}
    \forall \varepsilon'>0,\ \p\l( \l| \sum_{1\leq i<j\leq n}  \l[ r_{ij}(\X) -   r^{\rm Tr}_{ij}(\X) \r] \r| \geq n\varepsilon'\r) &= \p\l(q^2(\X) \geq n^2\varepsilon'^2\r) = \frac{O(1)}{n^2\varepsilon'^2}.
\end{align*}
The result of (\ref{chapter2:lemma.5.8.1}) follows then from combining the above with (\ref{chapter2:eq:P-limit}). This ends the proof of the lemma.
\end{proof}
In light of the above Lemma, we next show that property (\ref{chapter2:lemma.5.8.1}) extends to the sparse disorder $\Y \distrib \G(n, t/n)$. 

\begin{proposition}\label{chapter2:lemma:5.9}
For every $\varepsilon>0$, it holds w.h.p. as $n\to\infty$ followed  by $d\to\infty$
\begin{align}
    \frac{1}{n} \sum_{1\leq i<j \leq n}  r^{\rm Tr}_{ij}(\Y) \geq \mathcal{P}-\varepsilon \label{chapter2:lemma.5.9.1.d}.
    \end{align}    
\end{proposition}
\begin{proof}[Proof of Proposition~\ref{chapter2:lemma:5.9}]
Fix $c, \beta >0$ and recall $\phi,\  \psi_{\beta,c}$ from Lemma \ref{chapter2:lemma:pre:4}. For convenience we redefine $\psi_{\beta,c}$ as
\begin{align*}
    \psi_{\beta,c}(z)=\phi\l(\beta \l(\frac{z}{n} - c\r)\r).
\end{align*}
 Consider an interpolation path from $\X$ to $\Y$ given by $\Z^0,\ldots,\Z^N$ where $\Z^0=\X,\ \Z^N=\Y$ and $N\triangleq \binom{n}{2}$.  For $\ell\in[N]$ we will identify $\ell$ with its corresponding (enumerated) edge $(ij)$, and thus write $\X_\ell,\Y_\ell$ instead of $\X_{ij}, \Y_{ij}$. We have
\begin{align*}
    &\E\left[\psi_{\beta, c}\left(\sum_{1\leq i<j\leq n}  r^{\rm Tr}_{ij}(\Y)\right)\right] - \E\left[\psi_{\beta, c}\left(\sum_{1\leq i<j\leq n}  r^{\rm Tr}_{ij}(\X)\right)\right]\\
    &= \sum_{\ell=0}^{N-1} \E\l[\psi_{\beta, c}\left(\sum_{1\leq i<j\leq n}  r^{\rm Tr}_{ij}(\Z^{\ell+1})\right)\r] - \E\l[\psi_{\beta, c}\left(\sum_{1\leq i<j \leq n}  r^{\rm Tr}_{ij}(\Z^{\ell})\right) \r].
\end{align*}
By definition of  $\mathcal{T}_{n, 2}$,  
the monomials $\Z^{\pmb{\alpha}}$ in $\sum_{1\leq i<j \leq n}  r^{\rm Tr}_{ij}(\Z)$ satisfy $\max(\pmb{\alpha}) \leq 2$. In particular, we can write for $\ell\leq N-1$
\begin{align}
    \sum_{1\leq i<j\leq n}  r^{\rm Tr}_{ij}(\Z^\ell) &= u_\ell(\Z^\ell) + \X_{\ell} a_\ell(\Z^{\ell}) + \X_{\ell}^2 b_\ell(\Z^\ell) \label{chapter2:lemma.5.9.2},
\end{align}
where $u_\ell(\x),\ a_\ell(\x),\ b_\ell(\x)$ are LDPs in the variables $\x=(\x_{ij}, 1\leq i<j \leq n)$, and do not depend on the variable associated with
edge $\ell$, i.e., the LDPs do not depend on $\x_\ell$. Furthermore, since we switch the distribution of $[\Z^\ell]_{\ell}$ from $\X_\ell$ to $\Y_\ell$ to obtain $\Z^{\ell+1}$ (and keep everything else unchanged), it follows that
\begin{align}
    \sum_{1\leq i<j\leq n}  r^{\rm Tr}_{ij}(\Z^{\ell+1}) &= u_\ell(\Z^\ell) + \Y_{\ell} a_\ell(\Z^{\ell}) + \Y_{\ell}^2 b_\ell(\Z^\ell) \label{chapter2:lemma.5.9.3}.
\end{align}
Three important elementary properties of $u_\ell,a_\ell,b_\ell$ are
\begin{align}
    &\forall \ell\in [0,N-1],\ u_\ell(\x),\  \x_\ell a_\ell(\x),\ \x_\ell b_\ell(\x) \text{ are connected}, 
    \label{chapter2:lemma.5.9.uab:1}\\
    & \forall \ell\in [0,N-1],\  \x_\ell a_\ell(\x),\ \x_\ell b_\ell(\x) \text{ contain the endpoints of edge } \ell,\label{chapter2:lemma.5.9.uab:1.5} \\
    &\forall \ell\in [0,N-1],\   \|u_\ell\|_{\rm coef} + \|a_\ell\|_{\rm coef} + \|b_\ell\|_{\rm coef} \leq c',\label{chapter2:lemma.5.9.uab:2}
\end{align}
where $c'$ is some constant independent of $n$ and $d$. The connectedness of $u_\ell(\x),\ \x_\ell a_\ell(\x)$, $\ \x_\ell b_\ell(\x)$ in (\ref{chapter2:lemma.5.9.uab:1}) follows from first noting  that $u_{\ell}(\x),\ \x_\ell^2 a_\ell(\x),\ \x_\ell^2 b_\ell(\x)$ are connected (as $\sum  r^{\rm Tr}_{ij}$ is connected) and using the fact that $\x_\ell w (\x) $ is connected if and only if $(\x_\ell)^k w(\x)$ is connected for all $k\geq 1$, where $w$ is any LDP. The second property (\ref{chapter2:lemma.5.9.uab:1.5}) means that $ \b{x}_\ell a_{\ell}(\b{x}),\ \b{x}_\ell b_{\ell}(\b{x})\in \V_{\{i^*, j^*\}}$, where $(i^*, j^*)$ is the edge enumerated by $\ell$, and follows from the fact that these LDPs are connected and  $\{i^*, j^*\}\subseteq V(\b{x}_{\ell})$. Finally, property (\ref{chapter2:lemma.5.9.uab:2}) follows immediately from (\ref{chapter2:lemma.5.8.5}).

To simplify notation, we will write $u_\ell,\ a_\ell,\ b_\ell$ instead of $u_\ell(\Z^\ell),\  a_\ell(\Z^\ell),\ b_\ell(\Z^\ell)$ (i.e., we write $u_\ell$ to denote the value of the polynomial $u_\ell$ at $\Z^\ell$). Introduce for $\ell\leq N-1$
\begin{align*}
v_\ell(z)&=u_\ell + z a_\ell + z^2 b_\ell, \\
    \xi_{\beta, c}(z) &= \psi_{\beta,c}(v_\ell(z)).
\end{align*}
It follows from (\ref{chapter2:lemma.5.9.2}), (\ref{chapter2:lemma.5.9.3}) that
\begin{align*}
    &\sum_{\ell=0}^{N-1} \E\left[ \psi_{\beta, c}\left(\sum_{1\leq i<j\leq n}  r^{\rm Tr}_{ij}(\Z^{\ell+1})\right)\r] - \E\l[\psi_{\beta, c}\left(\sum_{1\leq i<j \leq n}  r^{\rm Tr}_{ij}(\Z^{\ell})\right) \right]\\
    &= \sum_{\ell=0}^{N-1} \E\l[ \xi_{\beta, c}(\Y_\ell)\r] -\E\l[\xi_{\beta, c}(\X_\ell) \r] \numberthis \label{chapter2:interpol:1}.
\end{align*}
We will drop the subscripts in $\xi_{\beta, c},\ \psi_{\beta,c}$ and instead simply write $\xi,\ \psi$ respectively. Using a third-order Taylor expansion, there exists $\X_\ell^* \in \l[- |\X_\ell|, |\X_\ell|\r] $ and $\Y_\ell^* \in \l[-|\Y_\ell|, |\Y_\ell|\r]$ such that
    \begin{align*}
        \xi(\X_\ell) &= \xi(0) + \X_\ell \xi^{(1)}(0)  + \frac{\X_\ell^2}{2} \xi^{(2)}(0)  + \frac{\X_\ell^3}{6} \xi^{(3)}(\X_\ell^*),\\
        \xi(\Y_\ell) &= \xi(0) + \Y_\ell \xi^{(1)}(0) + \frac{\Y_\ell^2}{2} \xi^{(2)}(0)  + \frac{\Y_\ell^3}{6} \xi^{(3)}(\Y_\ell^*).
    \end{align*}
Taking the difference of the above and the expectation yields
\begin{align*}
    \E\l[\xi(\Y_\ell)] - \E[\xi(\X_\ell)\r] = \frac{\E\l[\Y_\ell^3 \xi^{(3)}(\Y_\ell^*)\r] - \E\l[\X_\ell^3 \xi^{(3)}(\X_\ell^*)\r]}{6}. \numberthis \label{chapter2:taylor:diff:1}
\end{align*}
We set $\beta = \log n$ in the definition of $\psi$ for the remaining parts of the proof of Proposition \ref{chapter2:lemma:5.9}. Using  Lemma \ref{chapter2:lemma:5.10} in (\ref{chapter2:taylor:diff:1}), it follows that
\begin{align*}
    |\E[\xi(\Y_\ell)] - \E[\xi(\X_\ell)]| &\leq \gamma (\log n)^2 n^{-3}t^{-\frac{1}{2}},
\end{align*}
where $\gamma$ is a constant independent of $n, d$ and $\ell$. Plugging the above in (\ref{chapter2:interpol:1}) and summing over $\ell\leq N-1\sim n^2/2$ yields
\begin{align*}
    \l| \sum_{\ell=0}^{N-1} \E[\xi(\Y_\ell)] - \E[\xi(\X_\ell)]\r| &\leq  \gamma (\log n)^2 n^{-1}t^{-\frac{1}{2}},
\end{align*}
which  implies
\begin{align}
    \l| \E\left[\psi_{\beta, c}\left(\sum_{1\leq i<j\leq n}  r^{\rm Tr}_{ij}(\Y)\right)\right] - \E\left[\psi_{\beta, c}\left(\sum_{1\leq i<j\leq n} r^{\rm Tr}_{ij}(\X)\right)\right] \r| &=o(1)\label{chapter2:lemma.5.9.4}.
\end{align}
We now show  $\p\l( \frac{1}{n} \sum_{1\leq i<j \leq n} r_{ij}^{\rm Tr}(\Y) \geq \mathcal{P} -\varepsilon \r)= 1-o(1)$. To ease notation, introduce
\begin{align*}
    h :\quad \Z \mapsto \sum_{1\leq i<j \leq n} r^{\rm Tr}_{ij}(\Z). 
\end{align*}
Let $\varepsilon' = \frac{\varepsilon}{4}$ and set $c=\mathcal{P} - \frac{\varepsilon}{2}$. Since $\psi_{\beta, c}$ is a nonnegative increasing function, we have
\begin{align*}
    \E\l[\psi_{\beta,c}(h(\X))\r] &\geq  \E\l[\psi_{\beta, c}(h(\X)) \b{1}_{h(\X) \geq n(\mathcal{P} - \varepsilon') }\r]\\
    &\geq \psi_{\beta, c}\l(n(\mathcal{P} - \varepsilon') \r) \cdot \p\l( h(\X) \geq n(\mathcal{P} - \varepsilon') \r)\\
    &=\phi\l( \frac{\beta \varepsilon}{4} \r) \cdot \p\l( h(\X) \geq  n(\mathcal{P} - \varepsilon') \r)\\
    &= (1-o(1)) \cdot \p\l( h(\X) \geq n(\mathcal{P} - \varepsilon')\r)\numberthis\label{chapter2:lemma.5.9.5}\\
    &= 1-o(1) \numberthis\label{chapter2:lemma.5.9.6},
\end{align*}
where (\ref{chapter2:lemma.5.9.5}) follows from $\beta \varepsilon=\omega(1)$ due to $\beta=\log n$, and (\ref{chapter2:lemma.5.9.6}) follows from (\ref{chapter2:lemma.5.8.1}) in Lemma \ref{chapter2:lemma:5.8}. Since $\psi_{\beta,c}\leq 1$, then (\ref{chapter2:lemma.5.9.6}) implies
\begin{align*}
    \E\l[\psi_{\beta,c}(h(\X))\r] = 1 -o(1).
\end{align*}
Combining the above with (\ref{chapter2:lemma.5.9.4}), it follows that
\begin{align}
    \E\l[\psi_{\beta,c}(h(\Y))\r] = 1 -o(1) \label{chapter2:lemma.5.9.7}.
\end{align}
Since $\psi_{\beta, c}\leq 1$ and $\psi_{\beta, c}$ is increasing, we have
\begin{align*}
\E\l[\psi_{\beta,c}(h(\Y))\r] &=     \E\l[\psi_{\beta,c}(h(\Y)) \b{1}_{h(\Y) < n(\mathcal{P}-\varepsilon)}\r] +  \E\l[\psi_{\beta,c}(h(\Y)) \b{1}_{h(\Y) \geq n(\mathcal{P}-\varepsilon)}\r]\\
&\leq \psi_{\beta,c}\l(n(\mathcal{P}-\varepsilon) \r) + \p\l(h(\Y) \geq n(\mathcal{P}-\varepsilon) \r)\\
&= \phi\l(- \frac{\beta \varepsilon}{2}\r) + \p\l(h(\Y) \geq n(\mathcal{P}-\varepsilon) \r)\\
&= o(1) + \p\l(h(\Y) \geq n(\mathcal{P}-\varepsilon) \r) \numberthis \label{chapter2:lemma.5.9.8},
\end{align*}
where the last line follows from $\beta\varepsilon = \omega(1)$. Using (\ref{chapter2:lemma.5.9.8}) and (\ref{chapter2:lemma.5.9.7}), it follows that
\begin{align}
    &\p\l(h(\Y) \geq n(\mathcal{P}-\varepsilon)\r) = 1-o(1),
\end{align}
namely
\begin{align}
    \p\l(\frac{1}{n} \sum_{1\leq i<j \leq n}  r^{\rm Tr}_{ij}(\Y) \geq \mathcal{P}-\varepsilon\r) = 1 -o(1), \label{chapter2:lemma.5.9.9} 
\end{align}
which concludes the proof of Proposition \ref{chapter2:lemma:5.9}.
\end{proof}
To conclude the proof of (\ref{chapter2:eq:P-limit-Y.1}), let $q\triangleq \sum(r_{ij} -  r^{\rm Tr}_{ij})$ as in the proof of Lemma \ref{chapter2:lemma:5.8} and recall that we showed $(q^2)^{\rm Tr}=0$ and $\|q^2\|_{\rm coef}= O(1)$. Applying  (\ref{chapter2:lemma.4.1.2}) from Lemma \ref{chapter2:lemma:pre:6} to $q^2$ with $S\triangleq \emptyset$, it follows that
\begin{align}
    \l|\E \l[ q^2(\Y) \r] \r| &=  \l|\E \l[ q^2(\Y) - (q^2)^{\rm Tr}(\Y)\r] \r| =  O( n^2 t^{-1}).
\end{align}
Let $\varepsilon>0$. Using the above, it follows by Chebyshev's inequality
\begin{align*}
    \p\l(\l| \sum_{1\leq i<j\leq n} \l[ r_{ij}(\Y)- r^{\rm Tr}_{ij}(\Y)\r]\r| \geq n \varepsilon/2\r) &= \p(q^2(\Y)\geq n^2 \varepsilon^2/4) \leq \frac{O(1)}{ t \varepsilon^2} = o(1).
\end{align*}
Moreover, we have from Proposition \ref{chapter2:lemma:5.9} that $\frac{1}{n} \sum_{1\leq i<j \leq n}  r_{ij}^{\rm Tr}(\Y) \geq \mathcal{P}-\varepsilon/2$ w.h.p. Therefore
\begin{align*}
    \p\l(\frac{1}{n} \sum_{1\leq i<j \leq n}  r_{ij}(\Y) \geq \mathcal{P} - \varepsilon\r) = 1 -o(1),
\end{align*}
which completes the proof of (\ref{chapter2:eq:P-limit-Y.1}) as 
$\frac{1}{2n} p(\Y)^\top \Y p(\Y)  = \frac{1}{n} \sum_{1\leq i<j \leq n}  r_{ij}(\Y)$.

\end{proof}

\subsection{Proof of $(\ref{chapter2:eq:box-limit}) \implies  (\ref{chapter2:eq:box-limit-Y.1})$}

\begin{proof}%[Proof of (\ref{chapter2:eq:box-limit-Y})]
Let $\chi :\R \to \R$ be given by $\chi(x) = d(x, [-1,1])^2 = \max(0, |x|-1)^2$. For $x, y\in \R$, we have
\begin{align*}
    |\chi(x) - \chi(y)| &= |d(x, [-1, 1]) - d(y, [-1,1])|\cdot \l|d(x, [-1, 1]) + d(y, [-1,1])\r|\\
    &\leq |x-y| (1 + |x| + |y|),
\end{align*}
where we used the fact that $x\mapsto d(x, [-1,1])$ is $1$-Lipschitz, and $|d(x, [-1,1])| \leq |x|$ in the last line. Therefore, $\chi$ is pseudo-Lipschitz of order $k=2$. Moreover, we can rewrite (\ref{chapter2:eq:box-limit})  as
\begin{align*}
    \plim \frac{1}{n} \sum_{i=1}^{n} \chi(p_i(\b{X})) = 0,
\end{align*}
using the above and applying Theorem \ref{chapter2:lemma:state_ev} with the test function $\chi$, it follows that
\begin{align*}
    \plimnd \frac{d(p(\Y), [-1, 1])^2}{n} = \plimnd \frac{1}{n} \sum_{i=1}^{n} \chi(p_i(\b{Y})) = 0,
\end{align*}
which readily shows  (\ref{chapter2:eq:box-limit-Y.1}) and concludes the proof of $(\ref{chapter2:eq:box-limit}) \implies (\ref{chapter2:eq:box-limit-Y.1})$.
\end{proof}
This ends the proof of \ref{chapter2:thm:main-formal.a} in Theorem \ref{chapter2:thm:main-formal}.

\subsection{Proof of \ref{chapter2:thm:main-formal.b}}\label{chapter2:section:rounding}

In this subsection, we extend a simple rounding procedure from \cite{montanari2019optimization} to the LDP output $p(\Y)$ in order to obtain a point $\pmb{\sigma}(p(\Y), \Y)\in \{-1,1\}^n$. On the model $\G(n, \tau_n/n)$ with $\tau_n \gg \log n$, Lemma~\ref{chapter2:lemma:random:matrix:sparse} ensures that $\|\Y\| = O(1)$, and  the analysis of the rounding scheme is relatively direct in this case. However, when $\tau_n$ does not satisfy $\tau_n \gg \log n$, and in particular in the sparse regime $\G(n, d/n)$, the operator norm $\|\Y\|$ grows with $n$, which requires a substantially more delicate argument to establish optimality for the rounded solution.

To handle these sparser regimes, we use a regularization argument. First, we remove the edges incident to a small set of problematic vertices with  high degrees, as these vertices are responsible for the divergence of $\|\Y\|$. Second, we establish near optimality of the LDP output $p(\Y)$ on the regularized graph disorder, relying crucially on delocalization properties of LDP outputs. Third, we prove near optimality of the projection of the LDP solution into $[-1,1]^n$ on the regularized disorder. Fourth, we show that the Hamiltonian evaluated at this projected solution with respect to $\Y$ versus the regularized disorder differs only by a negligible error. Finally, we sequentially round any coordinate of the projected LDP output lying in $(-1,1)$ to $\pm 1$ without decreasing the objective. Combining the errors from all these steps yields the desired optimality bound on $H(\pmb{\sigma}(p(\Y), \Y), \Y)$. We present the proof of \ref{chapter2:thm:main-formal.b} below.

    \begin{proof}[Proof of {\normalfont \ref{chapter2:thm:main-formal.b}}]
    Suppose \ref{chapter2:thm:main-formal.2} holds and let $\Y \distrib \G(n, t/n)$, with $d\leq t\leq \tau_n$ where $\tau_n$ satisfies $\omega(1)\leq \tau_n \leq o(n)$.    Recall that $\pmb{\sigma}(p(\Y), \Y)$ is computed by first projecting the terms $p_i(\Y)$ that lie outside $[-1,1]$ into $[-1,1]$, then sequentially setting the terms $p_i(\Y)$ that lie in $(-1,1)$ to $\pm 1$ while increasing the   objective value. Note in particular that $\pmb{\sigma}(p(\Y), \Y)$ can be computed in polynomial time complexity from $p(\Y)$. Let $\pi(\Y)$ be the projection of $p(\Y)$ onto  $[-1, 1]^n$. In the setting of Lemma \ref{chapter2:lemma:3}, let $\Y^{\rm prune}$ be the regularized matrix obtained after removing edges incident to vertices with degree above $4t$. It follows from Lemma \ref{chapter2:lemma:3} that there exist positive constants $c_2$ and  $c_1$ such that
    \begin{align*}
     \E\l[\l| p(\Y)^\top \Y p(\Y) -  p(\Y)^\top \Y^{\rm prune} p(\Y) \r| \r] &\leq c_2 \max_{i\in [n]} \|p_i^2\|_{\rm coef} n t^{2\Delta + 1} e^{-c_1 t}, \numberthis \label{chapter2:rounding.proof.1} \\
     \E\l[\l| \pi(\Y)^\top \Y \pi(\Y) -  \pi(\Y)^\top \Y^{\rm prune} \pi(\Y) \r| \r] &\leq c_2 nt e^{-c_1 t}. \numberthis \label{chapter2:rounding.proof.2}
\end{align*}
    Using Lemma \ref{chapter2:lemma:pre:1}, we have $\max_{i\in [n]} \|p_i^2\|_{\rm coef} \leq 2^{2\Delta} \max_{i\in [n]} \|p_i\|^2_{\rm coef} = O(1)$. Therefore, we have using (\ref{chapter2:rounding.proof.1}) and Markov's inequality
    \begin{align*}
        \p\l(\l| p(\Y)^\top \Y p(\Y) -  p(\Y)^\top \Y^{\rm prune} p(\Y) \r| \geq \frac{2n}{t}\r)\leq  O\l(t^{2\Delta + 2}e^{-c_1 t}\r).
    \end{align*}
    Hence, the following holds w.h.p. as $n\to \infty$ followed by $d\to\infty$
    \begin{align*}
          \frac{1}{2n}  p(\Y)^\top \Y^{\rm prune} p(\Y) \geq \frac{1}{2n}  p(\Y)^\top \Y p(\Y) - \frac{1}{t}. \numberthis \label{chapter2:rounding.proof.3}
    \end{align*}
    Fix $0 < \varepsilon'< 1$. From  (\ref{chapter2:eq:box-limit-Y.1}) it follows that $\|p(\Y) - \pi(\Y)\|_2 \leq \varepsilon' \sqrt{n}$ w.h.p. as $n\to\infty$ followed by $d\to \infty$. We then have
    \begin{align*}
        &\frac{1}{2n}| p(\Y) ^\top \Y^{\rm prune} p(\Y)  - \pi(\Y)^\top \Y^{\rm prune} \pi(\Y) | \\
        &\leq \frac{1}{2n} |(p(\Y) -\pi(\Y))^\top \Y^{\rm prune} (p(\Y)  - \pi(\Y))| + \frac{1}{n} |\pi(\Y)^\top \Y^{\rm prune} (p(\Y)  - \pi(\Y))|\\
        &\leq \frac{1}{2n} \|\Y^{\rm prune}\| \cdot \|p(\Y)  - \pi(\Y)\|_2^2 + \frac{1}{n} \|\Y^{\rm prune}\| \cdot \|p(\Y)  - \pi(\Y)\|_2 \cdot \|\pi(\Y)\|_2. \numberthis \label{chapter2:rounding.proof.4}
    \end{align*}
    Using Lemma \ref{chapter2:lemma:pruned_norm}, there exists a constant $\theta$ such that $\|\Y^{\rm prune}\|\leq \theta$ w.h.p. as $n\to \infty$ followed by $d\to\infty$. Combining the latter operator bound with (\ref{chapter2:rounding.proof.4}), we have w.h.p. as $n\to \infty$ followed by $d\to\infty$
    \begin{align*}
        \frac{1}{2n}| p(\Y) ^\top \Y^{\rm prune} p(\Y)  - \pi(\Y)^\top \Y^{\rm prune} \pi(\Y) | &\leq \frac{\theta}{2n} \|p(\Y)  - \pi(\Y)\|_2^2 \\
        &+ \frac{\theta}{n}  \|p(\Y)  - \pi(\Y)\|_2 \|\pi(\Y)\|_2\\
        &\leq \frac{\theta\varepsilon'^2}{2} + \theta\varepsilon'\\
        &\leq 2 \theta \varepsilon', \numberthis \label{chapter2:rounding.proof.5}
    \end{align*}
    where we used $\|\pi(\Y)\|_2\leq\sqrt{n}$ and  $\|p(\Y) - \pi(\Y)\|_2 \leq \varepsilon' \sqrt{n}$. Using (\ref{chapter2:rounding.proof.2}) and Markov's inequality, it follows that
    \begin{align*}
        \p\l( \l| \pi(\Y)^\top \Y \pi(\Y) -  \pi(\Y)^\top \Y^{\rm prune} \pi(\Y) \r| \geq   \frac{2n}{t} \r) \leq O(t^2 e^{-c_1 t}).
    \end{align*}
    Therefore, the following holds w.h.p. as $n\to \infty$ followed by $d\to\infty$
    \begin{align*}
        \frac{1}{2n}  \pi(\Y)^\top \Y \pi(\Y) \geq \frac{1}{2n}  \pi(\Y)^\top \Y^{\rm prune} \pi(\Y) - \frac{1}{t}. \numberthis \label{chapter2:rounding.proof.6}
    \end{align*}
    Henceforth, we have w.h.p. as $n\to \infty$ followed by $d\to\infty$
    \begin{align*}
        \frac{1}{2n}  \pi(\Y)^\top \Y \pi(\Y) &\geq \frac{1}{2n}  \pi(\Y)^\top \Y^{\rm prune} \pi(\Y) - \frac{1}{t}\\
        &\geq \frac{1}{2n}  p(\Y)^\top \Y^{\rm prune} p(\Y) - 2\theta \varepsilon' - \frac{1}{t} \numberthis \label{chapter2:rounding.proof.7}\\
        &\geq \frac{1}{2n}  p(\Y) ^\top \Y p(\Y) - 2\theta \varepsilon' - \frac{2}{t} \numberthis  \label{chapter2:rounding.proof.8}\\
        &\geq \mathcal{P} - \varepsilon' - 2\theta \varepsilon' - \frac{2}{t}, \numberthis \label{chapter2:rounding.proof.9}
    \end{align*}
    where we used (\ref{chapter2:rounding.proof.5}) in line (\ref{chapter2:rounding.proof.7}), (\ref{chapter2:rounding.proof.3}) in line (\ref{chapter2:rounding.proof.8}), and (\ref{chapter2:eq:P-limit-Y.1}) in line (\ref{chapter2:rounding.proof.9}). Since the diagonal terms of $\Y$ are null, it follows that the restriction of $\pi(\Y) \mapsto \pi(\Y)^\top \Y \pi(\Y)$ to any single entry $\pi(\Y)_{i}, i\in [n] $ is linear. In particular, if $\pi(\Y)_{i} \in (-1,1)$, we can increase the objective value  $\pi(\Y)^\top \Y \pi(\Y)$ by  moving $\pi(\Y)_{i}$ either to $-1$ or $1$.  Apply this rounding scheme iteratively  for all entries inside $(-1, 1)$. As defined in Section~\ref{chapter2:section:notation}, we denote the obtained vector after all rounding steps by $\pmb{\sigma}(p(\Y), \Y) \in \{-1, 1\}^n$. Since each rounding step improves the objective, it follows that 
    \begin{align*}
         \frac{1}{2n}\pmb{\sigma}(p(\Y), \Y)^\top \Y \pmb{\sigma}(p(\Y), \Y)  &\geq \frac{1}{2n}  \pi(\Y)^\top \Y \pi(\Y) \\
         &\geq \mathcal{P} - \varepsilon' - 2\theta \varepsilon' - \frac{2}{t}\\
         &\geq \mathcal{P} - \varepsilon' - 2\theta \varepsilon' - \frac{2}{d},
    \end{align*}
    where we used (\ref{chapter2:rounding.proof.9}) and $d\leq t$ in the last line. Taking $\varepsilon'$ small enough such that $\varepsilon'+ 2 \theta \varepsilon' + \frac{2}{d} \leq \varepsilon$ yields the result of \ref{chapter2:thm:main-formal.d} and concludes the proof of  \ref{chapter2:thm:main-formal.b}.
\end{proof}

\section{{\rm IAMP} Representation and Proof of Theorem \ref{chapter2:thm:2-formal}}\label{chapter2:section:3}
The goal of this section is to prove Theorem~\ref{chapter2:thm:2-formal} by showing that the output of Incremental AMP can be approximated in $\ell_2$ norm by a vector of low-degree polynomials (LDPs) with the structural properties required in this paper, and achieving near-optimal objective value. While generic AMP--to--polynomial approximation results are available in the literature, they do not directly guarantee the specific form we need, namely uniformly bounded constant coefficients, and the connectivity and exponent constraints encoded by $\mathcal{T}_{n,2,\Delta}$.

The purpose of the following subsections is as follows. In Section~\ref{chapter2:section:iamp_poly}, we show that the IAMP denoisers can be replaced by polynomial functions under minimal regularity assumptions, and we establish that the resulting modified scheme remains near-optimal. We denote this intermediate algorithm by ${\rm IAMP}'$. In Section~\ref{chapter2:section:onsager_limit}, we further modify ${\rm IAMP}'$ by replacing its Onsager coefficients $b_{t,j}$, as defined in (\ref{chapter2:amp}), by their limits in probability as $n\to\infty$, leading to a second scheme denoted ${\rm IAMP}''$. This modification is crucial for enforcing the connectivity  property in Theorem~\ref{chapter2:thm:2-formal}. The algorithm ${\rm IAMP}''$ is no longer a standard AMP, and therefore classical state evolution results do not apply directly. To address this issue, we rework a state evolution argument that is sufficient for our purposes in Section~\ref{chapter2:subsection:5.4}, which is deferred to the end of this section. Finally, in Section~\ref{chapter2:section:proof:thm:2-formal}, we combine these ingredients to complete the proof of Theorem~\ref{chapter2:thm:2-formal}.

%The purpose of the following subsections is therefore twofold: first, to obtain a polynomial approximation of IAMP under minimal assumptions, and second, to modify and project this approximation so that it satisfies the precise structural constraints needed to apply the universality result of Section 2.

\subsection{Approximation of {\rm IAMP} by Polynomials}\label{chapter2:section:iamp_poly}
%It is well known in the literature that AMP algorithms can be approximated by Low-Degree Polynomials  modulo some regularity assumptions on the denoiser functions $f^t$~\cite{montanari2022equivalence,ivkov2024semidefinite}. However, this is mostly an informal statement, and technical work must be done on a per-problem basis to rigorously justify the approximation.  We begin by restating the state evolution property satisfied by the AMP iterates.
It is well known in the literature that AMP algorithms can be approximated by replacing their denoiser functions with polynomials, under suitable regularity assumptions on the denoisers $f^t$~\cite{montanari2022equivalence,ivkov2024semidefinite}.  However, this is mostly an informal statement, and technical work must be done on a per-problem basis to rigorously justify the approximation. The purpose of this section is to perform this replacement for Incremental AMP in our setting and to show that doing so preserves the relevant performance guarantees. We also note that LDP structure is not addressed at this stage.  We begin by restating the state evolution property satisfied by the AMP iterates.
\begin{proposition}[Proposition 2.1 in \cite{montanari2021optimization}]\label{chapter2:state:evolution}
Assume the denoisers $f^t$ are pseudo-Lipschitz of order $m$. Let $(U_j)_{j\geq 1}$ be a centered Gaussian process independent of $U_0$ with covariance $\b{Q} = (\b{Q}_{kj})_{k,j\geq 1}$ given  by (\ref{chapter2:eq:covariance-Q}).
Assume $\b{Q}_{\leq k} \triangleq (\b{Q}_{ij})_{1\leq i,j\leq k}$ is strictly positive definite for all $k \leq T$. Then for any $k,\ell \in \mathbb{Z}_{\geq 0}$, and any function $\psi: \mathbb{R}^{k+1}\to \mathbb{R}$ that is pseudo-Lipschitz of order $\ell$, we have
\begin{align*}
    \plim \frac{1}{n} \sum_{i=1}^{n} \psi(\u^0_i, \ldots, \u^k_i) = \E[ \psi(U_0,\ldots,U_k)].
\end{align*}
\end{proposition}

Next we restate a lemma from the appendix of \cite{ivkov2023semidefinite} (which is the preprint version of \cite{ivkov2024semidefinite}) showing that general AMP can be well approximated by LDP under certain regularity assumptions stated below.

\begin{lemma}[Lemma B.4 in \cite{ivkov2023semidefinite}]\label{chapter2:lemma:ivkov}
Fix $\eta>0$ and $T\in \mathbb{Z}_{\geq 0}$. Suppose that
\begin{itemize}
    \item $\{f^t:\mathbb{R}^{t+1} \to \mathbb{R}\}$ is a sequence of $L$-Lipschitz denoiser functions that produce IAMP iterates $\b{u}^0,\dots,\b{u}^T$.
    \item For each $i\in[n]$, $\frac{\partial f^t}{\partial \b{u}_i}$ is either pseudo-Lipschitz or an indicator.
    \item The covariance matrix $\b{Q}_{\leq t}$ for $t\in [T]$ satisfies $\b{Q}_{\leq t}\succeq \b{I}_{t}$ and $\| \b{Q}_{\leq t}\|_{\infty}\leq 2$.
\end{itemize}
Then, there exists a sequence of polynomial denoisers $\{q^t: \R^{t+1}\to \R\}$ producing IAMP iterates $\hat{\u}^0, \ldots, \hat{\u}^T$ such that
\begin{align*}
    \plim \frac{1}{\sqrt{n}} \|\u^t - \hat{\u}^t\|_2 \leq \eta,
\end{align*}
for all $t\leq T$. Furthermore, there exist constants $\Delta=\Delta(\eta, T, L), c=c(\eta, T, L)$, such that the polynomials $q^t$ have degree bounded by $\Delta$, and coefficients bounded in absolute value by $c$.
\end{lemma}
As a  Corollary of the above lemma, we obtain a polynomial representation result for the {\rm IAMP} algorithm described in \cite{montanari2019optimization}.
\begin{corollary}\label{chapter2:prop:1}
    Suppose $\X\distrib {\rm GOE}^0(n)$. Let $\eta >0$ and $T\in \N$. Consider the {\rm IAMP} algorithm in \cite{montanari2019optimization}, and let its denoiser functions be  $f^t : \R^{t+1} \to \R$, and the {\rm IAMP} iterates be $\b{u}^0,\ldots,\b{u}^T$. Then, there exist $\Delta=\Delta(\eta, T), c=c(\eta, T)$, and a sequence of polynomial denoisers $\{q^t: \R^{t+1} \to \R\}$ satisfying ${\rm Deg} (q^t) \leq \Delta, \|q^t\|_{\rm coef}\leq c$, and  producing AMP iterates $\hat{\b{u}}^0, \ldots, \hat{\b{u}}^T$ such that the following holds
    $$\plim \frac{\|\b{u}^t - \hat{\b{u}}^t\|_2}{\sqrt{n}} \leq \eta,$$
    for all $t\leq T$.
\end{corollary}
\begin{proof}
    The proof is an  application of Lemma \ref{chapter2:lemma:ivkov} to the {\rm IAMP} iterates in \cite{montanari2019optimization}. The conditions required to apply the Lemma are: (1) $f^t$ are Lipschitz functions, (2) $f^t$ are weakly differentiable with pseudo-Lipschitz or indicator partial derivatives, and (3) the covariance matrix $\b{Q}_{\leq t}$  satisfies $\b{Q}_{\leq t} \succeq \b{I}_t, \|\b{Q}_{\leq t}\|_{\infty}\leq 2$. The first condition is satisfied by the {\rm IAMP} described in equations (9) and (10) in \cite{montanari2019optimization}: $f^t$ are compositions/sums/products of bounded Lipschitz continuous functions. The second condition is only used to show state evolution for $b_{t, j}$ in the proof of Lemma \ref{chapter2:lemma:ivkov}  in \cite{ivkov2023semidefinite} (see Claim B.7 in \cite{ivkov2023semidefinite}), however, state evolution for $b_{t,j}$ is  directly shown in Lemma II.3 in \cite{montanari2019optimization}. Finally, the covariance matrix $\b{Q}_{\leq t}$ is shown to be diagonal in Lemma II.2 in \cite{montanari2019optimization} with entries upper/lower bounded by positive constants. We then claim that we can rescale the denoiser functions to satisfy condition (3) and detail this argument next. Given positive coefficients $a_0, \ldots, a_{T+1} \in \mathbb{R}_{>0}$, introduce for $t\leq T$
    \begin{align*}
        \tilde{f}^t: \mathbb{R}^{t+1} \to \mathbb{R},  \quad (x_0, \ldots, x_t) \mapsto a_{t+1} f^t\l(x_0/a_0, \ldots, x_t/a_t\r).
    \end{align*}
    Let $\tilde{\b{u}}^0 = a_0\b{u}^0$ and denote by $\tilde{\b{u}}^t$ the iterates of the {\rm IAMP} with adjusted denoisers $\tilde{f}^t$ and initialization $\tilde{\b{u}}^0$. Furthermore denote by $\tilde{b}_{t, j}$ the adjusted $b_{t,j}$ coefficients for $\tilde{f}^t$.    We first show by  induction that $\forall t\leq T, \tilde{\b{u}}^t = a_t \b{u}^t$. If $t=0$ the result follows by definition of $\tilde{\b{u}}^0$. Assume the result for all $t'\in [t]$ for some $t<T$, then
    \begin{align*}
        \tilde{\b{u}}^{t+1} &= \mathbf{X} \tilde{f}^t (\tilde{\b{u}}^0, \ldots,\tilde{\b{u}}^t) - \sum_{j=1}^{t} \tilde{b}_{t,j} \tilde{f}^{j-1} (\tilde{\b{u}}^0,\ldots,\tilde{\b{u}}^{j-1})\\
        &= a_{t+1} \mathbf{X} f^t(\tilde{\b{u}}^0/a_0, \ldots,\tilde{\b{u}}^t/a_t) - \sum_{j=1}^{t} \tilde{b}_{t,j} a_{j} f^{j-1} (\tilde{\b{u}}^0/a_0,\ldots,\tilde{\b{u}}^{j-1}/a_{j-1})\\
        &= a_{t+1} \l[ \mathbf{X} f^t(\b{u}^0, \ldots,\b{u}^t) -  \sum_{j=1}^{t} \frac{\tilde{b}_{t,j} a_{j}}{a_{t+1}} f^{j-1} (\b{u}^0,\ldots,\b{u}^{j-1}) \r].
    \end{align*}
    For $j\in [ t]$ we have
    \begin{align*}
        \tilde{b}_{t, j} &= \frac{1}{n} \sum_{i=1}^{n} \frac{\partial \tilde{f}^t}{\partial \tilde{\b{u}}_i^j}(\tilde{\b{u}}_i^0,\ldots,\tilde{\b{u}}_i^t)\\
        &=\frac{1}{n} \sum_{i=1}^{n}  \frac{a_{t+1}}{a_j}\frac{\partial f^t}{\partial\b{u}_i^j}(\tilde{\b{u}}_i^0/a_0,\ldots,\tilde{\b{u}}_i^t/a_t)\\
        &= \frac{a_{t+1}}{a_j} b_{t, j}.
    \end{align*}
    Henceforth
    \begin{align*}
        \tilde{\b{u}}^{t+1} &= a_{t+1} \l[ \mathbf{X} f^t(\b{u}^0, \ldots,\b{u}^t) -  \sum_{j=1}^{t} b_{t,j}  f^{j-1} (\b{u}^0,\ldots,\b{u}^{j-1}) \r]\\
        &= a_{t+1} {\b{u}}^{t+1},
    \end{align*}
    which concludes the induction proof. Let $\b{Q}, \tilde{\b{Q}}$ be the covariance matrices associated with the {\rm IAMP} obtained from denoisers $f^t$ and $\tilde{f}^t$ respectively, and given recursively by
    \begin{align*}
        \b{Q}_{k+1, j+1} &= \E[f^k(U_0, \ldots, U_k) f^{j}(U_0, \ldots, U_j)],\\
        \tilde{\b{Q}}_{k+1, j+1} &= \E[\tilde{f}^k(\tilde{U}_0, \ldots, \tilde{U}_k) \tilde{f}^{j}(\tilde{U}_0, \ldots, \tilde{U}_j)],
    \end{align*}
    where $(U_j)_{j\geq 1}$ ($(\tilde{U}_j)_{j\geq 1}$ resp.) is a centered Gaussian process independent of $U_0$ ($\tilde{U}_{0}$ resp.). As shown in Lemma II.2 in \cite{montanari2019optimization}, the matrix $\b{Q}$ is diagonal. Furthermore, the proof extends verbatim for $\tilde{\b{Q}}$. We next show that $\tilde{\b{Q}}_{t+1, t+1} = (a_{t+1})^2 \b{Q}_{t+1, t+1}$ for all $t<T$. It is straightforward to see that we can take $\tilde{U}_t = a_t U_t$ for $t\leq T$, since $\tilde U_0 \distrib a_0 U_0$ from our initialization $\tilde{\b{u}}^0 = a_0 \b{u}^0 $. Therefore $(\tilde{U}_j)_{j\geq 1}$ is a centered Gaussian process independent of $\tilde{U}_0$ with diagonal covariance matrix, and we have for $0\leq t < T$,
    \begin{align*}
         \tilde{\b{Q}}_{t+1, t+1} &=\E[\tilde{f}^t(\tilde{U}_0,\ldots, \tilde{U}_t)\tilde{f}^t(\tilde{U}_0,\ldots, \tilde{U}_t)] \\
         &= (a_{t+1})^2 \E[f^t(U_0,\ldots, U_t) f^t( U_0,\ldots, U_t)]\\
         &= (a_{t+1})^2 \b{Q}_{t+1, t+1}.
    \end{align*}
    Henceforth $\tilde{\b{Q}}_{\leq T} = {\rm Diag}((a_1)^2 \b{Q}_{1, 1}, \ldots, (a_T)^2 \b{Q}_{T, T})$. It suffices then to pick $a_1, \ldots, a_T$ so that $\tilde{\b{Q}}_{\leq T}$ is the identity matrix, and thus satisfies the two requirements  $\tilde{\b{Q}}_{\leq t} \succeq \b{I}_t$ and $\|\tilde{\b{Q}}_{\leq t}\|_{\infty}\leq 2$ for $t\leq T$, in order to apply Lemma \ref{chapter2:lemma:ivkov}. This concludes the proof.
\end{proof}
The candidate for near-optimum associated to the AMP with denoisers $q^t$ from Lemma \ref{chapter2:lemma:ivkov} is given similarly to (\ref{chapter2:candidate}) by
\begin{align*}
    \hat{\b{v}} = \sqrt{\delta} \sum_{k=1}^{\lfloor \bar q  / \delta \rfloor} q^k(\hat{\b{u}}^0, \ldots,\hat{\b{u}}^k).
\end{align*}
The next proposition shows that $\hat{\b{v}}$ is close in $\ell_2$ distance to  $\b{v}$ given by (\ref{chapter2:candidate}), i.e., the candidate for near-optimum associated with the {\rm IAMP} algorithm in \cite{montanari2019optimization}.

\begin{proposition}\label{chapter2:prop:1.1}
    Fix $\theta, \delta>0$ and $T\in \mathbb{Z}_{\geq 0}$. In the setting of Lemma \ref{chapter2:lemma:ivkov}, the polynomials $q^t$ can be chosen to satisfy the following 
    \begin{align*}
        \plim \frac{\|\b{v}-\hat{\b{v}}\|_2}{\sqrt{n}} \leq \theta.
    \end{align*}
\end{proposition}
\begin{proof}
    Let $\eta \in (0, 1)$. By Lemma \ref{chapter2:lemma:ivkov}, we can chose the polynomials $q^t$ so that $\plim \frac{\|\u^t - \hat{\u}^t\|_2}{\sqrt{n}}\leq \eta, \forall t\leq T$.    We have by the triangle inequality
    \begin{align*}
        \frac{\|\b{v}-\hat{\b{v}}\|_2}{\sqrt{n}} &\leq \sqrt{\delta} \sum_{k=1}^{\lfloor \bar q  / \delta \rfloor} \frac{ \| f^k(\b{u}^0, \ldots,\b{u}^k) - q^k(\hat{\b{u}}^0, \ldots,\hat{\b{u}}^k)\|_2}{\sqrt{n}},
    \end{align*}
    therefore it suffices to show that each of the summands above is bounded by $O(\eta)$ w.h.p. and then take $\eta$ small enough. Let $k\in [ \lfloor \bar q  / \delta \rfloor]$. We have
    \begin{align*}
        \|f^k(\b{u}^0, \ldots,\b{u}^k) - q^k(\hat{\b{u}}^0, \ldots,\hat{\b{u}}^k)\|_2 &\leq \underbrace{\|f^k(\b{u}^0, \ldots,\b{u}^k) - f^k(\hat{\b{u}}^0, \ldots,\hat{\b{u}}^k)\|_2}_{A} \\
        &+ \underbrace{\|f^k(\hat{\b{u}}^0, \ldots,\hat{\b{u}}^k) - q^k(\hat{\b{u}}^0, \ldots,\hat{\b{u}}^k)\|_2}_{B}.
    \end{align*}
    Let $\b{U}, \hat{\b{U}}$ be matrices in $\mathbb{R}^{n\times (k+1)}$ with columns $(\b{u}^0, \ldots,\b{u}^k)$, and $(\hat{\b{u}}^0, \ldots,\hat{\b{u}}^k)$ respectively. Since $f^k$ is Lipschitz with constant Lipschitz coefficient $L$, it follows that
    \begin{align*}
        A &\leq L \|\b{U} - \hat{\b{U}}\|_{\rm fro} = L \sqrt{\sum_{t=0}^{k} \|\b{u}^t - \hat{\b{u}}^t\|_2^2}.
    \end{align*}
    From Lemma \ref{chapter2:lemma:ivkov}, we have $\|\b{u}^t - \hat{\b{u}}^t\|_2 \leq \eta \sqrt{n}, \forall t\in [T]$ w.h.p. as $n\to \infty$, it follows that $A \leq L\eta \sqrt{nk}$ w.h.p. We now bound $B$. We have by Proposition~\ref{chapter2:state:evolution} 
    (using the fact that pseudo-Lipschitz functions are closed under sums/products)

    \begin{align*}
        \plim \frac{B^2}{n} &= \plim \frac{1}{n}\sum_{i=1}^{n} | f^k(\hat{\b{u}}_i^0, \ldots,\hat{\b{u}}_i^k) - q^k(\hat{\b{u}}_i^0, \ldots,\hat{\b{u}}_i^k)|^2\\
        &= \E \l[(f^k(\hat{U}^0, \ldots, \hat{U}^k) - q^k(\hat{U}^0, \ldots, \hat{U}^k))^2\r]\\
        &\leq g_2(k)^2,
    \end{align*}
    where $g_2$ is an increasing function introduced in Claim B.6 in \cite{ivkov2023semidefinite} and satisfies $g_2(T)\leq c\eta^2$, where $c$ is a constant depending on $T, L$. Hence
    \begin{align*}
        \plim \frac{B^2}{n} \leq c^2 \eta^4,
    \end{align*}
    which readily implies $B\leq O(\eta \sqrt{n})$ w.h.p. Taking $\eta$ small enough, we conclude that  $\plim \frac{\|\b{v}-\hat{\b{v}}\|_2}{\sqrt{n}} \leq \theta$, which ends the proof.
\end{proof}

\subsection{Approximation of AMP with Connected  Low-Degree Polynomials}\label{chapter2:section:onsager_limit}
The purpose of this section is to establish parts of the LDP structure required in Theorem~\ref{chapter2:thm:2-formal}. Starting from the polynomial denoiser representation obtained in Section~\ref{chapter2:section:iamp_poly}, we modify the resulting scheme by replacing the random Onsager coefficients by their limits in probability. This replacement allows us to derive the LDP structure of the iterates.

Following the statement of Theorem \ref{chapter2:thm:2-formal}, we may, without loss of generality, assume  that the  initialization point $\b{u}^0$ has entries bounded by a constant $M>0$ independent of $n$. This assumption appears restrictive for ${\rm IAMP}$, since  \cite{montanari2019optimization} requires a random initialization $\b{u}^0 \sim \delta \mathcal{N}(0, \b{I}_n)$ where $\delta$ is a fixed parameter driving the optimality gap of the {\rm IAMP} output. Nonetheless, this condition is implicitly satisfied by  the {\rm IAMP} algorithm as it includes a truncation step on the iterates, i.e.,  the denoisers depend on $\b{u}^0$ only through $[\b{u}^0]_M$ with $[u]_M\triangleq \max(-M, \min(u, M)) $ for some fixed large enough $M>0$. Indeed, let $\bar{\b{u}}^k = [\b{u}^k]_M, \forall k\in [T]$.  
Observe that for all $t\leq T$ the denoisers $f^t$ given in (9) and  (10) in \cite{montanari2019optimization} only depend on the iterates $\b{u}^k$ through $\bar{\b{u}}^k$, i.e.
\begin{align*}
    \forall t\leq T, f^t(\b{u}^0, \ldots, \b{u}^t) = f^t(\bar{\b{u}}^0, \ldots, \bar{\b{u}}^t),
\end{align*}
thus switching the initialization $\b{u}^0$ with $\bar{\b{u}}^0$ has no effect on the {\rm IAMP} iterations, and furthermore preserves the covariance matrix $\b{Q}$ since it holds for $k,j \geq 0$
\begin{align*}
    &\E_{U_0,\ldots,U_{\max(j,k)}}[f^k(U_0, \ldots,U_k) f^j(U_0, \ldots, U_j)] \\
    &= \E_{U_0,\ldots,U_{\max(j,k)}}[f^k([U_0]_M, \ldots,U_k) f^j([U_0]_M, \ldots, U_j)]\\
    &= \E_{[U_0]_M,\ldots,U_{\max(j,k)}}[f^k([U_0]_M, \ldots,U_k) f^j([U_0]_M, \ldots, U_j)].
\end{align*}
Hence, adding the truncation step has the effect of adding a secondary limit $M\to \infty$ after the limits on $n,d$ on all results mentioned in this paper. 

Corollary~\ref{chapter2:prop:1} implies that the IAMP iterates can be approximated by LDPs. In order to complete our second main result, Theorem~\ref{chapter2:thm:2-formal}, we need to show that  the LDPs induced by the denoisers $q^t$ satisfy additional assumptions, namely, 
the connectivity, and the norm/degree bounds.
This is  our next goal.

In the setting of Lemma \ref{chapter2:lemma:ivkov} and Proposition \ref{chapter2:prop:1.1},  we modify the AMP iterations with polynomial denoisers $q^t$ by replacing the coefficients $b_{t,j}$ by their respective limits in probability denoted $\bar{b}_{t,j} \triangleq \plim  b_{t,j}$. That is, we consider the following iterations

\begin{align}
 \tilde{\b{u}}^{t+1} &= \mathbf{X} q^t (\tilde{\b{u}}^0, \ldots,\tilde{\b{u}}^t) - \sum_{j=1}^{t} \bar{b}_{t,j} q^{j-1} (\tilde{\b{u}}^0,\ldots,\tilde{\b{u}}^{j-1}), \label{chapter2:amp:2}\\
 b_{t, j} &= \frac{1}{n} \sum_{i=1}^{n} \frac{\partial q^t}{\partial \hat{\b{u}}_i^j}(\hat{\b{u}}_i^0,\ldots,\hat{\b{u}}_i^t) \notag,
\end{align}
where $\hat{\b{u}}_i^0,\ldots,\hat{\b{u}}_i^T$ are the iterates of the (standard) AMP with denoisers $q^t$. We note that the existence of the limits $\bar{b}_{t,j}$ is justified by applying Proposition \ref{chapter2:state:evolution} to the partial derivatives of the polynomial denoisers $q^t$. We show next that the above modification leads to effectively the same algorithm, and that the dynamics of $\ut^t$ are roughly the same as $\hat{\u}^t$.

\begin{proposition}\label{chapter2:conjecture}
    Fix $\delta>0$ and $T\in \mathbb{Z}_{\geq 0}$. In the setting of Lemma \ref{chapter2:lemma:ivkov},  let $\hat{\b{u}}^0,\ldots,\hat{\b{u}}^T$ be the AMP iterates obtained with polynomial denoisers $q^t, t\in [T]$, and let $\tilde{\b{u}}^0, \ldots, \tilde{\b{u}}^T$ be the AMP iterates obtained by replacing $b_{t,j}$ with $\bar {b}_{t,j} \triangleq \plim b_{t,j}$ as shown in (\ref{chapter2:amp:2}). Let the near-optimum candidates be given by
    \begin{align*}
    \hat{\b{v}} &= \sqrt{\delta} \sum_{k=1}^{\lfloor \bar q  / \delta \rfloor} q^k(\hat{\b{u}}^0, \ldots,\hat{\b{u}}^k), \numberthis \label{chapter2:sigma:hat}\\
    \tilde{\b{v}} &= \sqrt{\delta} \sum_{k=1}^{\lfloor \bar q  / \delta \rfloor} q^k(\tilde{\b{u}}^0, \ldots,\tilde{\b{u}}^k). \numberthis \label{chapter2:sigma:tilde}
\end{align*}
Then
    \begin{align*}
        \plim \frac{\|\hat{\b{v}}-\tilde{\b{v}}\|_2}{\sqrt{n}} =0.
    \end{align*}

\end{proposition}
The proof of Proposition \ref{chapter2:conjecture} is given in Section \ref{chapter2:subsection:5.4}. Using Proposition \ref{chapter2:conjecture}, we next establish the key representation result we require.

\begin{proposition}\label{chapter2:prop:2}
    Let $\delta>0, T\in \N$ and consider the AMP iteration (\ref{chapter2:amp:2}) with polynomial denoisers $q^t: \R^{t+1} \to \R$ with constant degrees $\Delta^t$, and coefficients bounded in absolute value by a constant. Let $\tilde{\b{u}}^0, \ldots, \tilde{\b{u}}^T$ be the iterates produced, where $\tilde{\b{u}}^0$ will be assumed to satisfy $\|\tilde{\b{u}}^0\|_{\infty}=O(1)$ w.h.p. Then there exist LDPs $w_i^t$ and $h_i$ such that
     for all $t\in[0,T]$ and $i\in[n]$
    \begin{enumerate}
        \item  
        $\tilde{\b{u}}_i^t =  w_i^t(\X)$, and $\tilde{\b{v}}_i =  h_i(\X)$.
        \item $ \max_i\|w_i^t\|_{\rm coef}=O(1)$, and $\max_i\|h_i\|_{\rm coef}=O(1)$. 
        \item $w_i^t,\ h_i$ are connected LDPs containing node $i$ (i.e., $w_i^t \in \V_i$ and $h_i \in \V_i$). 
    \end{enumerate}
\end{proposition}
\begin{proof}
    We first show the required properties for the iterates $\ut^t$. The claim for $\tilde{\b{v}}$ will follow similarly.  Note that by Proposition~\ref{chapter2:state:evolution}, the quantities $|\bar{b}_{t,j}|$ are finite constants for all $t\in[0,T]$ and $j\in [t]$, and in particular are uniformly bounded by a constant independent of $n$. The proof proceeds by induction on $t\leq T$. If $t=0$ then $w_i^0$ are uniformly bounded constant polynomials, thus they satisfy all claims in the proposition by definition. Assume the result for all $t' \in [0,t]$ where $t<T$ is fixed. By symmetry, it suffices to check that it holds at $t+1$ for $i=1$. We have
    $$\tilde{\b{u}}_1^{t+1} = \underbrace{\sum_{\ell=1}^{n} \X_{1\ell} q^t (\tilde{\b{u}}_\ell^0, \ldots,\tilde{\b{u}}_\ell^t)}_{A_{1, t}} - \underbrace{\sum_{j=1}^{t} \bar{b}_{t,j} q^{j-1} (\tilde{\b{u}}_1^0,\ldots,\tilde{\b{u}}_1^{j-1})}_{O_{1, t}}.$$
     We first show that $O_{1,t}$ is a connected LDP in $\X$ with constant coefficients. Each term $q^{j-1} (\tilde{\b{u}}_1^0,\ldots,\tilde{\b{u}}_1^{j-1})$ is a polynomial  in $\tilde{\b{u}}_1^0,\ldots,\tilde{\b{u}}_1^{t}$ with constant degree and constant coefficients. Furthermore, by the inductive hypothesis, each of $\tilde{\b{u}}_1^0,\ldots,\tilde{\b{u}}_1^{t}$ is a connected LDP in $\X$ with $O(1)$ coefficients containing node $1$. Therefore, it follows that $q^{j-1} (\tilde{\b{u}}_1^0,\ldots,\tilde{\b{u}}_1^{j-1})$ is an LDP in $\X$ with $O(1)$ coefficients. Moreover, the latter LDP is connected and contains node $1$ (i.e., $q^{j-1} (\tilde{\b{u}}_1^0,\ldots,\tilde{\b{u}}_1^{j-1} \in \V_1 )$), since it is the product of connected monomials sharing node $1$.  As $\bar{b}_{t,j}$ are constants uniformly bounded in $n$, it follows that each summand in $O_{1,t}$ is a connected LDP with $O(1)$ coefficients containing node $1$ (i.e., $O_{1,t}\in \V_1$). As $O_{1,t}$ is a finite sum of such LDPs, it readily follows that $O_{1, t}$ is a connected LDP in $\X$ with $O(1)$ coefficients.

 We now deal with the term $A_{1, t}$. Let $r_\ell(\X) = \X_{1\ell} q^t (\tilde{\b{u}}_\ell^0, \ldots,\tilde{\b{u}}_\ell^t)$ for $\ell\geq 2$ (we recall here that $\X_{\ell \ell}=0,\forall \ell \in [n]$). Note that $(r_\ell, \ell\geq 2)$ are LDPs and $\forall \ell\in [2,n]$, we have $\X_{1\ell} \in \V_{\{1, \ell\}}$. Furthermore, since each of $\tilde{\b{u}}_\ell^0, \ldots,\tilde{\b{u}}_\ell^t$ is a connected LDP in $\X$ with $O(1)$ coefficients containing node $\ell$ (i.e., $\forall 0\leq j\leq t,\ \tilde{\b{u}}_\ell^j \in \V_{\ell}$), it follows that $q^t (\tilde{\b{u}}_\ell^0, \ldots,\tilde{\b{u}}_\ell^t)$ is a connected LDP in $\X$ with $O(1)$ coefficients in $\X$ containing node $\ell$ (i.e., $ q^t (\tilde{\b{u}}_\ell^0, \ldots,\tilde{\b{u}}_\ell^t \in \V_{\ell})$). Therefore $r_{\ell}(\X)$ is a connected LDP in $\X$ with $O(1)$ coefficients containing nodes $\{1, \ell\}$ (i.e., $ r_\ell \in \V_{\{1, \ell\}}$) as the product of $\X_{1\ell}$ (an edge-tree with endpoints $1,\ell$) and $q^t(\tilde{\b{u}}_\ell^0, \ldots,\tilde{\b{u}}_\ell^t)$ (LDP containing node $\ell$). It follows readily that $A_{1,t}$ is a connected LDP in $\X$ containing node $1$ (i.e., $A_{1,t}\in \V_1$). It remains to show $\|A_{1,t}\|_{\rm coef} = \|\sum_{2\leq \ell\leq n} r_\ell\|_{\rm coef}=O(1)$.    Using Lemma \ref{chapter2:lemma:pre:2} with sets $S_{\ell}\triangleq \{1,\ell\}$ for $2\leq \ell \leq n$ and $\gamma=1, c=1$, it follows that 
    \begin{align*}
        \l\|A_{1,t}\r\|_{\rm coef} \leq (1+\Delta^t) \max_{2\leq \ell\leq n} \|r_\ell\|_{\rm coef}= O(1).
    \end{align*}
    Therefore $\tilde{\b{u}}_1^{t+1}$ is a connected LDP in $\X$  containing node $1$ (i.e, $\tilde{\b{u}}_1^{t+1}\in \V_1$) with $O(1)$ coefficients. Finally, note that any bounds on the coefficients of the LDP $\tilde{\b{u}}_1^{t+1}$ extend generically to the LDPs $\tilde{\b{u}}_i^{t+1}$ for all $i\in [n]$, so that $\max_i \|\tilde{\b{u}}_i^{t+1}\|_{\rm coef}=O(1)$. This completes the induction and ends the proof for $\ut^t$.

    We now show similarly the representation result on $\tilde{\b{v}}$. By symmetry, it suffices to verify the result for $i=1$. We have $\tilde{\b{v}}_1 = \sqrt{\delta} \sum_{k=1}^{\lfloor \bar q  / \delta \rfloor} q^k(\tilde{\b{u}}_1^0, \ldots,\tilde{\b{u}}_1^k)$. As shown above, we have that $\tilde{\b{u}}^t_1, t\in [k]$ are connected LDPs in $\X$ with constant coefficients and contain node $1$ (i.e., $\tilde{\b{u}}^t_1\in \V_1$). Since $q^k$ are polynomials with constant coefficients, it follows readily that each of the summands $q^k(\tilde{\b{u}}_1^0, \ldots,\tilde{\b{u}}_1^k)$ is an LDP in $\X$ with constant coefficients. Furthermore, since all the LDPs $\tilde{\b{u}}^t_1, t\in [k]$ are connected and share node $1$, it follows that all monomials in $q^k(\tilde{\b{u}}_1^0, \ldots,\tilde{\b{u}}_1^k)$ are connected and contain node $1$ (i.e., $q^k(\tilde{\b{u}}_1^0, \ldots,\tilde{\b{u}}_1^k)\in \V_1$). Therefore, the claims in items (1) and (3) in the proposition for $h_1$ hold. Finally, since $\tilde{\b{v}}_1$ is a finite sum of LDPs with constant coefficients, it follows that $\|h_1\|_{\rm coef}=O(1)$, and moreover $\max_i \|h_i\|_{\rm coef}=O(1)$ as the bound on $\|h_1\|_{\rm coef}$ extends to all $h_i$,  which shows the claim of item (2)  and concludes the proof.

\end{proof}

\subsection{Proof of Theorem \ref{chapter2:thm:2-formal}}\label{chapter2:section:proof:thm:2-formal}
In this section, we prove Theorem~\ref{chapter2:thm:2-formal}. The proof combines the approximation results obtained in Sections~\ref{chapter2:section:iamp_poly} and~\ref{chapter2:section:onsager_limit}. In particular, Section~\ref{chapter2:section:onsager_limit} yields a representation of the IAMP output by connected LDPs. Here, we further project this representation onto tree-structured LDPs and show, in Proposition~\ref{chapter2:prop:3}, that this additional projection incurs a negligible error. The total approximation error is controlled by telescoping the successive modifications introduced, thus completing the proof of Theorem~\ref{chapter2:thm:2-formal}.

As a corollary of Proposition \ref{chapter2:prop:2} and Proposition \ref{chapter2:conjecture}, we can construct for every fixed $\varepsilon>0$ an AMP algorithm  with $T=T(\varepsilon)$ iterations, and polynomial denoisers $q^t$ satisfying the assumptions of Lemma \ref{chapter2:lemma:ivkov}, and we can encode the near-optimum candidate  as $h(\X) \triangleq (h_1(\X), \ldots, h_n(\X))$  where $h_i$ satisfy the assumptions of Proposition \ref{chapter2:prop:2}. We next provide a justification of the quality of the output of $p \triangleq h^{\rm Tr}$.
\begin{proposition}\label{chapter2:prop:3}
    Let $\X\distrib {\rm GOE}^0(n)$ and $\varepsilon>0$. There exists $\delta = \delta(\varepsilon) >0$ and $T = T(\varepsilon)\in \N$ such that if $\tilde{\b{v}}=h(\X)$ in the setting of Proposition \ref{chapter2:prop:2}, and $p\triangleq h^{\rm Tr}$, then the following holds w.h.p. as $n\to \infty$ 
    \begin{equation}
        \begin{aligned}
            \frac{1}{2n}| \b{v}^\top \X \b{v}  - p(\X)^\top \X p(\X) | &\leq \varepsilon ,\\
            \frac{d(p(\X), [-1,1]^n)}{\sqrt{n}} &\leq \varepsilon,
        \end{aligned}
    \end{equation}
    where $\b{v}$ is the output candidate of {\rm IAMP} given in (\ref{chapter2:candidate}). 
\end{proposition}
\begin{proof}
    Fix $\varepsilon' >0 $, let $\theta>0$ as in Proposition \ref{chapter2:prop:1.1}, and let $\hat{\b{v}}$ be given as in (\ref{chapter2:sigma:hat}). Then, w.h.p. as $n\to \infty$
    \begin{align*}
        \frac{d(p(\X), [-1,1]^n)}{\sqrt{n}} 
        &\leq \frac{\| p(\X) -  \tilde{\b{v}} \|_2}{\sqrt{n}} + \frac{\|\tilde{\b{v}} - \hat{\b{v}}\|_2 }{\sqrt{n}}+ \frac{\|\hat{\b{v}} - \b{v}\|_2 }{\sqrt{n}} + \frac{d(\b{v}, [-1,1]^n)}{\sqrt{n}}\\
        &\leq \frac{\|p(\X) - \tilde{\b{v}} \|_2}{\sqrt{n}} + \kappa_n + \theta + \varepsilon' \numberthis \label{chapter2:eq:v:tilde:box},
    \end{align*}
    where $\kappa_n$ is a random variable such that $\plim \kappa_n =0$, and the last line follows from applying Proposition \ref{chapter2:conjecture}, Proposition \ref{chapter2:prop:1.1} and (\ref{chapter2:eq:v:box}) respectively. We have
    \begin{align*}
        \E\l[\|p(\X)- \tilde{\b{v}}\|_2^2\r] &= \sum_{i=1}^{n} \E\l[(h^{\rm Tr}_i(\X) - h_i(\X))^2\r].
    \end{align*}
    Let $g_i \triangleq h_i^{\rm Tr} - h_i$. Since $h_i$ is connected and $h_i \in \V_i$ it follows that $g_i$ is connected and $g_i \in \V_i$. Since $g_i^{\rm Tr}=0$ and $g_i$ is connected, we have by Lemma \ref{chapter2:lemma:pre:3} that  $[(g_i)^2]^{\rm Tr}=0$. Using Lemma \ref{chapter2:lemma:pre:6} on $(g_i)^2$ with $S\triangleq \{i\}$  yields
    \begin{align*}
        \E[g_i(\X)^2] &\leq \frac{c\|(g_i)^2\|_{\rm coef}}{n}\leq \frac{c2^{2{\rm Deg}(h_i)}\|h_i\|^2_{\rm coef}}{n},
    \end{align*}
    where $c$ is a constant, and we used Lemma \ref{chapter2:lemma:pre:1} in the last inequality. Therefore $\E\l[\|p(\X)- \tilde{\b{v}}\|_2^2\r] $ $\leq c\sum_{i=1}^{n} 2^{2{\rm Deg}(h_i)}\|h_i\|^2_{\rm coef} / n$. Since $\max_i \|h_i\|_{\rm coef}$ $=O(1)$, and $\max_i {\rm Deg}(h_i)=O(1)$, thus $\E\l[\|p(\X)- \tilde{\b{v}}\|_2^2\r] = O(1)$. Using Chebyshev's inequality, we then have for all $\eta>0$
    \begin{align*}
        \p(\|p(\X)- \tilde{\b{v}}\|_2 \geq \eta \sqrt{n}) &= \p(\|p(\X)-\tilde{\b{v}} \|_2^2 \geq \eta^2 n) \leq \frac{O(1)}{\eta^2n}.
    \end{align*}
    Combining the above with  (\ref{chapter2:eq:v:tilde:box}), it holds w.h.p.
    \begin{align*}
        \frac{d(p(\X), [-1,1]^n)}{\sqrt{n}} &\leq \eta + \kappa_n + \theta + \varepsilon'. \numberthis \label{chapter2:eq:v:tilde:box:2}
    \end{align*}
    Let $\b{u} = p(\X) - \b{v}$, so that $\|\b{u}\|_2/\sqrt{n} \leq \eta + \kappa_n + \theta$ w.h.p. as shown above. We have
    \begin{align*}
        |   p(\X)^\top \X p(\X)  - \b{v}^\top \X \b{v} | &= |2  \b{u}^\top \X \b{v}  +  \b{u}^\top \X \b{u} | \\
        &\leq \left( 2 \|\b{u}\|_2 \|\b{v}\|_2  + \|\b{u}\|_2^2 \right) \|\X\|. 
    \end{align*}
    From Proposition (\ref{chapter2:eq:v:box}) it follows that $\|\b{v}\|_2\leq (1+\varepsilon')\sqrt{n}$. Therefore $\frac{2 \|\b{u}\|_2 \|\b{v}\|_2  + \|\b{u}\|_2^2}{n} \leq 2 (\eta+\kappa_n + \theta)(1 + \varepsilon') + (\eta+\kappa_n + \theta)^2$ w.h.p. Using Lemma \ref{chapter2:lemma:random:matrix:dense}, we have $\|\X\| = O(1)$ w.h.p. Therefore, there exists a constant $C>0$ such that
    \begin{align}
        | p(\X)^\top \X p(\X)  - \b{v}^\top \X \b{v}  | &\leq C n \left(2 (\eta+\kappa_n + \theta)(1 + \varepsilon') + (\eta+\kappa_n + \theta)^2 \right). \label{chapter2:eq:v:tilde:obj}
    \end{align}
    The claim of the proposition follows readily by taking $\eta,\theta, \varepsilon'$ small enough combined with $\plim \kappa_n = 0$ in (\ref{chapter2:eq:v:tilde:box:2}) and (\ref{chapter2:eq:v:tilde:obj}).
\end{proof}
This concludes the proof of Theorem \ref{chapter2:thm:2-formal}.

\subsection{Proof of Proposition \ref{chapter2:conjecture}}\label{chapter2:subsection:5.4}
This section proves Proposition~\ref{chapter2:conjecture}, which compares the AMP iterates with polynomial denoisers to the corresponding iterates with constant Onsager coefficients. The proof recursively controls the discrepancy between the two sequences of iterates and establishes state evolution for the scheme with constant Onsager coefficients. We first show the following Lemma on projection for conditional distributions of Gaussian matrices.
\begin{lemma}\label{chapter2:lemma:A}
    Let $\X \distrib {\rm GOE}^0(n)$. Let $\b{A} \in \R^{d\times n}$ and $\b{D}\in \R^{d\times n}$. Suppose $d\leq n$ and ${\rm Rank}(\b{A})=d$. Let $\X'\distrib {\rm GOE}^0(n)$ be an independent copy of $\X$. Then, the distribution of $\X$ conditional on $\b{A}\X = \b{D}$ is equal to the distribution of $\b{A}^\top (\b{A}\b{A}^\top)^{-1}\b{D}+{\rm Proj}_{\{\b{z}\in \R^n \mid \b{A} \b{z}=0\}}(\X')$, where ${\rm Proj}_{\mathcal{V}}(\b{Y})$ denotes the matrix obtained by projecting each column of $\b{Y}$ onto the vector space $\mathcal{V}$.
\end{lemma}
\begin{proof}
Let $\b{B}\in \R^{(n-d)\times n}$ be a matrix with orthonormal rows, such that the rows of $\b{B}$ are also orthogonal to the rows of $\b{A}$. In particular $[\b{A}^\top | \b{B}^\top]$ is invertible. Conditionally on $\b{A}\X=\b{D}$, we have
\begin{align*}
    \X &= \b{A}^\top (\b{A} \b{A}^\top)^{-1}\b{A} \X +\b{B}^\top \b{B} \X\\
        &= \b{A}^\top (\b{A} \b{A}^\top)^{-1}\b{D} +  \b{B}^\top \b{B} \X,
\end{align*}
where we used $\b{A}^\top (\b{A} \b{A}^\top)^{-1}\b{A} + \b{B}^\top \b{B} = \b{I}_n$.  Let $\mathcal{V}$ be the vector space spanned by the rows of $\b{B}$. Note that $\b{B}^\top \b{B} \X = {\rm Proj}_{\mathcal{V}}(\X)$, and by Cochran's Theorem, it holds that conditionally on $\b{A}\X=\b{D}$, ${\rm Proj}_{\mathcal{V}}(\X) \distrib {\rm Proj}_{\mathcal{V}}(\X')$. This concludes the proof.
\end{proof}

\begin{proof}[Proof of Proposition \ref{chapter2:conjecture}]
Given a sequence of random variables $X_i, i\geq 1$ and a constant $C$, we write $\plimsup X_n < C$ if and only if there exists $\varepsilon>0$ such that $\lim_{n\to \infty} \p(X_n \geq C-\varepsilon) =0$. This definition is introduced only for brevity, as all uses of $\plimsup X_n < C$ simply encode that $X_n < C'$ with high probability for our purposes, where $C'$ is a constant depending on $C$. We first show the following Proposition. 

\begin{proposition}\label{chapter2:prop:appendix}
    Let $\X \distrib {\rm GOE}^0(n)$. Let $q^t:\mathbb{R}^{t+1}\to \mathbb{R}, t=0,..,T$ be multivariate polynomials with constant coefficients and degrees $\Delta_t=O(1)$. Let $\b{u}^0 \in \mathbb{R}^n$ be an initialization independent of $\X$  satisfying $\|\b{u}^0\|_{\infty}\leq O(1)$ w.h.p. and  $\forall \ell\in \mathbb{Z}_{\geq 0}, \hat p_{\b{u}} \xrightarrow{W_\ell} p_{U_0}$ (where $p_{U_0}$ is any distribution on $\mathbb{R}$ with finite  moments of all orders, $\hat p_{\b{u}}$ is the empirical distribution on $\{\b{u}_1, \ldots, \b{u}_n\}$, and the limit is in Wasserstein distance).     Consider the iterates $\u^t, \tilde{\u}^t$ for $t\in[T]$ given recursively by
    \begin{align*}
        \b{u}^{t+1} &= \X q^t(\b{u}^0,\ldots,\b{u}^{t}) - \sum_{j=1}^{t} b_{t,j} q^{j-1}(\b{u}^0,\ldots,\b{u}^{j-1}),\\
        \tilde{\b{u}}^{t+1} &= \X q^t(\tilde{\b{u}}^0,\ldots,\tilde{\b{u}}^{t}) - \sum_{j=1}^{t} \bar{b}_{t,j} q^{j-1}(\tilde{\b{u}}^0,\ldots,\tilde{\b{u}}^{j-1}).
    \end{align*}
    where $t<T$, $\tilde{\b{u}}^0 = \b{u}^0$ and
    \begin{align*}
        b_{t, j} &= \frac{1}{n} \sum_{i=1}^{n} \frac{\partial q^t}{\partial \b{u}_i^j}(\b{u}_i^0,\ldots,\b{u}_i^t),\\
        \bar{b}_{t, j} &= \plim b_{t, j},
    \end{align*}
    where the existence of the limit points $\bar b_{t,j}$ is justified by Proposition \ref{chapter2:state:evolution}. Let $(U_j)_{j\geq 1}$ be a centered Gaussian process independent of $U_0$ as defined in Proposition \ref{chapter2:state:evolution}, and let $\b{Q}$ be its covariance matrix. Suppose
    \begin{align*}
        \exists c_1, c_2>0, \forall t\in [T], \b{Q}_{\leq t} \succeq c_1 \b{I}_{t}, \text{and } \|\b{Q}\|_{\infty}\leq c_2.\numberthis \label{chapter2:Q:bound}
    \end{align*}
    Then, the following properties hold for all $t\in [0,T]$
    \begin{itemize}
        \item $\A_1(t):$ For all $\ell \in \mathbb{Z}_{\geq 0}$, there exist constants $c_{t,\ell}$ such that
    \begin{align*}
         \plimsup \frac{1}{n} \|\b{u}^t\|_\ell^\ell \leq c_{t,\ell}, \quad \plimsup \frac{1}{n} \|\tilde{\b{u}}^t\|_\ell^\ell \leq  c_{t,\ell}.
    \end{align*}
        \item $\A_2(t):$
        \begin{align*}
            \plimsup \frac{1}{\sqrt{n}} \|\b{u}^t - \tilde{\b{u}}^t\|_2 &= 0.
        \end{align*}
        \item $\A_3(t)$: For all $m\in \mathbb{Z}_{\geq 0}$ and any function $\psi:\mathbb{R}^{t+1}\to \mathbb{R}\in {\rm PL}(m)$
        \begin{align*}
             \plim \l| \frac{1}{n} \sum_{i=1}^{n} \psi(\tilde{\b{u}}^0_i,\ldots,\tilde{\b{u}}^{t}_i) - \frac{1}{n} \sum_{i=1}^{n} \psi(\b{u}^0_i,\ldots,\b{u}^{t}_i)  \r| = 0.
        \end{align*}
    \end{itemize}

\end{proposition}
\begin{proof}
The proof is essentially identical to the proof of Proposition 2.1 in \cite{montanari2021optimization} with minor adjustments. We provide a full proof here for completeness. Using the same notation as in \cite{montanari2021optimization}, we denote by $\b{q}^t, \tilde{\b{q}}^t$ the vectors $q^t(\b{u}^0,\ldots,\b{u}^t), q^t(\tilde{\b{u}}^0,\ldots,\tilde{\b{u}}^t)$ respectively. Let $\b{F}_t, \tilde{\b{F}}_t$ be the matrices in $\mathbb{R}^{n\times (t+1)}$ with columns $\b{q}^0,\ldots,\b{q}^t$, and $\tilde{\b{q}}^0,\ldots,\tilde{\b{q}}^t$ respectively. Let $\b{P}_t, \tilde{\b{P}}_t$ be the projection matrices on the column span of $\b{F}_t, \tilde{\b{F}}_t$ respectively, and let $\P_t^\perp = \b{I}_n -\P_t$ and $\Pt_t^\perp = \b{I}_n - \Pt_{t}$. Furthermore, let $\b{Q}$ be the covariance matrix associated with the AMP iterates $\b{u}^t$. Applying Proposition \ref{chapter2:state:evolution} to the terms ${\q^j}^\top \q^i$, we have 
\begin{align*}
    \forall t\in[T], \plim \frac{1}{n} \b{F}_{t-1}^\top \b{F}_{t-1} = \b{Q}_{\leq t}. \numberthis \label{chapter2:Q:limit}
\end{align*}
We prove the three properties inductively on $t$. Claims $\A_2(0), \A_3(0)$ hold trivially as $\u^0 = \ut^0$. Furthermore, $\A_1(0)$ holds as a direct application of Proposition \ref{chapter2:state:evolution}, namely $\plim \frac{1}{n} \sum_{i=1}^{n}|\u^0_i|^\ell = \E[|U_0|^\ell]<\infty$. Suppose that the properties $\A_1(t'),\A_2(t'), \A_{3}(t')$ hold for all $t'\in[0,t]$ for some $t<T$. 

We first show $\A_1(t+1)$. For $\ell\geq 0$, let $\psi:\R^{t+2}\to \R, (u^0,\ldots,u^{t+1})\mapsto |u^{t+1}|^\ell$. Clearly $\psi$ is in ${\rm PL}(O(\ell))$, thus by Proposition \ref{chapter2:state:evolution} $ \plim \frac{1}{n}\sum_{i=1}^{n}\psi(\u^0_i, \ldots, \u^{t+1}_i) = \E[\psi(U_0,\ldots,U_{t+1})] < \infty$. This proves $\plimsup \frac{1}{n} \|\b{u}^{t+1}\|_\ell^\ell<\infty$. Thus, we only need to show $\A_1(t+1)$ for the iterates $\tilde{\u}^t$. Let $\rrt= \sum_{j=1}^{t} \bar{b}_{t,j} \qt^{j-1}$. Following \cite{montanari2021optimization}, we write 
\begin{align*}
    \ut^{t+1} &= \X \qt^t - \rrt^t\\
    &= \Pt_{t-1}^\perp \X \Pt_{t-1}^\perp \qt^t+ \Pt_{t-1}^\perp \X \Pt_{t-1} \qt^t + \Pt_{t-1} \X \qt^t - \rrt^t\\
    &\distrib \Pt_{t-1}^\perp \hat{\X} \Pt_{t-1}^\perp \qt^t+ \Pt_{t-1}^\perp \X \Pt_{t-1} \qt^t + \Pt_{t-1} \X \qt^t - \rrt^t \numberthis \label{chapter2:line:projection}\\
    &= \hat{\X} \Pt_{t-1}^\perp \qt^t - \Pt_{t-1}\hat{\X}\Pt_{t-1}^\perp \qt^t -\rrt^t + \Pt_{t-1}^\perp \X \Pt_{t-1}\qt^t + \Pt_{t-1}\X \qt^t\\
    &\triangleq \b{v}_1 - \b{v}_2 - \b{v}_3+\b{v}_4 +\b{v}_5.
\end{align*}
where we used  Lemma \ref{chapter2:lemma:A} in line (\ref{chapter2:line:projection}) and $\hat{\X}$ is distributed as $\X$ but independent of the $\sigma$-algebra $\sigma(\{\qt^j, \ut^j\}_{j\leq t})$. Furthermore, note that using $\A_{3}(j), j\leq t$, and (\ref{chapter2:Q:limit})
\begin{align*}
    \plim \frac{1}{n} \Ft_{t-1}^\top \Ft_{t-1} = \plim \frac{1}{n} \b{F}_{t-1}^\top \b{F}_{t-1} = \b{Q}_{\leq t}.
\end{align*}
Therefore $ \Ft_{t-1}^\top \Ft_{t-1}$ is invertible w.h.p. and the following holds
\begin{align*}
\Pt_{t-1} = \Ft_{t-1}(\Ft_{t-1}^\top\Ft_{t-1})^{-1}\Ft_{t-1}^\top \numberthis \label{chapter2:matrix:projection}
\end{align*}
We next bound $\plimsup \frac{1}{n^{1/\ell}}\|\b{v}_k\|_\ell, k \in [5]$.
\begin{enumerate}
    \item We first deal with $\b{v}_1$. Denote $\b{w} \triangleq \Pt_{t-1}^\perp \qt^t$. Let $\b{D} = {\rm Diag}(z_1,\ldots,z_n)$ where $z_i$ are independent normal variables $\mathcal{N}(0, 2/n)$, so that $\hat \X + \b{D} \distrib {\rm GOE}(n)$, i.e., the Gaussian Orthogonal Ensemble. 
    Since $\hat \X + \b{D}$ is independent of $\b{w}$,  it follows by simple verification that $(\hat \X + \b{D})\b{w} \distrib (\|\b{w}\|_2/\sqrt{n}) \b{g} + g_0 \b{w}/\sqrt{n}$ where $(g_0,\b{g}) \distrib \mathcal{N}(0, \b{I}_{n+1})$. We then have
    \begin{align*}
        \|\b{v}_1\|_\ell &= \| (\hat \X + \b{D})\b{w} - \b{D} \b{w}\|_\ell\\
        &\leq \|(\hat \X +\b{D})\b{w}\|_\ell + \|\b{D} \b{w}\|_\ell\\
        &\leq \frac{\|\b{w}\|_2}{\sqrt{n}}\|\b{g}\|_\ell + \frac{|g_0|}{\sqrt{n}} \|\b{w}\|_\ell + \|\b{D}\b{w}\|_\ell \numberthis \label{chapter2:A.1}.
    \end{align*}
    To bound the $\plimsup$ of the above, we first bound $\|\b{w}\|_\ell$. Note that $\b{w} = \qt^t - \Pt_{t-1} \qt^t$, and
    \begin{align*}
      \Pt_{t-1} \qt^t = \sum_{j=0}^{t-1} a_{tj} \qt^j, \quad a_{tj} \triangleq \sum_{k=0}^{t-1} (\Ft_{t-1}^\top \Ft_{t-1})^{-1}_{jk} \langle \qt^{k}, \qt^{t}\rangle.
    \end{align*}
    Note that
    \begin{align*}
        |a_{tj}| &\leq \sum_{k=0}^{t-1} \l|n (\Ft_{t-1}^\top \Ft_{t-1})^{-1}_{jk} \r| \l|\l\langle \frac{\qt^{k}}{\sqrt{n}}, \frac{\qt^{t}}{\sqrt{n}}\r\rangle \r|.
    \end{align*}
    By $\A_{1}(k), k\leq t$ and using the fact that $\qt^j, j\leq t$ are polynomials (of constant degrees/coefficients) in the iterates $\ut^j$, it follows that $\plimsup |\langle \frac{\qt^{k}}{\sqrt{n}}, \frac{\qt^{t}}{\sqrt{n}}\rangle| = O(1)$. Furthermore, by (\ref{chapter2:matrix:projection}) we have $\plim  n (\Ft_{t-1}^\top \Ft_{t-1})^{-1}_{jk} = ({\b{Q}}_{\leq t})^{-1}_{jk} = O(1)$. Therefore, it follows that $\plimsup |a_{tj}|=O(1)$.  Hence
    \begin{align*}
        \plimsup \frac{1}{n^{1/\ell}}\|\b{w}\|_\ell &\leq \plimsup \frac{1}{n^{1/\ell}}\|\qt^t\|_\ell + \plimsup \frac{1}{n^{1/\ell}} \|\Pt_{t-1} \qt^t\|_\ell\\
        & \leq O(1) +\plimsup \sum_{j=0}^{t-1} \l(|a_{tj}| \frac{1}{n^{1/\ell}}\|\qt^j\|_\ell\r)\\
        &= O(1),
    \end{align*}
    where the last two inequalities follow from $\A_{1}(k), k\leq t$ combined with $\plimsup |a_{tj}|=O(1)$. Furthermore, we have
    \begin{align*}
       \plimsup \frac{1}{n^{1/\ell}} \|\b{D}\b{w}\|_\ell &= \plimsup \frac{1}{n^{1/\ell}}\l(\sum_{i=1}^{n} |z_i|^\ell |\b{w}_i|^\ell \r)^{\frac{1}{\ell}}\\
       &\leq \plimsup \max_{1\leq i \leq n} |z_i| \frac{1}{n^{1/\ell}} \|\b{w}\|_\ell\\
       &= 0,
    \end{align*}
    where the last line follows from $\plimsup \frac{1}{n^{1/\ell}} \|\b{w}\|_\ell$ $= O(1)$ and the fact  $\max_{1\leq i \leq n} |z_i|$ $ = \Theta(\sqrt{\log(n)/n})$ with probability $1-n^{-\Theta(1)}$. Putting everything together in (\ref{chapter2:A.1}) yields
    \begin{align*}
        \plimsup \frac{1}{n^{1/\ell}}\|\b{v}_1\|_\ell &\leq  \plimsup\frac{1}{n^{1/\ell}} \frac{\|\b{w}\|_2}{\sqrt{n}}\|\b{g}\|_\ell + \plimsup\frac{1}{n^{1/\ell}}\frac{|g_0|}{\sqrt{n}} \|\b{w}\|_\ell \\
        &+ \plimsup\frac{1}{n^{1/\ell}}\|\b{D}\b{w}\|_\ell\\
        &=  \plimsup  \frac{\|\b{w}\|_2}{\sqrt{n}} \frac{\|\b{g}\|_\ell}{n^{1/\ell}} + \plimsup\frac{|g_0|}{\sqrt{n}} \frac{\|\b{w}\|_\ell}{n^{1/\ell}},
    \end{align*}
    note that $\plimsup \frac{\|\b{w}\|_2}{\sqrt{n}}=O(1), \plimsup \frac{\|\b{w}\|_\ell}{n^{1/\ell}} = O(1)$ from previous arguments. Moreover, it holds that $\plimsup \frac{|g_0|}{\sqrt{n}}=0$, and by virtue of CLT we have $\frac{\|\b{g}\|_\ell}{n^{1/\ell}} = O(1)$. Henceforth $\plimsup \frac{1}{n^{1/\ell}}\|\b{v}_1\|_\ell\leq O(1)$. This ends the analysis of $\| \b{v}_1 \|_\ell$.

    \item We now deal with $\b{v}_2$ similarly to the term $\b{v}_1$. Specifically, let $\b{w}\triangleq \Pt_{t-1}^\perp \qt^t$
    \begin{align*}
        \b{v_2} &= \Pt_{t-1}\l(  (\hat \X  + \b{D})\b{w} - \b{D} \b{w}\r)\\
        &= \Pt_{t-1} \l(  (\|\b{w}\|_2/\sqrt{n}) \b{g} + g_0 \b{w}/\sqrt{n} - \b{D} \b{w} \r)\\
        &= (\|\b{w}\|_2/\sqrt{n}) \Pt_{t-1}\b{g} - \Pt_{t-1} \b{D} \b{w},
    \end{align*}
    where we used $\Pt_{t-1} \b{w} = 0$ in the last line as $\b{w} = \Pt_{t-1}^\perp \qt^t$. From the analysis of $\b{v}_1$, we have $\plimsup \|\b{w}\|_2 / \sqrt{n} = O(1)$. Furthermore
    \begin{align*}
        \Pt_{t-1} \b{g} &= \frac{1}{\sqrt{n}} \sum_{j=0}^{t-1} \tilde{g}_j \qt^j,
    \end{align*}
    where $(\tilde{g}_0,\ldots,\tilde{g}_{t-1})|_{\Ft_{t-1}} \distrib \mathcal{N}(0, (\Ft_{t-1}^\top \Ft_{t-1} /n)^{-1})$. By the induction hypothesis $\A_{1}(j),$ $ j\leq t$, we have $\plimsup \|\qt^{j}\|_\ell / n^{1/\ell} = O(1)$. Therefore
    \begin{align*}
        \plimsup \frac{1}{n^{1/\ell}} \|\Pt_{t-1} \b{g}\|_\ell &\leq \plimsup \frac{O(1)}{\sqrt{n}} \sum_{j=0}^{t-1} |\tilde{g}_j|\\
        &\leq \plimsup \frac{O(t)}{\sqrt{n}} \|\tilde{\b{g}}\|_2\\
        &= 0,
    \end{align*}
    where the last line follows from noting that $\E[\|\tilde{\b{g}}\|_2^2| \Ft_{t-1}] = {\rm Tr}((\Ft_{t-1}^{\top} \Ft_{t-1}/n)^{-1}) \underset{n\to \infty}{\to} {\rm Tr} (\tilde{\b{Q}}_{\leq t}^{-1})$, which is bounded, then combining the latter with Markov inequality. Therefore
    \begin{align*}
        \plimsup \frac{1}{n^{1/\ell}} \|\b{v}_2\|_\ell &\leq \plimsup \frac{1}{n^{1/\ell}} \|\Pt_{t-1} \b{D} \b{w}\|_\ell.
    \end{align*}
    Note that
    \begin{align*}
        \Pt_{t-1} \b{D} \b{w} &= \sum_{j=0}^{t-1} a_{tj} \qt^j, \quad a_{tj} \triangleq \sum_{k=0}^{t-1} (\Ft_{t-1}^\top \Ft_{t-1})^{-1}_{jk} \langle \qt^{k}, \b{D}\b{w}\rangle.
    \end{align*}
    As shown in the analysis of $\b{v}_1$, it suffices to show $|a_{tj}|=O(1)$ to show  $\plimsup \allowbreak \frac{1}{n^{1/\ell}} \|\Pt_{t-1} \b{D} \b{w}\|_\ell=O(1)$. We have
    \begin{align*}
        a_{tj} =\sum_{k=0}^{t-1} (\Ft_{t-1}^\top \Ft_{t-1}/n)^{-1}_{jk} \l\langle \frac{\qt^{k}}{\sqrt{n}}, \frac{\b{D}\b{w}}{\sqrt{n}}\r\rangle.
    \end{align*}
    As shown above $(\Ft_{t-1}^\top \Ft_{t-1}/n)^{-1}_{jk} = O(1)$, and by the induction hypothesis $\allowbreak \plimsup \allowbreak \|\qt^{k}\|_2 / \sqrt{n} = O(1)$, and as shown in the analysis of $\b{v}_1$, we have $\plimsup \|\b{D}\b{w}\|_2 /\sqrt{n} = 0$. It follows that $\plimsup |a_{t,j}|=0$. Henceforth, we have $\plimsup \frac{1}{n^{1/\ell}} \|\Pt_{t-1} \b{D} \b{w}\|_\ell=0$, which readily yields the bound $\plimsup \frac{1}{n^{1/\ell}}\|\b{v}_2\|_\ell = O(1)$. 
    \item We now deal with $\b{v}_3$. We have
    \begin{align*}
        \plimsup \frac{1}{n^{1/\ell}} \|\b{v}_3\|_\ell &\leq \plimsup  \sum_{j=1}^{t} |\bar{b}_{t,j}|\frac{1}{n^{1/\ell}}\|\qt^{j-1}\|_\ell.
    \end{align*}
    We have $|\bar{b}_{tj}|=O(1)$, and by the inductive hypothesis $\plimsup {n^{1/\ell}}\|\qt^{j-1}\|_\ell=O(1)$, therefore $\plimsup \frac{1}{n^{1/\ell}} \|\b{v}_3\|_\ell = O(1)$.

    \item We next deal with $\b{v}_4$. Let $\tilde{\b{U}}_k \triangleq [\ut^0|\ldots|\ut^{k+1}]$ and $\tilde{\b{R}}_k \triangleq [0|\rrt^1|\ldots|\rrt^k]$, and $\tilde{\b{Y}}_k = \tilde{\b{U}}_k + \tilde{\b{R}}_k$ for $k\geq 0$. Note that $\X \Ft_{t-1} = \tilde{\b{Y}}_{t-1}$. Introduce $\b{b}^t\triangleq (\Ft_{t-1}^\top \Ft_{t-1})^{-1}\Ft_{t-1}^\top \qt^{t} \in \mathbb{R}^{t}$. We have
    \begin{align*}
        \b{v}_4 &= \Pt_{t-1}^\perp \X \Pt_{t-1}\qt^t\\
                 &= (\b{I}_n - \Pt_{t-1}) \X \Pt_{t-1}\qt^t\\ 
                 &= (\b{I}_n - \Pt_{t-1})  \X \Ft_{t-1}(\Ft_{t-1}^\top \Ft_{t-1})^{-1}\Ft_{t-1}^\top \qt^t \\
                 &=\tilde{\b{Y}}_{t-1} \b{b}^t - \Pt_{t-1} \tilde{\b{Y}}_{t-1} \b{b}^t\\
                 &\triangleq \b{v}_{4,a} + \b{v}_{4,b}.
    \end{align*}
    Using similar arguments to previous ones, we have $\plimsup \|\b{b}^t\|_{\infty}=O(1)$. Furthermore, it follows from the inductive hypothesis $\A_1(j), j\leq t$ that $\plimsup \frac{1}{n^{1/\ell}} \|\ut^j\|_\ell=O(1)$  and $\plimsup \frac{1}{n^{1/\ell}} \|\rrt^j\|_\ell = O(1)$ for all $j\leq t$. Therefore
    \begin{align*}
        \plimsup \frac{1}{n^{1/\ell}} \|\b{v}_{4,a}\|_\ell &\leq \plimsup \|\b{b}^t\|_{\infty} \sum_{j=1}^{t}\frac{1}{n^{1/\ell}}\l(\|\ut^j\|_\ell + \|\rrt^{j-1}\|_\ell\r)\\
        &= O(1).
    \end{align*}
    We next bound $\b{v}_{4,b}$. Note that
    \begin{align*}
        \b{v}_{4,b} = \sum_{j=1}^{t} h_j \qt^{j-1}, \quad (h_1,\ldots,h_t) = (\Ft_{t-1}^\top \Ft_{t-1})^{-1}\Ft_{t-1}^\top \tilde{\b{Y}}_{t-1} \b{b}^t.
    \end{align*}
    Using the inductive hypothesis, we show similarly $\plim \max_{1\leq j \leq t} |h_j|=O(1)$, we then conclude using $\plimsup \frac{1}{n^{1/\ell}} \|\qt^j\|_\ell = O(1)$ that $\plimsup \frac{1}{n^{1/\ell}} \|\b{v}_{4,b}\|=O(1)$. Which ends the analysis of $\b{v}_4$.
    \item The analysis of $\b{v}_5$ is similar to $\b{v}_4$. Indeed, note that
    \begin{align*}
        \b{v}_5 &=  \Pt_{t-1}\X \qt^t\\
        &= \Ft_{t-1}(\Ft_{t-1}^\top \Ft_{t-1})^{-1}\Ft_{t-1}^\top \X \qt^t\\
        &= \Ft_{t-1}(\Ft_{t-1}^\top \Ft_{t-1})^{-1} (\X \Ft_{t-1})^\top\qt^t\\
        &= \Ft_{t-1}(\Ft_{t-1}^\top \Ft_{t-1})^{-1} \tilde{\b{Y}}_{t-1}^\top\qt^t.
    \end{align*}
    It suffices then to recall $\plimsup \|n(\Ft_{t-1}^\top \Ft_{t-1})^{-1} \|_{\infty}=O(1)$, and use the inductive hypothesis $\A_1(j), j\leq t$ to show $\plimsup \|\tilde{\b{Y}}_{t-1}^\top\qt^t\|_{\infty}=O(1)$, then conclude that $\plimsup \|\b{v}_5\|_\ell/n^{1/\ell}=O(1)$.
\end{enumerate}
This ends the proof of $\A_1(t+1)$. We next show $\A_2(t+1)$. We have
\begin{align*}
    &\frac{1}{\sqrt{n}}\|\ut^{t+1} - \u^{t+1}\|_2 \\
    &\leq \|\X\| \frac{1}{\sqrt{n}}\|\qt^t - \q^t\|_2 +\sum_{j=1}^{t} |b_{t,j} - \bar{b}_{t,j}| \frac{1}{\sqrt{n}} \|\q^{j-1}\|_2 + \sum_{j=1}^{t} |\bar{b}_{t,j}| \frac{1}{\sqrt{n}}\|\qt^{j-1} - \q^{j-1}\|_2\\
    &\triangleq a_1 + a_2 + a_3.
\end{align*}
\begin{enumerate}
    \item  We first deal with $a_1$. We have from Lemma \ref{chapter2:lemma:random:matrix:dense} that $\|\X\|\leq O(1)$ w.h.p. Therefore, we have w.h.p.
    \begin{align*}
        a_1 \leq O(1)  \frac{1}{\sqrt{n}}\|\qt^t - \q^t\|_2.
    \end{align*}
    Denote $\b{s}_i = (\u^0_i,\ldots,\u^t_i)$ and $\tilde{\b{s}}_i = (\ut^0_i,\ldots,\ut^t_i)$, so that
    \begin{align*}
        \frac{1}{n}\|\qt^t - \q^t\|_2^2 &= \frac{1}{n} \sum_{i=1}^{n} (q^t(\tilde{\b{s}}_i) - q^t(\b{s}_i))^2.
    \end{align*}
    Since $q^t$ is  polynomial, it belongs to ${\rm PL}(m)$ for some $m\geq 0$. Therefore
    \begin{align*}
        &\frac{1}{n} \sum_{i=1}^{n} (q^t(\tilde{\b{s}}_i) - q^t(\b{s}_i))^2 \\
        &\leq \frac{C}{n} \sum_{i=1}^{n}(1 + \|\tilde{\b{s}}_i\|_2^{m-1} + \|\b{s}_i\|_2^{m-1})^2 \|\tilde{\b{s}}_i - \b{s}_i\|_2^2\\
        &\leq \frac{C}{n} \sum_{i=1}^{n}(1 + \|\tilde{\b{s}}_i\|_2 + \|\b{s}_i\|_2)^{2m-1} \|\tilde{\b{s}}_i - \b{s}_i\|_2\\
        &\leq C \l( \frac{1}{n}\sum_{i=1}^{n}(1 + \|\tilde{\b{s}}_i\|_2 + \|\b{s}_i\|_2)^{4m-2}\r)^{1/2} \l(\sum_{i=1}^{n} \frac{1}{n}\|\tilde{\b{s}}_i - \b{s}_i\|_2^2 \r)^{1/2}\\
        &= C\l( \frac{1}{n}\sum_{i=1}^{n}(1 + \|\tilde{\b{s}}_i\|_2 + \|\b{s}_i\|_2)^{4m-2}\r)^{1/2} \l(\sum_{j=1}^{t} \frac{1}{n}\|\ut^j - \u^j\|_2^2 \r)^{1/2}.
    \end{align*}
    Using the inductive hypothesis $\A_2(j), j\leq t$, we have $\plimsup \l(\sum_{j=1}^{t} \frac{1}{n}\|\ut^j - \u^j\|_2^2 \r)^{1/2} \allowbreak= 0$.  Moreover, using $\A_1(j), j\leq t$, we have $\plimsup \l( \frac{1}{n}\sum_{i=1}^{n}(1 + \|\tilde{\b{s}}_i\|_2 + \|\b{s}_i\|_2)^{4m-2}\r)^{1/2} \allowbreak= O(1)$. Henceforth $\plim \frac{1}{n} \sum_{i=1}^{n} (q^t(\tilde{\b{s}}_i) - q^t(\b{s}_i))^2=0$, which yields $\plimsup  a_1 = 0$.
    
    \item We now deal with $a_2$. From the inductive hypothesis $\A_1(t)$, we have $\plimsup \frac{1}{\sqrt{n}} \|\q^t\|_2 = O(1)$, and by Proposition \ref{chapter2:state:evolution} we have $\plimsup b_{tj} = \bar{b}_{tj}$. Therefore $\plimsup a_2 = 0$.
    \item The term $a_3$ is dealt with identically to $a_1$ and $a_2$.
\end{enumerate}
This concludes the proof of $\A_2(t+1)$. It remains to prove $\A_3(t+1)$, which we do next. Let $\psi \in {\rm PL}(m)$, and use the same notation above. Then
\begin{samepage}
\begin{align*}
    &\l|\frac{1}{n} \sum_{i=1}^{n} \psi(\tilde{\b{s}}_i) - \frac{1}{n} \sum_{i=1}^{n}\psi(\b{s}_i)\r|\\
    &\leq \frac{C}{n} \sum_{i=1}^{n}(1 + \|\tilde{\b{s}}_i\|_2 + \|\b{s}_i\|_2)^{m-1} \|\tilde{\b{s}}_i - \b{s}_i\|_2\\
    &\leq C \l( \frac{1}{n}\sum_{i=1}^{n}(1 + \|\tilde{\b{s}}_i\|_2 + \|\b{s}_i\|_2)^{2m-2}\r)^{1/2} \l(\sum_{i=1}^{n} \frac{1}{n}\|\tilde{\b{s}}_i - \b{s}_i\|_2^2 \r)^{1/2}\\
    &= C \l( \frac{1}{n}\sum_{i=1}^{n}(1 + \|\tilde{\b{s}}_i\|_2 + \|\b{s}_i\|_2)^{2m-2}\r)^{1/2}\l(\sum_{j=1}^{t} \frac{1}{n}\|\ut^j - \u^j\|_2^2 \r)^{1/2}.
\end{align*}
\end{samepage}
Using the same arguments as in the proof of $\A_2(t+1)$, we conclude that 
$$\plimsup \l|\frac{1}{n} \sum_{i=1}^{n} \psi(\tilde{\b{s}}_i) - \frac{1}{n} \sum_{i=1}^{n}\psi(\b{s}_i)\r|=0.$$ 
Which ends the proof of $\A_3(t+1)$, and concludes the proof of Proposition \ref{chapter2:prop:appendix}.

\end{proof}
We now turn to the proof of Proposition \ref{chapter2:conjecture}.  Let $\b{Q}^{\rm IAMP}$ be the covariance matrix associated with the {\rm IAMP} algorithm in \cite{montanari2019optimization}. Let $\hat{\b{v}}, \tilde{\b{v}}$ as in (\ref{chapter2:sigma:hat}), (\ref{chapter2:sigma:tilde}) and let their associated  iterates be $\hat{\u}^t, \ut^t$, and let $\hat{\b{Q}}$ be the covariance matrix of the AMP iterates $\hat{\u}^t$. In order to apply Proposition \ref{chapter2:prop:appendix}, on the iterates $\hat{\u}^t, \ut^t$, we first need to verify (\ref{chapter2:Q:bound}) for $\hat{\b{Q}}$, which we do next. Property (\ref{chapter2:Q:limit}) is shown to hold for $\b{Q}^{\rm IAMP}$ in \cite{montanari2019optimization}, as the matrix $\b{Q}^{\rm IAMP}$ is diagonal with positive diagonal terms bounded above and below by constants. Moreover, it is shown in \cite{ivkov2023semidefinite}(Claim B.5) that $\|\hat{\b{Q}}_{\leq t} - \b{Q}^{\rm IAMP}_{\leq t}\|_{\rm fro}$ can be taken arbitrarily small for any $t\in [T]$ and fixed number of iterations $T$. Hence, taking $\sup_{t\leq T} \|\hat{\b{Q}}_{\leq t} - \b{Q}^{\rm IAMP}_{\leq t}\|_{\rm fro}$ small enough, the matrix $\hat{\b{Q}}$ satisfies (\ref{chapter2:Q:bound}). We then have
\begin{align*}
    \frac{\|\hat{\b{v}}- \tilde{\b{v}}\|_2}{\sqrt{n}} &\leq \sqrt{\delta} \sum_{t=1}^{\lfloor \bar q  / \delta \rfloor} \frac{\|q^t(\hat{\u}^0,\ldots, \hat{\u}^t) - q^t(\ut^0,\ldots,\ut^t)\|_2}{\sqrt{n}}.
\end{align*}
Following the notation in the proof of Proposition \ref{chapter2:prop:appendix}, denote by $\hat{\q}^t$ the vector $q^t(\hat{\u}^0,\ldots,\hat{\u}^t)$, and let $\hat{\b{s}}_i = (\hat{\u}^0_i,\ldots,\hat{\u}^t_i)$. Let $t\in [ \lfloor \bar q  / \delta \rfloor]$. Since $q^t$ is a polynomial with constant degree and coefficients, we have $\q^t\in {\rm PL}(m)$ for some constant $m\geq 0$. We then have

\begin{align*}
    &\frac{\|q^t(\hat{\u}^0,\ldots, \hat{\u}^t) - q^t(\ut^0,\ldots,\ut^t)\|_2^2}{n} \\
    &\leq \frac{C}{n} \sum_{i=1}^{n}(1 + \|{\hat{\b{s}}}_i\|_2 + \|\b{\tilde{s}}_i\|_2)^{m-1} \|\hat{\b{s}}_i - \tilde{\b{s}}_i\|_2\\
    &\leq C \l( \frac{1}{n}\sum_{i=1}^{n}(1 + \|\hat{\b{s}}_i\|_2 + \|\tilde{\b{s}}_i\|_2)^{2m-2}\r)^{1/2} \l(\sum_{i=1}^{n} \frac{1}{n}\|\hat{\b{s}}_i - \tilde{\b{s}}_i\|_2^2 \r)^{1/2}\\
    &= C \l( \frac{1}{n}\sum_{i=1}^{n}(1 + \|\hat{\b{s}}_i\|_2 + \|\tilde{\b{s}}_i\|_2)^{2m-2}\r)^{1/2}\l(\sum_{j=1}^{t} \frac{1}{n}\|\hat{\u}^j - \ut^j\|_2^2 \r)^{1/2}.
\end{align*}
Applying $\A_1(j),\A_2(j),j\leq t$ from Proposition \ref{chapter2:prop:appendix} to the above yields
\begin{align*}
    \plim \frac{\|q^t(\hat{\u}^0,\ldots, \hat{\u}^t) - q^t(\ut^0,\ldots,\ut^t)\|_2^2}{n} = 0,
\end{align*}
and therefore
\begin{align*}
    \plim \frac{\|\hat{\b{v}}- \tilde{\b{v}}\|_2}{\sqrt{n}} =0,
\end{align*}
which ends the proof of Proposition \ref{chapter2:conjecture}.

\end{proof}

\printbibliography

\end{document}